\documentclass[a4paper,12pt]{article}
\usepackage[utf8]{inputenc}
\usepackage{natbib}
\usepackage{amsfonts, amsmath, amssymb, amsthm}
\usepackage{caption}
\usepackage{subcaption}
\usepackage{longtable}
\usepackage{graphicx}
\usepackage{fancyhdr}
\usepackage{lastpage}
\usepackage{verbatim}
\usepackage{multicol, threeparttable}

\usepackage{fullpage}
\usepackage{breqn}

\usepackage{setspace}

\AtBeginDocument{%
  \setlength{\abovedisplayskip}{6pt}
  \setlength{\belowdisplayskip}{6pt}
  \setlength{\abovedisplayshortskip}{6pt}
  \setlength{\belowdisplayshortskip}{6pt}
}

\usepackage[colorlinks,filecolor=magenta, citecolor=cyan,urlcolor=blue,bookmarks=false,hypertexnames=true]{hyperref} 

\usepackage[a4paper,margin=1in]{geometry}

\usepackage[dvipsnames]{xcolor}

\newtheorem{definition}{Definition}

\newcommand{\fignote}[1]{%
    \captionsetup{font=footnotesize, justification=justified, singlelinecheck=false}%
    \caption*{\textit{Note:} #1}%
}

\title{\vspace{-2.5cm}{\color{MidnightBlue} \textbf{How much inflation can fiscal policy create? Separating household heterogeneity and liquidity}}}

\author{Matthias H\"ansel\thanks{Department of Economics and Business Economics, Aarhus University. I thank Lars Ljungqvist and David Domeij for guidance and support as well as Klaus Adam, Antoine Camous, Davide Debortoli, Tore Ellingsen, Richard Foltyn, Philipp Hochmuth, Mathias Klein, Peter Kress, Kieran Larkin, Ralph Luetticke, Markus Pettersson, Maximilien Pointurier, Zoltan Racz, Victor R\'ios-Rull, Ettore Savoia, an anonymous referee and the audiences at various conferences and seminars for helpful comments. Funding provided by the Jan Wallander and Tom Hedelius Foundation during the time of my PhD is gratefully acknowledged. 
This paper supersedes my earlier job market paper titled ``Idiosyncratic Risk, Government Debt and Inflation'' \citep{hansel2024}.
Contact:  m.haensel@econ.au.dk, Universitetsbyen 51, Building 1815, 8000 Aarhus.}
}
\date{May 30, 2026}
    
\begin{document}

\maketitle

\begin{abstract}
\noindent
A key determinant of monetary-fiscal interactions in Heterogeneous Agent New Keynesian (HANK) models is the liquidity value of public debt and its effect on interest rate dynamics. Yet, while household heterogeneity shapes this channel, it doesn't pin it down. In both analytically tractable and quantitative 2-asset HANK models, asset market assumptions unrelated to standard micro moments give rise to disparate implications of fiscal policy for inflation, as well as model determinacy and fiscal self-financing. To address this issue, I propose a simple model extension and discipline it with macro-level evidence on the relationship between public debt and treasury returns. This moderates the inflationary impact of public debt dynamics but does not render it negligible. After large fiscal shocks, it can still generate persistently elevated ``last mile'' inflation.

\vspace{.5cm} \noindent \textit{Keywords:} Monetary policy, Fiscal Policy, Inflation, HANK
\vspace{.5cm} \\ \noindent \textit{JEL Classification:}  E31, E52, E63
\end{abstract}
\bigskip

\pagenumbering{gobble}
\newpage
\pagenumbering{arabic}

\section{Introduction}

Recent Heterogeneous Agent New Keynesian (HANK) models generate rich monetary-fiscal interactions due to their departure from Ricardian equivalence, a property that gained them further interest in light of the post-Covid inflation shock. 
Compared to standard Representative Agent (RA) frameworks, these models introduce two related but distinct features: constrained households with high marginal propensities to consume (MPCs) out of transfers, and an explicit usefulness of government debt for insuring idiosyncratic risk \citep[see, e.g.,][]{kaplanEtAl2018b}.

In this paper, I argue that the second margin, the liquidity value of public debt, should be subject to additional discipline for conducting joint analyses of monetary and fiscal policy. HANK models are typically calibrated to be consistent with various micro moments, e.g., regarding the wealth distribution or MPCs. However, while the fact that households hold relatively little liquid wealth is consistent with fundamental liquidity scarcity, it does not require it.
But this distinction is crucial for the aggregate effects of fiscal policy.

To illustrate this point, I first analyze a Tractable HANK (THANK) model \`a la \cite{bilbiie2025} under two different assumptions on the supply of liquid assets. In the baseline case, I assume that the asset market is \emph{separated} such that public debt constitutes the only net supply of liquid assets, while stocks can only be held as illiquid assets. In a counterfactual, I assume that the asset market is \emph{integrated}, meaning that stocks are also useful to provide liquidity, as e.g. in the two-asset HANK model by \cite{auclertEtAljpe}. 
Either case can generate \emph{identical} micro moments in the model's stationary equilibrium.

The aggregate implications of these two cases, however, differ substantially. The baseline model displays the extended scope for monetary-fiscal interactions identified by the recent HANK literature: Following a debt-financed fiscal stimulus, we observe persistently elevated output and, even more noticeably, inflation. If monetary policy is not too hawkish, the shock can even be self-financing as in \cite{angeletos_selffinance}, in that this boom automatically generates enough tax revenue to pay for itself without further fiscal adjustments. Furthermore, the model's determinacy properties deviate from the standard RA case, whereby a unique stable equilibrium can be supported by a wider set of policy rules outside the regimes identified by \cite{leeper1991}. Despite initially identical micro-heterogeneity, none of that is the case for the counterfactual with integrated asset markets.

Subsequently, I turn to a quantitative 2-asset HANK model in which, following \cite{kaplanEtAl2018} and \cite{bayerEtAlaer}, households' illiquid assets consist of capital.
While it may be tempting to ascribe the strong implications of the asset market structure to the relatively simple set-up of the THANK model,  one can find close equivalents in this substantially richer framework. 

 Why does the asset market structure matter so much? 
 The reason is that it determines how public debt affects aggregate demand and thus the ``natural'' and ``neutral'' rates of interest.\footnote{The ``natural'' rate of interest is typically defined as the nominal rate consistent with a central bank's inflation target in the medium- to long term. In contrast, the ``neutral rate'' is sometimes designated as the interest rate that would be consistent with target inflation in the short or medium run \citep[see e.g.][]{obstfeld2023}. I will similarly use the term ``natural rate'' for the long-run (steady state) equilibrium rate and ``neutral rate'' when it comes to shorter-run dynamics.}
  If government bonds are traded on the liquid asset market that is \textit{separated} from the ones for other assets, a higher stock of public debt will necessarily result in better insurance of households against idiosyncratic risk. This drives up aggregate demand in the short run and the natural rate in the long run. In fact, in either model, higher government debt supply then increases the natural rate and in turn liquid bond rates \textit{much more} in the long run than suggested by relevant empirical evidence \citep[e.g.,][]{laubach2009}. 
  In contrast, if the asset market is \textit{integrated}, a higher supply of public debt crowds out the demand for other assets, so that aggregate demand and asset returns do not need to adjust much. In that case, the model-implied responsiveness of the natural rate to public debt is \emph{much weaker} than suggested by the aforementioned evidence.

  While the asset market structure cannot be pinned down using household-side information alone, its importance nevertheless connects to micro moments. In the context of the quantitative 2-asset model, generating the above-mentioned results requires a substantial return gap between liquid and illiquid assets. This is a key margin to combine high MPCs with a reasonable distribution of aggregate wealth in HANK models, as already highlighted in the previous literature \citep[c.f.][]{kaplanEtAl2022}. Thus, consistency with certain micro moments induces a strong liquidity margin of Ricardian non-equivalence, \emph{conditionally} on certain asset market assumptions.
  Auxiliary exercises further indicate that a richer specification of illiquid asset adjustment cannot easily overcome this link.\footnote{Convex costs for illiquid asset adjustment can in principle weaken the link between the average MPC and an excess natural rate sensitivity. But as I find in these auxiliary analyses, doing so comes at the cost of a poor fit with other relevant micro moments.}

  The issue remains to discipline the HANK model's asset market structure and in turn the liquidity effect of public debt. I do so by using a simple model extension that allows to move in between the polar cases of separated and integrated asset markets \textit{without} changing the model's steady state. It can be calibrated using a single macro-level statistic, the long-run effect of public debt on treasury returns, of which various estimates are available in the literature \citep[see, e.g.,][]{rachelEtAl2019}. Reassuringly, this seems to result in reasonable short-run dynamics of public debt's liquidity value, as a comparison with empirical IRFs suggests.\footnote{ While my extension also facilitates calibrations targeting empirical IRFs, the long-run approach usefully disentangles the liquidity value from short-run economic forces.}

  Ultimately, is public debt's liquidity value still a relevant channel of monetary-fiscal interaction in the properly disciplined HANK model? While its effects appear small for moderate fiscal shocks under conventional monetary policy, it matters noticeably for the HANK's determinacy properties and can amount to a relevant magnitude either if monetary policy is dovish or in the aftermath of big fiscal expansions.
 To illustrate the latter, I extend the 2-asset HANK model with an Effective Lower Bound (ELB) and filter series of business cycle shocks that make it align with several aggregate US time series between 2020 and 2024. In the calibrated baseline model, the respective ``debt inflation'' does not play a big role for generating the inflation peak in 2022, but causes the persistent ``last mile'' inflation in 2023 and afterwards. 
 With fully integrated asset markets, the HANK model would assign a bigger role to loose monetary policy and have predicted less inflation going forward.

\subsection{Relation to the literature}\label{sec:literature}

This paper connects to a long tradition in macroeconomics studying monetary-fiscal interactions, going back to the seminal works of \cite{sargentEtAl1981} and \cite{leeper1991}. 
\cite{leeperEtAl2016} and \cite{cochrane2023} offer summaries of this literature, including its modern incarnation as Fiscal Theory of the Price Level (FTPL). As already alluded to above, these works typically required public debt to be ``unfunded'' to become a concern for monetary policy. 
A relevant recent contribution of this strand is \cite{bianchiEtAl2023}, who find unfunded fiscal policy important to explain inflation persistence in the US and study the post-pandemic inflation.

Naturally, my work connects most closely to the sprawling HANK literature. In particular, the literature has found these frameworks to have substantially different implications of fiscal policy for real and nominal outcomes compared to the RA benchmark.
Firstly, the presence of constrained households with high MPCs can substantially amplify the aggregate effects of expansionary fiscal policy, see for example \cite{hagedornEtAl2019} and \cite{auclertEtAljpe}. 
These authors, together with \cite{hagedorn2021} and \cite{angeletosEtAl2024} also emphasize that these frameworks give rise to different patterns of equilibrium determinacy. \cite{angeletosEtAl2024} further argue that in response to fiscal shocks, HANK models can generate as much inflation as RA models under the FTPL.
Compared to these works, my paper explains that in addition to households' consumption behavior, the market for liquid assets can be crucial for HANK model's inflation responses and determinacy properties. It can (and arguably should) be considered separately from MPCs.

Some HANK papers actually focus on how public debt's liquidity value matters for the aggregate effects of fiscal policy. This includes \cite{bayerEtAl2023b}, who focus on real outcomes, and the independent contemporary work of \cite{camposEtAl2024}, who analyze some of its implications for monetary policy.
Additionally, \cite{cantoreEtAl2025} study a New Keynesian model with limited heterogeneity and analyse how the presence of an endogenous public debt liquidity premium affects the transmission of different business cycle shocks.
My work clarifies that their results depend importantly on the assumption of certain asset market structures in addition to household heterogeneity itself.\footnote{While not strictly ``HANK'' papers, \cite{aiyagariEtAl1998}, \cite{challeEtAl2011} and \cite{aguiarEtAl2024aer} study flexible-price incomplete markets model in which the presence of idiosyncratic risk gives rise to important liquidity effects of public debt. My work similarly suggests that in order to assess the magnitude of their respective outcomes, it is valuable to consider the empirical counterparts of public debt's accompanying interest rate effects.}

My insights on the importance of the asset market structure for liquid rate dynamics additionally bear resemblance to the findings by \cite{ZochEtal2023}, who study a 2-asset HANK model with explicit financial intermediation. Comparing their structure with alternative settings, they find the calibration of the asset market to be important for the real effects of different policy shocks. Yet, they do not study inflation outcomes and assume real returns on liquid assets to be fixed.\footnote{Their insights may thus be the reverse of mine: If changing liquidity supply would necessitate substantial interest rate adjustments for asset market clearing but this is prevented, then other parts of the economy have to adjust strongly.}

Finally, my work also connects to a set of previous studies analyzing fiscal policy in other settings deviating from Ricardian Equivalence. 
Related inflationary effects of ``funded'' government debt were also noticed by \cite{ascariEtAL2013} and \cite{aguiarEtAl2023} in the context of Overlapping Generations (OLG)-models with nominal rigidities and by \cite{linnemannEtAl2010} in a RA New Keynesian framework assuming that public debt provides transaction services. Besides the different micro-foundations, my work also employs a richer, more quantitatively oriented model. 
\\~\\
The remainder of the paper is structured as follows: Section \ref{sec:thank} presents the THANK model and the respective analyses. Section \ref{sec:two_asset_hank} describes the core features of the quantitative 2-asset HANK model, which will be used in Section \ref{sec:fisc_imps} to demonstrate equivalents to the THANK results. Section \ref{sec:liq_inflation} expands on the mechanism and proposes a calibration of the asset market structure. Afterwards, Section \ref{sec:mon_imps} evaluates the implications of the asset market calibration. Section \ref{sec:conclusion} concludes.

\section{Monetary-fiscal interactions in a tractable HANK}\label{sec:thank}

To illustrate the importance of the asset market structure for monetary-fiscal interactions, I begin by studying a simple ``Tractable HANK'' (THANK) model based on \cite{bilbiie2025}. Within this model, I will consider two different asset market structures, which, however, can generate an observationally equivalent steady state (SS). 

\subsection{Households and labor supply}

The model features a unit mass of households discounting the future at rate $\beta$. In contrast to textbook business cycle models, these switch between two different types called ``Savers'' ($S$) and  Hand-to-Mouth ($H$). 
In particular, $H$ and $S$ households stay within their group with probabilities $h$ and $s$, respectively. The remaining mass of households switch to the other group. These probabilities are time-invariant, so the mass of $H$ households is always $\lambda := (1-s)/(2-s-h)$. 

$S$ and $H$ households differ in the following way: $S$ households can save by investing in a liquid nominal asset $A$ and an illiquid stock $I$, while $H$ households cannot save at all. Additionally, the $S$ household switching to $H$ can take their share of liquid savings with them, which the $H$ agents can consume (but not reinvest). As in \cite{bilbiie2025}, economic decisions for all households are made by a ``family head'' who pools resources within every group of households and maximizes average utility given the constraints of the respective groups.
The utility maximization problem of the family head can then be written recursively as 
\begin{align}
    W(A_t, I_t) &= \max_{C_t^S, C_t^H, A_{t+1}, I_{t+1}} \lambda u(C_t^H) + (1-\lambda)u(C_t^S) + \beta \mathbb{E}_t W(A_{t+1}, I_{t+1}) \label{teq:W}\\
    \text{s.t. }~~ &C_t^S + A_{t+1} + I_{t+1} = (1-\tau^y)w_t N_t + R_t^I I_t + s R_t^A A_t - \tau_t + T_t \nonumber\\
    &C_t^H = (1-\tau^y)w_t N_t\omega + \frac{(1-\lambda) (1-s)}{\lambda}R_t^A A_{t} + \bar{\tau}/\lambda + T_t. \nonumber
\end{align}
Both resource constraints for $S$ and $H$ households are stated in real terms. $\omega\in(0,1)$ is a labor productivity parameter for $H$. $\tau_y$ is a proportional tax on labor income that will stay constant for the dynamic analysis below. $\tau_t$ is a lump-sum tax on the $S$ type, while $\bar{\tau}$ is a lump-sum welfare benefit paid to the $H$ agents. While $\tau_t$ will be more relevant for fiscal consolidation in the analysis below, having a proportional tax $\tau^y$ will be important for analyzing the mentioned self-financing result. $T_t$ is a lump-sum transfer to all households that will always equal $0$ except in Subsections \ref{subsec:t_fiscshock} and \ref{subsec:t_selffinance}. Note that the returns to liquid asset holdings are scaled by $s$ and $\frac{(1-\lambda) (1-s)}{\lambda}$ due to the switching. 

From maximization problem \eqref{teq:W}, we can derive the following Euler equations for the optimal consumption and saving decisions of the family head:
\begin{align}
    (C_t^S)^{-\sigma} &= \beta \mathbb{E}_t R_{t+1}^A\left[s(C_{t+1}^S)^{-\sigma} +(1-s)(C_{t+1}^H)^{-\sigma} \right]  \label{teq:EU_A}\\
    (C_t^S)^{-\sigma} &= \beta  \mathbb{E}_t R_{t+1}^I(C_{t+1}^S)^{-\sigma} \label{teq:EU_I}
\end{align}
We note that the liquid asset Euler equation features the switching probabilities and thus a precautionary motive. Given that the asset $I$ is useless for insuring consumption in the $H$ state, its Euler equation is standard.

Regarding labor supply, I again follow the THANK literature and assume for tractability that hours worked are chosen by a representative union such that a standard labor supply condition holds for the aggregate economy, i.e.,
\begin{align}
    C_t^{-\sigma} (1-\tau^y) w_t = \varsigma N_t^\phi, \label{teq:N}
\end{align}
where $C_t = \lambda C_t^H + (1-\lambda)C_t^S$ is aggregate consumption. Aggregate labor supply is then given by $L_t = (1-\lambda + \lambda \omega)N_t$.

\subsection{Supply side}

The production side is as in textbook New Keynesian models. It features a unit mass of monopolistically competitive firms that produce differentiated goods using labor only. The production function is given by $Y_t = L_t$. Firms set prices in a staggered fashion à la Calvo, which will give rise to a standard New Keynesian Phillips curve in the analyses below. Given that this part of the model is well-known, I omit related derivations. Note that in the SS, the real wage is given by $w_{ss} = 1/\mu$ where $\mu$ is the SS markup. The firms are owned by a diversified mutual fund and their profits paid out as dividends $d_t$ to the holders of the mutual fund's shares $M$, see Subsection \ref{subsec:Tasset} below for more details.

\subsection{Government}

The government consists of a fiscal authority and a monetary authority. 
Outside of SS, the monetary authority sets the nominal interest rate $i_t$ on government debt according to a standard Taylor rule $\hat{i}_t = \phi_\pi \pi_t$.
In the long run (i.e., in the SS), it adjusts the intercept of its policy rule so that $\pi_{ss}=0$.

The fiscal authority, in turn, levies the proportional income tax $\tau^y$ and the lump-sum tax $\tau_t$ on the rich household. The government budget constraint is given by
\begin{align*}
B_{t+1} = R_t^B B_t + T_t + \bar{\tau} + G - \tau^y w_t L_t - (1-\lambda)\tau_t 
\end{align*}
where $B_t$ denotes the stock of government bonds and $G$ exogenous government spending. Below, I shall assume that in the long run, the government follows a fiscal rule that keeps $\bar{\tau}$, $G$ and $B$ to be constant ratios of GDP, i.e. $B_{ss} = \mathcal{B} Y_{ss}$, $\bar{\tau} =  \mathcal{T} Y_{ss}$ and $G = \mathcal{G} Y_{ss}$. The nominal rate interest on public debt is set by the central bank so that $R_t^B := (1+i_{t-1})/(1+\pi_{t})$. The constant tax rate $\tau^y$ is set so that $\tau^y w_{ss} L_{ss} = \bar{\tau}+G$. Debt service and a shock to $\bar{\tau}$ considered below are financed via the lump-sum tax $\tau_t$.\footnote{An argument for that assumption could be that the levels of income taxes may be harder to change in the short run than other revenue sources. Keeping the distortionary tax rate $\tau^y$ constant makes the analyses more tractable, yet studying self-financing necessitates some form of proportional tax to allow tax revenues to depend on economic activity. Hence, the introduction of a positive $\tau^y$ that finances the long-run level of the benefit and government consumption.} For the dynamic analysis around the steady state, we will similarly assume $G$ to be constant and $T_t = 0$ except for the analysis on the effects of the one-time transfer shock below.

\subsection{Asset market}\label{subsec:Tasset}

The asset market structure is the crucial element of this analysis. To sharply characterize its importance, I initially focus on two polar cases that I call the \emph{separated} and \emph{integrated} asset markets. In both, households don't hold public debt directly. Instead, their liquid assets will be supplied by a risk-neutral competitive Liquid Asset Fund (LAF), which, in the two cases, differs in its capabilities.
As I will argue in Section \ref{sec:liq_inflation} below, the ``truth'' is likely in between.

\subsubsection{Baseline: Separated asset markets}

As a baseline, I consider the case in which the LAF can invest \emph{only} in government bonds. In turn, we have the liquid asset market clearing condition $(1-\lambda)A_t = B_t$ and no-arbitrage condition  
\begin{align}
    \mathbb{E}_t R^A_{t+1} = \varphi\mathbb{E}_t R^B_{t+1} = \varphi\mathbb{E}_t \frac{1+i_{t}}{1+\pi_{t+1}} . \label{teq:NA}
\end{align}
Above, $1-\varphi \in (0,1) $ is an iceberg cost of liquidity provision, reflecting that the LAF may incur costs to provide liquidity to households. While assuming $\varphi=1$ would actually facilitate the analysis for this model version, the parameter will be useful by allowing the same non-stochastic SS for either asset market structure as I explain below. In turn, the asset market clearing condition for illiquid assets is $(1-\lambda)I_t = M Q_{t-1}$ and $\mathbb{E}_t R^I_{t+1} = \frac{Q_{t+1}+d_{t+1}}{Q_t}$, where $Q$ denotes the price of the mutual fund stocks and $d_t$ the dividends paid out by the firms. Note that in this model version, the only role the illiquid Euler equation \eqref{teq:EU_I} plays is determining the price of the mutual fund stocks and that allocation-wise, we could equivalently assume profits to be paid out lump-sum to the $S$ households. 
Also notice that in equilibrium, this set-up is equivalent to one in which households hold their liquid assets in government bonds directly (but potentially incur a cost for doing so) instead of them being ``passed through'' the LAF.

\subsubsection{Alternative: Integrated asset markets}

In the alternative asset market structure, I allow the LAF to not only hold government bonds but also to invest in or short-sell the mutual fund shares (firm stocks). In that case, in addition to \eqref{teq:NA}, the following no-arbitrage conditions must also hold in equilibrium:
\begin{align}
 \mathbb{E}_t  \frac{1+i_{t}}{1+\pi_{t+1}} = \mathbb{E}_t \frac{Q_{t+1}+d_{t+1}}{Q_t}.
\end{align}
Additionally, there is now a single asset market clearing condition instead of two separate ones for liquid and illiquid assets. It is given by $(1-\lambda) A_t = B_t + (M Q_{t-1} - (1-\lambda)I_t) $
and implies that liquid asset holdings generically can (but do not have to) deviate from the stock of public debt.\footnote{I note that since the mutual fund shares are in positive net supply, having $(1-\lambda)A_t > B_t$ need not imply that the $S$ households hold a negative amount of illiquid assets. While the equilibrium might involve the LAF short-selling shares, this seems arguably not unrealistic for a financial intermediary.}

Beyond this THANK model, I take \emph{separated} asset markets to refer to market arrangements in HA models in which there is a limited net supply of non-productive liquid assets (including public debt), the market for which has to clear independently. This means that liquid assets cannot directly be held as or crowd out other illiquid assets. A common assumption \citep[as e.g. in][]{kaplanEtAl2018} is that public debt is the only net supply of liquidity available to households, while illiquid assets consists only of capital or equity.

In contrast, I consider models as featuring an \emph{integrated} asset market if there is only a single aggregate asset market clearing condition despite the presence of liquid and illiquid assets from the households' perspective. This is possible in the presence of exogenous return wedges such as $\varphi$ in my THANK model. A prominent example in the literature is the 2-asset HANK setup in \cite{auclertEtAljpe}.\footnote{As explained in their Appendix F.1 (``Allowing for valuation effects'') and G.2 (``Reducing to a single asset market clearing condition''), their 2-asset HANK model features only a single asset market clearing condition.} 

The key difference between both setups is the following: With a separated asset market, liquidity is fundamentally scarce and increases in public debt have to be accompanied by higher household liquid asset holdings in equilibrium.
In an integrated asset market, the potential supply of liquidity can be large even though households hold only little of it. If the supply of government debt goes up (down), it can crowd out (in) other types of assets so that households' liquidity holding do not need to change. In different set-ups in the literature, the additional asset beyond public debt is sometimes equity or capital. This THANK model considers the former case and the subsequent 2-asset HANK model the latter, with the asset market structure being crucial for equilibrium interest rates and inflation in either case.

\subsection{Steady State analysis}\label{sec:thank_ss}

The non-stochastic SS of the baseline model version with separated asset markets is easily characterized. In that case, we know $(1-\lambda)A_{ss} = B_{ss}$, so that $C_{ss}= (1-(1-\varphi)R^B \mathcal{B}-\mathcal{G})Y_{ss}$ and given the assumptions on the income tax level, $\tau^y = \frac{(\mathcal{T}+\mathcal{G})Y_{ss}}{w_{ss}L_{ss}} = (\mathcal{T}+\mathcal{G})\mu$. Using this in the labor supply condition, one can derive $Y$ and $N$ in terms of parameters. In turn, one can solve for the SS liquid return $R_t^A$. As I show in Appendix \ref{app:SS_derivations}, the resulting solution is not tractable in general, but it will generically depend on $\mathcal{B}$, $s$ and $\lambda$.
 However, a simple expression can be obtained for the instructive special case of log utility and both $\omega \rightarrow 0$ and $\mathcal{T} \rightarrow 0$, i.e. if $H$ households have no labor- or transfer income and can only consume the liquid wealth they receive from switching $S$ households. In that case, we get the following relation between the SS liquid return and the debt-to-GDP ratio $\mathcal{B}$:
\begin{align}  
R^A =  \frac{1}{\beta}\frac{1-\lambda}{(s-\lambda/\varphi)} - \frac{\lambda(1-\mathcal{G})}{(s-\lambda/\varphi)\mathcal{B}}  \label{eq:thank_RA}
\end{align}
 Focussing on the case $s>\lambda/\varphi$, it can easily be shown that, as in richer incomplete markets models, $R^A$ is increasing and concave in the debt-to-GDP ratio $\mathcal{B}$. Additionally, as I show in Appendix \ref{app:SS_derivations}, we have $\frac{\partial^2 R^A}{ \partial \mathcal{B} \partial s}<0$, i.e., the sensitivity of the liquid return to the debt-to-GDP ratio is decreasing in $s$. This means that in the baseline model version with separated asset markets, if households are less likely to switch to the constrained $H$ state, the liquid return (and thus the long-run natural rate) reacts less to the Debt-to-GDP ratio.
 Indeed, equation \eqref{eq:thank_RA} has the implication that conditional on the parameters $\mathcal{G}$ and $\varphi$, which can be pinned down from macro data, the only way to affect the sensitivity of $R^A$ with respect to $\mathcal{B}$ is by changing micro heterogeneity (through $s$ or $\lambda$).

Under the alternative \emph{integrated} model, the SS is determined somewhat differently. Recall that in this set-up, the LAF can buy (short-sell) stocks if liquid asset demand exceeds (falls short of) the stock of public debt. 
From the resulting additional no-arbitrage condition between the stocks and government bonds and Euler equation \eqref{teq:EU_I}, it immediately follows that $R^A = \varphi/\beta$, implying that the long-run natural rate is \emph{completely independent} of the debt-to-GDP ratio and household heterogeneity. The LAF can substitute stocks for bonds and the equilibrium return is determined by the illiquid asset Euler equation.
Given $R^A$, \eqref{teq:EU_A} then pins down the SS ratio and it is not hard to see that if one chooses $\varphi$ such that
\begin{align}
    \varphi = \frac{1}{((1-s)\eta^{-\sigma} + s)}, \label{teq:varphi}
\end{align}
where $\eta = \tilde{C}^H/\tilde{C}^S$ is the SS consumption ratio from the \emph{baseline} model version with separated assets markets, then identical SS allocations can be obtained under either asset market structure. Hence, the respective SS micro moments cannot distinguish between the two asset-market structures. This already illustrates that while the presence of idiosyncratic risk and household heterogeneity may importantly shape the relationship between public debt and the natural rate of interest, it does so only conditional on the asset market structure. Relating just to respective household moments is not informative about the latter.
However, for given micro moments, the relation between public debt and interest rates in the long run can potentially be informative about the asset market structure.

\subsection{The asset market and aggregate demand}

To explore the implications of the asset market structure for the monetary and fiscal policy in the dynamic economy, I will linearize the economies' equilibrium around a non-stochastic SS with $B_{ss}=0$, $\omega > 0$ and $\mathcal{T}>0$ (with $\varphi$ chosen so that the SS allocation is identical irrespective of the asset market). Furthermore, it will be assumed that the transfer $\bar{\tau}$ remains fixed outside the steady state. As I show in Appendix \ref{app:loglin}, the resulting log-linearized IS condition for the baseline model version with separated asset markets is given by
\begin{align}
   &\hat{y}_t = \frac{\Gamma}{\Theta} \mathbb{E}_t \hat{y}_{t+1} - \frac{1}{\sigma \Theta}(\hat{i}_t - \mathbb{E}_t \pi_{t+1}) + \frac{\Gamma_b}{\Theta}\tilde{b}_{t+1} + \frac{\Theta_b}{\Theta}\tilde{b}_t \label{teq:IS_seg}
\end{align}
 if the government sets transfer $T$ to 0. $\hat{y}_t$ denotes the log-deviation of output and $\tilde{b}_t := B_t/Y_{ss}$ the relative deviation of public debt (around $B_{ss}=0$). Equation \eqref{teq:IS_seg} is derived from the log-linearized equation \eqref{teq:EU_A}, where $\Theta,\Theta_b, \Gamma$ and $\Gamma_b$ are composite parameters depending on household heterogeneity.
 $\Theta$ and $\Theta_a$ reflect how the output level and liquid savings reflect the distribution of resources in period $t$, while $\Gamma$ and $\Gamma_b$ capture how they affect the $S$ households' expected marginal utility in $t+1$ (and thus an insurance motive).

 If the government sets $\tilde{b}_t = 0~\forall t$, \eqref{teq:IS_seg} becomes isomorphic to the THANK IS condition derived by \cite{bilbiie2025} in which the forward-looking terms of the standard IS condition are scaled by additional terms. I shall similarly restrict attention to parameterizations such that $\Theta > 0$, which implies $\Gamma>0$. Even in that case, it is generally ambiguous whether $\Gamma/\Theta$ is greater or smaller than $1$. In the parlance of \cite{bilbiie2025}, the relative size of these terms determine whether inequality is pro- or countercyclical and if the share of the time-invariant transfer in the $H$ agents' income is relatively large, then $\Gamma/\Theta < 1$ and the IS equation is \emph{discounted}. Many recent analyses of monetary-fiscal interactions use models where a discounted IS condition is present \citep[e.g.,][]{angeletos_selffinance}, and assuming $\Gamma/\Theta < 1$ will be helpful for some analytical arguments (but not strictly necessary for the key insights). Additionally, we notice that the time path of public debt $\tilde{b}_t$ enters the aggregate demand relation: This is because higher public debt increases consumption sharing between $S$ and $H$ agents.\footnote{From \eqref{teq:IS_seg}, it is also clear that in the THANK model, a sufficiently sophisticated central bank reaction function could in principle eliminate both IS discounting/amplification and any effects of public debt by making $\hat{i}_t$ depend in just the right way on $\mathbb{E}_t \hat{y}_{t+1}$ and $\tilde{b}$.\label{footnote:rate}}

Turning to the alternative model version with integrated asset markets, the IS condition initially appears more involved, since assuming $\tilde{b}_t=0$ does not imply $\tilde{a}_t=0$. In general, it will be of the form
\begin{align}
    \Theta\hat{y}_t = \Theta\mathbb{E}_t \hat{y}_{t+1} - \Theta_a(\tilde{a}_{t+1}-\tilde{a_t}) - \frac{1}{\sigma}\left(\hat{i}_t - \mathbb{E}_t \pi_{t+1}\right), \label{teq:IS_int}
\end{align}
where $\tilde{a}_t = A_t/Y_{ss}$ describes liquid savings. However, $\tilde{a}$ can be shown to evolve according to $\tilde{a}_{t+1} = \Lambda \mathbb{E}_t \hat{y}_{t+1}$. \eqref{teq:IS_int} can be derived from the illiquid asset Euler equation \eqref{teq:EU_I}, while \eqref{teq:EU_A} determines $\Lambda$, a combination of parameters such that $\Lambda > 0$ if $\Gamma/\Theta < 1$ and vice versa.  In turn, unless shocks hit the economy in period $t$ so that $\mathbb{E}_{t-1}\hat{y}_t \neq \hat{y}_t$, the linearized IS relation fulfills
\begin{align}
    \hat{y}_t = \mathbb{E}_t \hat{y}_{t+1} - \frac{1}{\sigma \Omega}\left(\hat{i}_t - \mathbb{E}_t \pi_{t+1}\right), \label{teq:int_IS}
\end{align}
where $\Omega := \Theta - \Theta_a \Lambda$ is a positive constant fulfilling $\Omega < \Theta$ if $\Gamma/\Theta < 1$ and vice versa. Again, for details consult Appendix \ref{app:loglin}. Clearly, \eqref{teq:int_IS} is isomorphic to the standard IS condition in a representative agent model, except for the scaling of the interest rate elasticity by an additional term. 
Such an adjusted interest rate elasticity is a feature that less sophisticated Two Agent New Keynesian (TANK) models also introduce \citep[compare][]{bilbiie2008}.
Hence, the relevance of public debt for aggregate demand and discounting (or amplification) of the IS consolidation are not consequences of household heterogeneity and idiosyncratic risk per se, but additionally rely on the asset market structure. Here, if the asset markets are integrated, the IS equation is not discounted at all, even though the model features the same degree of household heterogeneity and idiosyncratic risk in SS as the baseline model version with separated asset markets.\footnote{As becomes clear in Appendix \ref{app:loglin}, the absence of IS discounting/amplification is because in this THANK model, the integrated asset markets eliminate cyclical variation in consumption inequality (but not income inequality) inequality. While it is not the focus of this paper, it suggests that model assumptions on liquid asset supply may also be crucial for analyses on whether household heterogeneity can amplify or resolve the so-called ``forward guidance puzzle'' \citep[cf.][]{mckayEtAl2016}.}

Below, either version of the aggregate demand block will be combined with the standard Taylor rule $\hat{i}_t = \phi_\pi \pi_t$ and New Keynesian Phillips curve $\pi_t = \kappa_w \hat{w}_t + \beta \mathbb{E}_t \pi_{t+1}$. 
\\~\\
\textbf{Calibration of the THANK model} To illustrate the further analytical results, I will use the following calibration: Household preferences are characterized by standard values $\sigma = 1.5$, $\beta = 0.99$ and $\phi = 1.0$, while $\varsigma$ is chosen to normalize $Y_{ss}=1$. I set $s = 0.94$ following \cite{bilbiie2025} and set $h$ so that $\lambda = 0.15$, which generates a quarterly MPC to lump-sum transfers of around 15\% as in the model of Kaplan et al. (2018). As implied by the linearization around $B_{ss}=0$, $\mathcal{B}=0$ and regarding government consumption, I use $\mathcal{G}=0.15$. Finally, I calibrate $\omega$ and $\mathcal{T}$ jointly for the model(s) to be consistent with a return wedge of $\varphi=0.99$ and transfers amounting to $75\%$ of the $H$ agents' income.\footnote{Given $s$ and $\lambda$, the equilibrium interest rate and in turn the $\varphi$ fulfilling \eqref{teq:varphi} is determined by consumption heterogeneity between $H$ and $S$ agents. The chosen value implies a return gap between liquid and illiquid assets of around 4\% annually, comparable to the value in the 2-asset HANK model used below.} The latter determines the size of $\Gamma/\Theta$ and the target implements a moderately discounted IS equation in the baseline model version with $\Gamma/\Theta \approx 0.985$. However, the key insights are robust to calibrating $\mathcal{T}$ so that $\Gamma/\Theta$ is larger than $1$. The SS tax rate implied by $\mathcal{T}$ and $\mathcal{G}$ amounts to $\tau^{y}=0.285$.
Regarding the macro parameters, I consider a central bank reaction parameter of $\phi_\pi = 1.5$ and a NKPC slope of $\kappa_w = 0.085$.

\subsubsection{Response to a fiscal shock}\label{subsec:t_fiscshock}

Having set up and linearized the model, we can now consider the response of inflation and output to a fiscal shock. I do so assuming a simple scenario for tractability: In period $t=0$, the fiscal authority issues an amount of public debt of $\tilde{b}_1=B_1/Y_{ss}>0$, keeps $\tau_t$ at zero and rebates all resources not needed to cover time-invariant expenses $G$ and $\bar{\tau}$ as transfers. 
Subsequently, it uses the lump-sum tax $\tau_t$ to reduce public debt back to its SS level according to the exogenous law of motion $\tilde{b}_{t+1} = \rho_b \tilde{b}_{t}$, a tractable fiscal consolidation path. 
We find that in the different model versions, output and inflation for periods $t\geq 1$ are given by 
    \begin{align*}
      \hat{y}^{BL}_t &= \left(\frac{\Theta_b(1-\rho_b\chi) +\rho_b(1-\chi)\frac{\tilde{\alpha}}{1-\lambda} + \frac{\kappa_b}{\sigma}\frac{\phi_\pi - \rho_b }{1-\rho_b\beta}}{\Theta - \Gamma \rho_b + \frac{\kappa}{\sigma}\frac{\phi_\pi - \rho_b }{1-\rho_b\beta}}\right)\tilde{b}_t := \psi^{BL}_y \tilde{b}_t \\
      \pi_t^{BL} &= \left(\frac{\kappa \psi_y^{BL} - \kappa_b}{1-\rho_b \beta}\right)\tilde{b}_t := \psi^{BL}_\pi \tilde{b}_t \\
      \hat{y}^{INT}_t &= 0 \\
        \pi_t^{INT} &= 0 .
    \end{align*}
    Again, for derivations and definitions of the composite parameters $\Theta_b$, $\chi$, $\tilde{\alpha}$ and $\kappa_b$, see Appendix \ref{app:fisc_shock}. 
    With IS equation discounting $\Gamma < \Theta$, the baseline responses $\psi^{BL}_y$ and $\psi^{BL}_\pi$ are generically positive for any reasonable parameterization. This means that in the aftermath of the shock, output and inflation remain elevated as long as public debt remains above its SS level $\tilde{b}_{ss}=0$.\footnote{I implicitly assume that parameters are such that the system is determinate, conditions for which will be explored below. In many calibrations with $\Gamma > \Theta$, $\psi^{BL}_y$ and thus $\psi^{BL}_\pi$ are typically positive as well, since the term $\frac{\kappa}{\sigma}\frac{\phi_\pi - \rho_b }{1-\rho_b\beta}$ keeps the denominator to be $>0$ as long as $\phi_\pi>1$.} With separated asset markets, higher public debt necessarily leads to more redistribution towards the $H$ agents, which boosts aggregate demand and increases the neutral rate of interest.

    In stark contrast, in the model version with integrated asset market, inflation and output return to their SS level \emph{immediately} after the shock. This is because higher public debt no longer implies better consumption insurance for the $H$ agents and thus does not feed back to aggregate demand. Since households and firms are forward-looking, the public debt-driven demand expansion can further amplify the impact of the shock in the setup with separated markets, but this effect is absent with integrated asset markets. 

\begin{figure}
    \centering
    \caption{Model responses to fiscal stimulus}
    \begin{subfigure}[b]{0.48\textwidth}
        \centering
        \caption{Output response}
        \includegraphics[width=\textwidth]{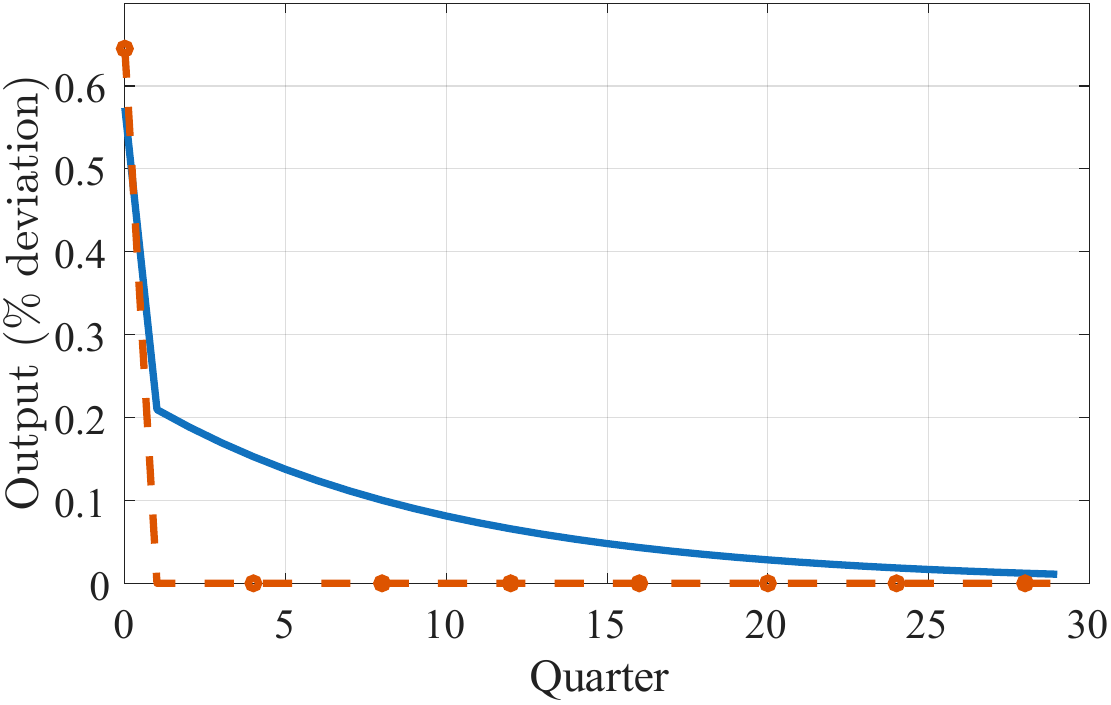}
        \label{fig:teq_y}
    \end{subfigure}
    \hfill
    \begin{subfigure}[b]{0.48\textwidth}
        \centering
        \caption{Inflation response}
        \includegraphics[width=\textwidth]{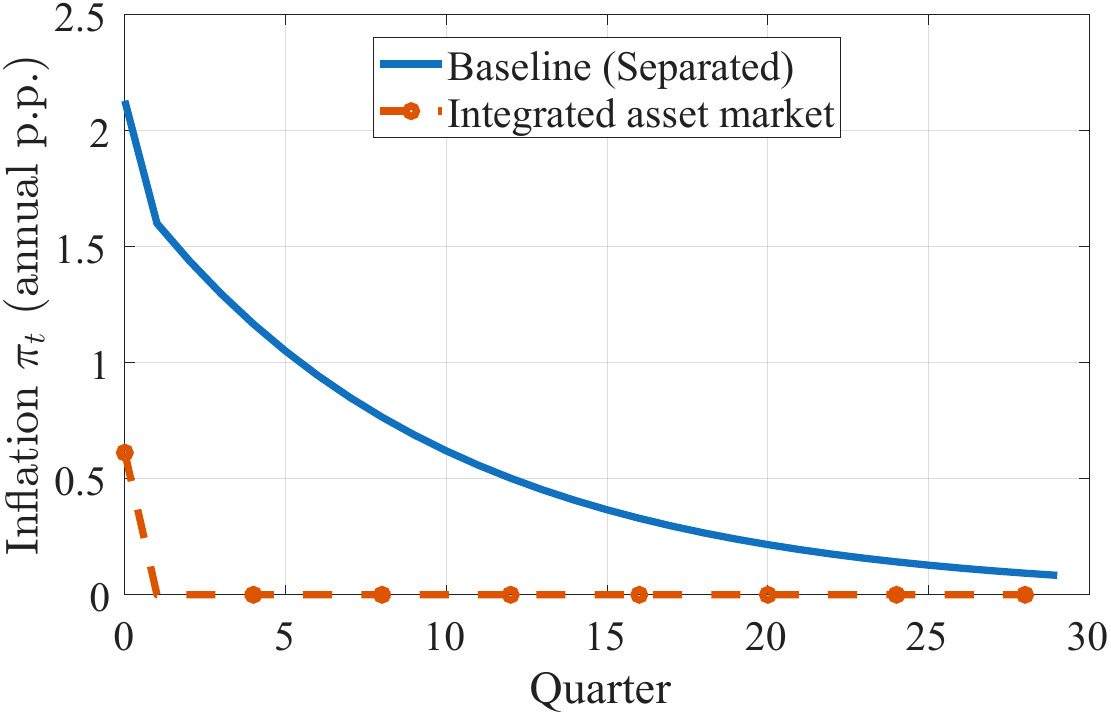}
        \label{fig:teq_pi}
    \end{subfigure}
    \label{teq:responses}
    \fignote{The figures display the output and inflation responses to the fiscal shock with $\tilde{b}_1 = 0.04$ and $\rho_b=0.9$. The solid lines display the response of the baseline model and the dash-dotted line the response of the alternative model version with integrated asset markets.}
\end{figure}

    What about the effect of the fiscal shock in period $t=0$? One can show that, as explained in Appendix \ref{app:fisc_shock}, we have $\pi_0^{BL}>\pi^{INT}_0$ for any reasonable calibration, i.e., the inflation response in the separated baseline model will exceed the one in the integrated alternative. The relative size of the initial output effect is, however, ambiguous and depends on the monetary reaction $\phi_\pi$: Intuitively, initial inflation in the baseline model is higher since price pressures are expected to remain elevated going forward. However, the higher initial inflation also implies a stronger monetary policy response, which can dampen the initial output effect.
    \\~\\
    \textbf{Quantitative illustration} To illustrate the above results, Figure \ref{fig:teq_y} and \ref{fig:teq_pi} show the response of output and inflation to the fiscal shock with $\tilde{b}_1 = 0.04$ and $\rho_b = 0.9$, i.e., the government issues an amount of public debt equal to 1\% of annual GDP in period $t=0$ and reduces the excess debt by 10\% every quarter from period $1$ forward. We see that the initial real output response is relatively similar in either model version, but only remains elevated in the model with separated asset markets. In addition, we notice that the initial inflation impact is much stronger in the baseline model and remains elevated as long as public debt is.

\subsubsection{Revenue effects of the fiscal shock}\label{subsec:t_selffinance}

\begin{figure}
    \centering
    \caption{Self-financing after the fiscal shock}
    \begin{subfigure}[b]{0.48\textwidth}
        \centering
        \caption{Baseline policy}
        \includegraphics[width=\textwidth]{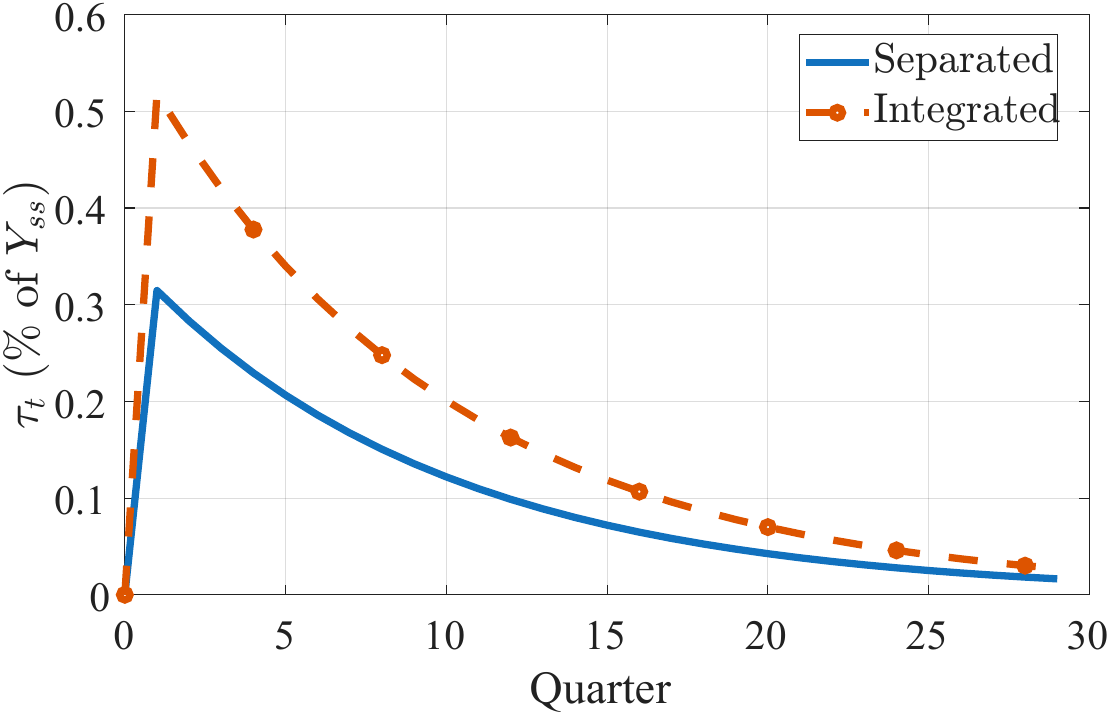}
        \label{fig:teq_tax1}
    \end{subfigure}
    \hfill
    \begin{subfigure}[b]{0.48\textwidth}
        \centering
        \caption{Alternative policy}
        \includegraphics[width=\textwidth]{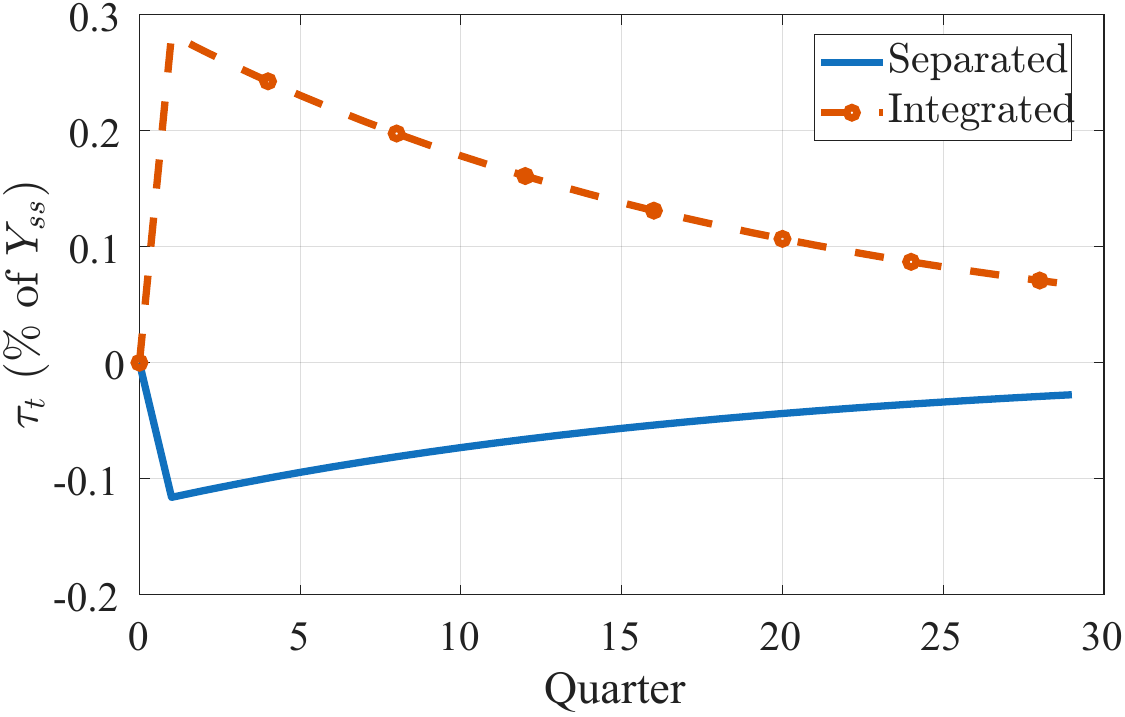}
        \label{fig:teq_tax2}
    \end{subfigure}
    \label{teq:tax}
    \fignote{The figures display the time path of lump-sum taxes $\tau_t$ after the fiscal shock with $\tilde{b}_1 = 0.04$. The solid lines display the response of the baseline model (``Separated'') and the dash-dotted line the response of the alternative model version with integrated asset markets. Subfigure \ref{fig:teq_tax1} displays the responses under the baseline policy $\phi_\pi = 1.5$ and $\rho_b = 0.9$, while Subfigure \ref{fig:teq_tax2} displays the responses under the alternative policy $\phi_\pi = 1.1$ and $\rho_b = 0.95$.}
\end{figure}

We can also briefly ask to what extent the fiscal shock may be self-financing, an issue made salient by \cite{angeletos_selffinance}. While I considered an exogenous law of motion for public debt to keep the analysis tractable, we can easily compute what amount of lump sum taxes $\tau_t$ the fiscal authority needs to raise in $t\geq 1$ to implement the intended path.
To do so, consider the government budget constraint linearized around the SS with $B_{ss}=0$
\begin{align*}
   \tilde{b}_{t+1} =\frac{1}{\beta}\tilde{b}_t + \tilde{t}_t -  (1-\lambda)\tilde{\tau}_t -  \phi_\tau \underbrace{\left[\left( \Phi + 1\right)\hat{y_t} - \frac{\sigma \tilde{\varphi}}{(1-\mathcal{G})}\tilde{a}_t\right]}_{= \hat{w}_t + \hat{n}_t},
\end{align*}
where $\tilde{t}_t := T_t/Y_{ss}$ and $\tilde{\tau}_t := \tau_t/Y_{ss}$. This expression already incorporates that income taxes and government consumption are kept constant over time. Again, see Appendix \ref{app:loglin} for details. To implement the fiscal consolidation plan $\tilde{b}_{t+1}=\rho_b \tilde{b}_t$ in $t\geq 1$ after the fiscal shock, the government will need to set the tax to 
\begin{align}
    \tilde{\tau}_t = \frac{1}{1-\lambda}\left[\left(\frac{1}{\beta}-\rho_b\right)\tilde{b}_t -  \phi_\tau \left(\left( \Phi + 1\right)\hat{y_t} - \frac{\sigma \tilde{\varphi}}{(1-\mathcal{G})}\tilde{a}_t\right)\right]. \label{teq:tax_paths}
\end{align}
Under separated asset markets, the persistently elevated output $\hat{y}_t = \psi^{BL}_y \tilde{b}_t$ will raise revenues through the income tax without any endogenous adjustments, which lowers $\tilde{\tau}_t$. This effect is absent in the model version with integrated asset markets and the difference can be substantial if $\psi^y_{BL}$ is large, which is the case if fiscal consolidation is slow or monetary policy not very hawkish.    
\\~\\
 \textbf{Quantitative illustration} Figure \ref{fig:teq_tax1} displays the time path of lump-sum taxes \eqref{teq:tax_paths} under the baseline policy with $\phi_\pi = 1.5$ and $\rho_b = 0.9$. We see that the levels necessary to induce the specified time path for fiscal consolidation are noticeably lower in the baseline model. As explained, this is since the debt-induced demand expansion generates higher labor income tax revenues in the baseline model, even though the income tax itself is kept constant. With integrated asset markets, this effect is absent. The difference is even more pronounced if we consider a scenario with less hawkish monetary policy and slower fiscal consolidation, as shown in Figure \ref{fig:teq_tax2} for the case $\phi_\pi = 1.1$ and $\rho_b = 0.95$. Here, the persistently elevated output levels in the baseline model version imply that the government can even reduce $\tilde{\tau}$ below zero while implementing the consolidation path, i.e. the fiscal expansion is fully self-financing in the sense of \cite{angeletos_selffinance}. Instead, noticeable increases of $\tilde{\tau}$ are necessary in the alternative model variant.

\subsection{Determinacy properties}\label{sec:thank_det}

As mentioned, HANK models also distinguish themselves from standard RA New Keynesian models in that the policy mix yielding a unique equilibrium is less clear-cut. To illustrate this in the THANK model, I deviate from the exogenous law of motion for public debt and instead introduce a feedback policy rule $(1-\lambda)\tilde{\tau}_t = \phi_b \tilde{b}_t$ with parameter $\phi_b$. This makes the amount of public debt endogenous and turns the economy into a 3-dimensional equilibrium system with one state- and two control variables.

The determinacy properties of this economy are studied in detail in Appendix \ref{app:determinacy}, which yields the following conclusions: In the model version with \emph{integrated} asset markets, a unique determinate equilibrium can exist only if
\begin{align*}
\left(1-\delta_b \right)\left(\phi_\pi-1 \right)> 0~~~\text{where}~~~~\delta_b:=\frac{1}{\beta}-\phi_b.
\end{align*}
This is an equivalent of the classic \cite{leeper1991} result that determinacy is possible in two distinct parameter regions: a) An ``active monetary'' one in which the monetary authority reacts more than one-to-one to inflation while the fiscal authority stabilises public debt (implied by $\delta_b<1$) and b) an ``active fiscal'' one with $\phi_\pi < 1 $ and $\delta_b > 1$.

\begin{figure}
    \centering
    \caption{Determinacy properties}
    \begin{subfigure}[b]{0.48\textwidth}
        \centering
        \caption{Baseline}
        \includegraphics[width=0.9\textwidth]{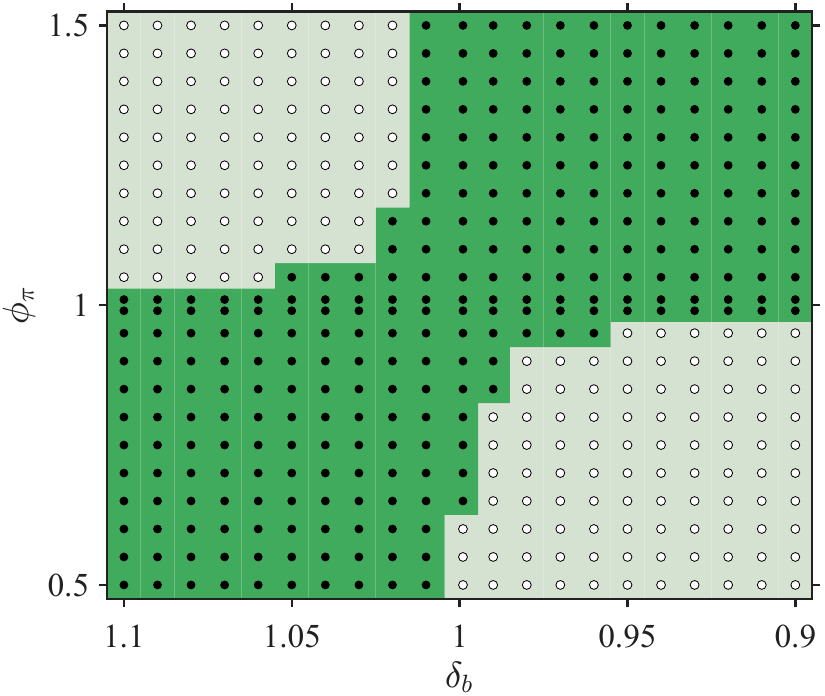}
        \label{fig:teq_det1}
    \end{subfigure}
    \hfill
    \begin{subfigure}[b]{0.48\textwidth}
        \centering
        \caption{Integrated asset markets}
        \includegraphics[width=0.9\textwidth]{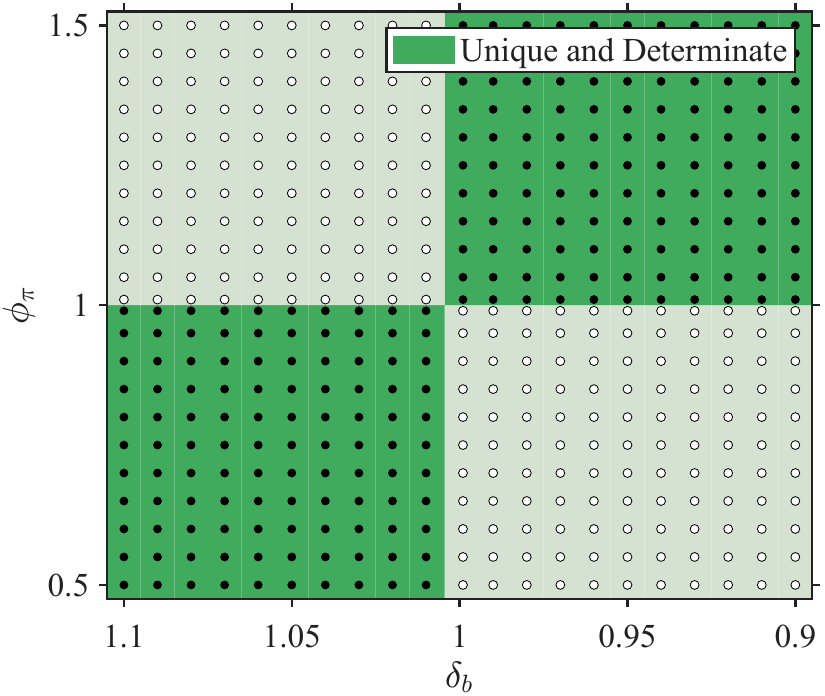}
        \label{fig:teq_det2}
    \end{subfigure}
    \label{teq:det}
    \fignote{The plot indicates whether the model provides for a unique stationary equilibrium for the different combinations of monetary and fiscal policy parameters $\phi_\pi$ and $\delta_b$. The dots indicate the evaluated parameter combinations.}
\end{figure}

In contrast, the corresponding condition for the baseline THANK model with separated markets is 
\begin{align}
   (1-\delta_b)(\phi_\pi - 1) > -\frac{\sigma(1-\beta)}{\kappa}\left[(1-\delta_b)(\Theta-\Gamma)+(\Theta_b +\Gamma_b)\phi_\tau (1+\Phi) \right]  \label{eq:detcond_seg}
\end{align}
which provides a much less clear separation also depending on the income taxes $\phi_\tau$ and various heterogeneity-dependent composite parameters from IS relation \eqref{teq:IS_seg}. Indeed, since $\Theta_b +\Gamma_b>0$, we note that the right-hand side of \eqref{eq:detcond_seg} is negative if $\delta_b < 1$ and the IS equation is discounted: determinacy is then possible even if $\phi_\pi  < 1$ and neither monetary nor fiscal policy are ``active'' in the sense of \cite{leeper1991}. Economically, the difference is due to public debt affecting aggregate demand (the IS curve) in this case.
\\~\\
 \textbf{Quantitative illustration} Under the calibration specified above, Figure \ref{teq:det} indicates which combinations of $\delta_b$ and $\phi_\pi$ yield a unique stationary equilibrium in either model version.\footnote{In contrast to the analytical expressions, which strictly provide only necessary conditions for determinacy, the results in Figure \ref{teq:det} are obtained by numerically checking the eigenvalues for the respective parameter combinations and thus based on sufficient conditions.} 
 The analysis confirms the analytical results above: With separated asset markets, there is no longer a clear separation between ``active monetary'' and ``active fiscal'' regions. With integrated markets, we obtain the well-known \cite{leeper1991} case.

\subsection{Taking stock}

This section studied monetary-fiscal interactions in a Tractable HANK model that of the type argued to capture the key features of richer HANK frameworks well. As such, its baseline version was able to generate key results from the recent literature on monetary-fiscal interactions in HANK models.  Yet, contrasting it with the alternative version featuring integrated asset markets, it became clear that these results are not a consequence of allowing for certain patterns of household heterogeneity and idiosyncratic risk per se, but rather doing so in combination with specific assumptions on liquid asset supply. At this point, a natural concern is whether this is due to some features of the THANK model, such as the illiquid asset returns being completely independent of idiosyncratic risk or the various assumptions needed to obtain tractability. However, as will become clear, one can obtain similar results also in rich quantitative models.

\section{The 2-asset HANK model}\label{sec:two_asset_hank}

For a more quantitatively oriented analysis, I employ a state-of-the-art 2-asset HANK model. Importantly, such models can generate a somewhat realistic level and distribution of aggregate wealth while being consistent with the consumption behavior that was found important for studying fiscal policy in previous work. While most features of my set-up are deliberately similar to frameworks in the previous literature, e.g., \cite{bayerEtAlaer} and \cite{auclertEtAljpe}, it again features a financial intermediary referred to as the \textit{liquid asset fund}. Now, it is set up so that one can flexibly change the \textit{potential} supply of liquid assets available to households, without affecting the initial model SS in any way. In the main text, I will focus on the household side, policy and this non-standard feature of the model, while the details on its standard parts are relegated to Appendix \ref{app:two_asset_hank}. Notice that some notation differs between the 2-asset HANK and THANK setups.

\subsection{Households}

There is a unit mass of households, which I again also refer to as ``agents'' interchangeably. These differ ex-post by several idiosyncratic states:

    Firstly, households vary in terms of their holdings of liquid and illiquid assets $a_{it}$ and $k_{it}$. $k_{it}$ represents holdings of capital and I require that $k_{it}\geq 0$ as well as $a_{it}\geq \underbar{a}$, with $\underbar{a}$ representing an exogenous borrowing/short-selling limit.
    Capital is illiquid in that a household can change its stock $k_{it}$ only infrequently: In particular, following \cite{bayerEtAlaer} and \cite{auclertEtAljpe}, I assume that the opportunity to do so arises randomly in an i.i.d. fashion, in that households only get to participate in the market for illiquid assets with probability $\lambda \in (0,1)$ every period.\footnote{In Appendix \ref{app:alt_illiquid}, I additionally discuss the implications of allowing for illiquid asset adjustment costs.}

    Secondly, as in \cite{bayerEtAlaer}, the agents can be workers ($\Xi_{it} = 0$) or ``entrepreneurs'' ($\Xi_{it} = 1$). The former participate in the regular labor market, while the latter don't supply labor to the market but receive the profits generated by the firms (to be described in Appendix \ref{app:two_asset_hank}), which are assumed to be shared equally among all households with $\Xi_{it} = 1$. Transitions to and out of the ``entrepreneur'' status are exogenous with probabilities $\zeta$ and $\iota$, implying a time-invariant mass of $m^\Xi := \frac{\zeta}{\zeta+\iota}$ agents in that state.

    Worker households ($\Xi_{it} = 0$) additionally differ by their idiosyncratic labor productivity or ``skill'' $s_{it}\in \mathcal{S}=\lbrace s_1,s_2,..., s_{ns} \rbrace$, which evolves stochastically according to a discrete Markov chain with transition probabilities $\pi^s(s_{it+1}\vert s_{it})$.
    Workers who are selected to become entrepreneurs lose their idiosyncratic $s_{it}$, while exiting entrepreneurs draw a new $s_{it}$ from the ergodic distribution of the Markov Chain.

\subsubsection{The Household problem}

Households gain utility from consumption $c$ and disutility from their amount of hours worked according to the preference structure
\begin{align}
\mathbb{E}_0 \sum^{\infty}_{t=0} \beta^t\prod_{\tau=0}^{t}\left(A_\tau\right) \left(\frac{c_{it}^{1-\xi}-1}{1-\xi} - \varsigma\frac{N_t^{1+\gamma}}{1+\gamma} \right)\text{.} \label{eq:CRRA}
\end{align}
The above formulation allows for a time-varying demand shock $A_t$ shifting all households' discount factor $\beta$. This is a vehicle to induce consumption restraints for an exercise in Section \ref{sec:mon_imps}.\footnote{This is in line with \cite{bardoczyEtAl2024}, who also use discount factor shocks in a HANK study relating to the US post-2020 period.} 
As, e.g., in \cite{auclertEtAljpe}, households are \emph{not} free to choose their own labor supply. Instead, they are required to work the number of hours demanded by their employers at the wage determined by a set of labor unions as detailed in Appendix \ref{app:two_asset_hank}. These will be equal in equilibrium, i.e., $N_{it}=N_t~\forall i\in [0,1]$.

An agent who gets to adjust her illiquid capital stock will face the budget constraint
\begin{align}
c_{it} + q_t k_{it+1} + a_{it+1}= y_{it}(s_{it},\Xi_{it})  + R_t^a (a_{it})a_{it} + (q_t+r^k_t)k_{it} + T_{it}   \label{BC_adjust}
\end{align}
while for non-adjusters, the constraint will be of the form
\begin{align}
  c_{it}  + a_{it+1}= y_{it}(s_{it},\Xi_{it})  +  R_t^a (a_{it})a_{it} + r^k_t k_{it} + T_{it}   \label{BC_nonadjust}  
\end{align}
since they maintain their illiquid holdings $k_{it+1}=k_{it}$.
Both budget constraints are already written in real terms. 
 $q_t$ denotes the time-$t$ price of capital goods, $T_{it}$ a transfer from the government, $r_t^k$ the real net return of capital goods and $ R_t^a (a_{it})$ the real gross return on liquid assets $a_{it}$. The latter depends on $a_{it}$ due to the presence of a borrowing penalty. In particular, we have
\begin{align}
R_t^a(a_{it}) = \begin{cases} R_t^l~~~~\text{if}~~a_{it}\geq 0 ;\\
 R_t^l + \Bar{R} ~~~~\text{if}~~a_{it}< 0, \end{cases}   \label{Ra_eq}
\end{align}
where $R_t^l$ is the real return on liquid savings, which will depend on the nominal central bank rate $R_t^R$ and inflation $\pi_t=\frac{P_t}{P_{t-1}}$ as specified below. $\Bar{R}$ is a real borrowing penalty.\footnote{My specification for the borrowing wedge implies that every unit of debt held by a household incurs a real resource cost of $\Bar{R}$, e.g. due to costly monitoring.}
Finally, $y_{it}$ represents a household's post-tax labor- or profit income which is given by
\begin{align}
 y_{it}(s_{it},\Xi_{it}) = \begin{cases}
 (1-\tau_t^w)\left(w_t s_{it} N_t \right)^{1-\tau^p}~~~~\text{if}~~~\Xi_{it}=0;\\
 (1-\tau_t^\Xi)\frac{\Pi_t}{m^\Xi}~~~~\text{if}~~\Xi_{it}=1. \\
 \end{cases} \label{inc_eq}
\end{align}
$\Pi_t$ denotes the aggregate amount of firm profits in the economy.
Labor income is subject to an isoelastic tax schedule in the vein of \cite{benabou2002} for which the parameters $\tau^w$ and $\tau^p$ determine the level and degree of progressivity, respectively. Similarly, $\tau^\Xi$ is the tax rate on entrepreneurs' profit income. Both level parameters may be adjusted by the government over time and thus have a time subscript. 

In Appendix \ref{app:ha_recursive}, I state the full recursive utility maximization problem of the households. 

\subsection{Production}

The model's supply side is similar to standard ``medium scale'' DSGE models: Production is vertically integrated. There is a final good that can either be consumed or used by capital goods producers to produce investment goods subject to adjustment costs. This final good is assembled by a representative final goods producer, that in turn requires differentiated inputs provided by a continuum of retailers. The retailers set prices in a monopolistically competitive fashion subject to nominal rigidities and require intermediate goods to produce their output. 
These are provided by a set of competitive intermediate goods producers that require capital and labor services as inputs. The latter are an aggregate of different labor varieties, the wage for which is decided by monopolistically competitive unions that are also subject to nominal rigidities. This adds nominal wage rigidity to the model. Given that these ingredients are well-known, I refer to Appendix \ref{app:two_asset_hank} for the details on the production side of the model.

\subsection{Government}

The government again consists of two branches, a monetary authority and a fiscal authority.

\subsubsection{Monetary Authority}

The monetary authority sets the nominal return $R^R$ on a reserve asset that is in zero net supply. Specifically, it is assumed to follow a Taylor rule of the form
\begin{align}
    R_{t+1}^R = \max\left\lbrace R^R_{SS}\left(\frac{R_{t}^R}{R^R_{SS}}\right)^{\rho^R}\left[\left(\frac{\pi_{t}}{\pi_{SS}}\right)^{\theta_\pi}\left(\frac{Y_{t}}{Y_{t-1}}\right)^{\theta_y} \right]^{1-\rho^R}\exp\left(\epsilon_t^R\right),~ R^{LB}\right\rbrace . \label{Taylor_eq}
\end{align}
which features an Effective Lower Bound (ELB) denoted as $R^{LB}$. 
The parameter $\rho^R$ introduces rate smoothing and if $\theta_y \neq 0$, the rule reacts to output growth in addition to inflation. $\epsilon_t^R$ represents an exogenous disturbance to the rule (``monetary policy shock'').
Since the calibrations will result in a stable SS with gross inflation $\pi_{SS}=1$, $R^R_{SS}$ always constitutes the true long-run natural rate of interest.

\subsubsection{Fiscal Authority}

The fiscal authority collects taxes, pays out transfers $T_{it}$ and engages in government consumption $G_t$. It also issues nominal long-term government bonds, which I introduce using a simple geometric maturity structure as in \cite{bayerEtAl2023b}. Bonds are long-lived. Every period, they pay one nominal unit of return and a random fraction $\delta^B$ of them retire without repaying the principal.\footnote{Equivalently, such a setting can be interpreted as featuring infinitely-lived bonds with geometrically declining coupon payments. See \cite{woodford2001}.} 
Denoting the nominal period $t$ price of a bond as $Q_t^B$, its expected nominal payoff is given by $\mathbb{E}_t (Q_{t+1}^B(1-\delta^B)+1)$.

To state the government's budget constraint in a convenient form, let us define $B_t^g$ to denote the value of public debt outstanding at the beginning of period $t$ in terms of its period $t-1$ real market value $Q_{t-1}/P_{t-1}$. 
The dynamics of public debt must then be consistent with 
\begin{align}
    B^g_{t+1} = (1-\delta^B)\frac{Q^B_{t}B^g_t}{Q^B_{t-1}\pi_{t}} + G_t + \int^1_0 T_{ti}di + \frac{B_t^g}{Q^B_{t-1} \pi_t} -  \Upsilon_t, \label{Gbr_eq}
\end{align}
i.e., the real period $t$ market value of public debt outstanding at the end of period $t$ equals the re-valued stock of public debt that did not retire plus the government's real spending obligations minus real tax revenues $\Upsilon_t$. The latter equal
\begin{align*}
    \Upsilon_t = \tau_t^\Xi \Pi_t + \int^1_{m^\Xi} \left(w_t s_{it} N_t - (1-\tau_t^w)\left(w_t s_{it}N_t \right)^{1-\tau^p}\right)di.
\end{align*}
As baseline, I again assume both government spending items $G_{t}$ and $T_{it}$ to be solely determined by exogenous shocks. Without any occurring, they remain fixed at $G_t = G_{SS}$ and $T_{it}=0$.

To ensure public debt stability in the face of various business cycle shocks, the fiscal authority is furthermore assumed to adjust taxes as $\tau^w_t = \tau_t \tau^w_{ss}$ and $\tau^\Xi_t = \tau_t \tau^\Xi_{ss}$. The tax level $\tau_t$ evolves according to
\begin{align}
   \left(\frac{\tau_t}{\tau_{ss}}\right) = \left(\frac{\tau_{t-1}}{\tau_{ss}}\right)^{\rho_\tau}\left(\frac{B^g_{t}}{B^g_{ss}}\right)^{(1-\rho_\tau)\psi_B} ~~, \label{eq:taxrule}
\end{align}
a functional form also used in \cite{bianchiEtAl2023}.
Adjusting all tax levels proportionally by the same factor aims to reduce the distributional impact of fiscal consolidation, in order to better isolate the role of the public debt level.
Otherwise, the fiscal authority issues any amount of bonds $B^g_{t+1}$ necessary to fulfill its budget constraint \eqref{Gbr_eq}. Intuitively, policy rule \eqref{eq:taxrule} means that the government will eventually raise taxes to pay back debt in surplus of its long-run target, but may do so only slowly. 

\subsection{Liquid Asset Provision}\label{sec:lafs}

While I assume a centralized market for claims to (illiquid) capital, households again obtain their liquid assets from a set of competitive \textit{liquid asset funds} (LAFs). While such entities were already present in THANK, I now use them as a vehicle to introduce my simple model extension.
Here, in contrast to households, these funds are able to trade claims to capital every period and also have access to a technology to short-sell any asset.
Their objective is to maximize expected real returns by investing the liquid savings $A_{t+1}^l$ they receive from the households in capital, government bonds and reserves.
In particular, the LAFs solve
\begin{align}
    \max_{B^l_{t+1},V^l_{t+1}} &\left\lbrace \mathbb{E}_t\left[(r_{t+1}^k + q_{t+1})\frac{A_{t+1}^l- B_{t+1}^l - V_{t+1}^l}{q_t}+ \frac{Q_{t+1}^B(1-\delta^B)+1}{\pi_{t+1}}\frac{B^l_{t+1}}{Q_t^B} + \frac{R^R_{t+1}}{\pi_{t+1}}V^l_{t+1}\right] \right.\nonumber\\
    & \left. - A_{t+1}^l\left(\varphi + \frac{\Psi}{2}\left(1-\frac{B_{t+1}^l+V^l_{t+1}}{A_{t+1}^l}\right)^2 \right)\right\rbrace\label{eq:LAF1},
\end{align}
where $A_{t+1}^l$ denotes the total amount of assets intermediated by the LAF and $B_{t+1}^l$ and $V_{t+1}^l$ the amount of government debt and reserves it chooses to acquire.
A fund faces costs for each unit of liquid asset it invests on behalf of the households. This involves a linear component $\varphi$ and a part $\frac{\Psi}{2}\left(1-\frac{B_{t+1}^l+V_{t+1}^l}{A_{t+1}^l}\right)^2$ that increases in the relative amount of the fund's asset positions that are \emph{not} in liquid government assets. 
This structure implies that the equilibrium government bond prices $Q^B$ must fulfill the no-arbitrage condition
\begin{align}
    \mathbb{E}_t \left(\frac{R^{R}_{t+1}}{\pi_{t+1}}\right)=\mathbb{E}_t\left(\frac{Q^B_{t+1}(1-\delta^B)+1}{\pi_{t+1} Q_t^B}\right).\label{eq:LAF_noarb}
\end{align}
Furthermore, the LAFs' aggregate portfolio choice can be determined from the corresponding F.O.C.
\begin{align}
    \mathbb{E}_t\left(\frac{r^k_{t+1}+q_{t+1}}{q_t}\right) - \Psi\left(1-\frac{B_{t+1}^l}{A_{t+1}^l}\right) =   \mathbb{E}_t\left(\frac{R^R_{t+1}}{\pi_{t+1}}\right)\label{eq:LAF2}
\end{align}
and the ex-post real return to households' liquid savings will be given by
\begin{align}
R_t^l = \frac{q_t + r_t^k}{q_{t-1}}\frac{A_t^l-B_t^l}{A_t^l}+\frac{Q^B_{t}(1-\delta^B)+1}{\pi_{t} Q_{t-1}^B}\frac{B^l_t}{A_t^l}-\varphi -\frac{\Psi}{2}\left(1-\frac{B_t^l}{A_t^l}\right)^2 , \label{eq:LAF3}
\end{align}
anticipating that in equilibrium $V_t^l=0$ as reserves are in 0 net supply.

A few words on the above structure are in order: 
Although the quadratic portfolio cost in \eqref{eq:LAF1} is reminiscent of some DSGE models featuring a financial sector \citep[e.g.,][]{geraliEtAl2010}, its aim is \textit{not} to provide an especially realistic model of financial intermediation.
Rather, it introduces a parsimonious way to flexibly move between various assumptions on liquid asset supply in the literature.
For this purpose, the parameter $\Psi$ has a simple interpretation as determining how easily capital assets can be used for liquidity provision. In case $\Psi \rightarrow \infty$, the model nests \textit{separated} markets for liquid and illiquid assets, as the high cost renders capital practically useless for the provision of liquid assets.
In contrast, for $\Psi\rightarrow 0$ it nests an \textit{integrated} market in which capital is a perfect substitute for government bonds for the purpose of liquidity provision and \eqref{eq:LAF2} collapses to a no-arbitrage condition.\footnote{In the context of two-asset HANK models, separated asset markets are for example used in \cite{kaplanEtAl2018} or \cite{bayerEtAlaer}, who assume that government bonds can \textit{only} be held as a liquid asset and capital \textit{only} as an illiquid asset --- see the asset market clearing conditions in Sections II.B and I.D of the respective papers. As mentioned, \cite{auclertEtAljpe} assume that government bonds and ``shares'' (claims on capital returns and monopoly rents) are perfect substitutes for liquidity provision, i.e., they assume an integrated asset market.\\
My LAF set-up results in equal returns on public debt and capital in steady state, as in \cite{auclertEtAljpe}. In settings with $\Psi>0$, the return on government bonds $B_t$ might in principle exceed the one on capital. This is, however, a mechanical consequence of how the liquidity cost is set up and does not preclude liquidity from being scarce if $\Psi$ is large. Making the linear liquidity cost $\varphi$ specific to the LAF's capital holdings instead would result in a mechanical liquidity premium on public debt without noticeably affecting model dynamics.}

In fact, such an LAF structure can also be introduced into the THANK model. In Appendix \ref{app:thank_LAF}, I show that this delivers a tractable interpolation between the separated IS condition \eqref{teq:IS_seg} and the integrated IS condition \eqref{teq:int_IS}, with weights determined by $\Psi$.

While micro-founded models of financial intermediation, building e.g. on \cite{gertlerEtAl2011},
could also generate imperfect usefulness of capital for liquidity provision, the above formulation has several benefits: 
Importantly, if the model's SS is again calibrated so that household net liquid asset holdings equal the net supply of government bonds, one can move ``in-between'' the above-mentioned assumptions on asset market structure \emph{without} changing its SS. In this regard, notice that if $B^g_{ss} = B^l_{ss} = A^l_{ss}$, all terms involving $\Psi$ will equal 0 in SS regardless of the value of this parameter. 
\\
Besides, my simple structure makes it particularly transparent how varying the usefulness of capital for liquidity provision is achieved and requires just one additional parameter $\Psi$ to be calibrated. Under $B^g_{ss} = A^l_{ss}$, condition \eqref{eq:LAF_noarb} implies that the dynamics of $R^l_{t}$ as in \eqref{eq:LAF3} equal the return on government bonds at first order (up to the wedge $\varphi$). This additionally clarifies that $\Psi$ affects the economy by moderating the dynamics of treasury returns through \eqref{eq:LAF_noarb} and \eqref{eq:LAF2}, not by introducing an additional interest rate $R_t^l$.

\subsection{Market clearing conditions and equilibrium}

The Definition of Equilibrium is standard, but tedious, given that the quantitative model features multiple markets and also requires keeping track of the evolution of the aggregate distribution.
In turn, I relegate these details to Appendix \ref{app:equilibrium}.

\subsection{Numerical Approach}

To approximate the dynamic equilibrium of the model, I use established techniques that conduct first-order perturbation around the economy's non-stochastic steady state. Specifically, I employ the Sequence Space Jacobian (SSJ)-method proposed by \cite{auclertssj} and the State Space method employed by \cite{bayerEtAlaer}. 
Both have comparative advantages for different purposes: For example, the Sequence Space method can more conveniently handle a binding ELB relevant for the analysis described in Section \ref{sec:filtering}, as one can impose the lower bound on nominal rates via monetary news shocks as in \cite{mckayEtAl2021}. 
In contrast, the State Space method allows to easily check whether the model admits a unique and stable solution for a given parameterization via the Blanchard-Kahn-conditions and proved conveniently fast for exercises requiring the model to be solved many times.
Further details on the numerical implementation are provided in Appendix \ref{app:numerics}.

\subsection{Calibration}\label{sec:calib}

A period is interpreted to be a quarter. I aim for the model to be consistent with the most relevant features of the US economy and the key empirical moments emphasized by the HANK literature. This includes a plausible income- and wealth distribution, the presence of poor and wealthy ``Hand-to-Mouth'' (HtM) households and in turn a fairly high aggregate MPC.
To do so, I first set a range of parameters exogenously, relying on the previous literature: In addition to standard preference- and technology parameters, this includes some parameters exclusively affecting the dynamic model response to aggregate shocks, for which I rely on previous papers estimating HANK models. While this part is relegated to Appendix \ref{app:ext_calib}, it is important to note that I parameterize the models' monetary policy rule with inflation coefficient of $\theta_\pi = 1.5$, placing me firmly inside the parameter range associated with active monetary policy. Furthermore, the robustness of the main results with regard to various external calibration choices is analyzed in Appendix \ref{app:robustness}.

 The remaining parameters are chosen to match various targets in the non-stochastic steady state. To clarify how they come about, I present for each parameter the moment I use to identify it.
 While in principle any of these parameters will somewhat affect any of the stationary equilibrium's target moments, achieving a good fit with the target relies primarily on the stated parameter.

 I discipline $(\beta,\zeta,\lambda, \bar{R})$ using four moments also targeted in the seminal works of \cite{kaplanEtAl2018} and \cite{bayerEtAlaer}.
 I choose the household discount factor $\beta$ to match a ratio of average SS capital holdings to output of $11.22$ as in \cite{bayerEtAlaer}, resulting in $\beta = 0.9836$.
 The probability $\zeta$ determines the amount of ``super rich'' entrepreneur households and I use it to target a Top 10\% wealth share of 70\%, as computed by \cite{kruegerEtAl2016} using SCF data. This requires a value of approx. $0.0005$.

 $\lambda$ determines the (il-)liquidity of capital and thus how many liquid assets agents wish to additionally hold for self-insurance purposes: I use it to target net liquid asset holdings by households to equal 1.8 times quarterly GDP. Firstly, this is in line with the amount of \emph{domestically} held public debt in the US before the start of the 2020 Covid pandemic, arguably the relevant measure in my closed-economy model. However, the target is also close to the average \emph{overall} debt-to-GDP ratio for the US over the period 1970-2019 and can thus also be interpreted in this way.\footnote{The statement regarding \emph{domestically-held debt} is based on subtracting FRED series FDHBFIN (Federal Debt held by foreign and international investors) from FYGFDPUN (Federal Debt Held by the Public), while the overall debt-to-GDP ratio is taken from FYGFGDQ188S.} 
 
 Regarding household borrowing, I follow \cite{kaplanEtAl2018} and assume the borrowing limit to equal the average quarterly (post-tax) labor income. I also set the borrowing penalty $\bar{R}$ so that 16\% of households end up having $a < 0$ in SS.
 The return wedge $\varphi$ is chosen so that the real net return to liquid savings is $0$, which requires setting $\varphi$ equal to the SS return on capital, equal to $0.0093$ in the Baseline model --- a return of $3.75\%$ in annual terms. As will be further elaborated below, allowing for a relatively high return gap between liquid and illiquid assets is an important calibration margin for generating a substantial fraction of HtM households with high MPCs.
 Initially, I entertain  the assumptions on asset market segmentation from the previous literature, i.e. $\Psi=0$ and $\Psi\rightarrow\infty$. 
 Section \ref{subsec:discipline} below will eventually propose an explicit calibration strategy.

 As I show in the additional Appendix \ref{app:dist_moments}, the model generates household micro moments in line with the evidence typically emphasized in the HANK literature. In particular, it achieves a fairly good fit for the distribution of income and wealth, and plausible shares of ``poor'' and ``wealthy'' Hand-to-Mouth households (in the sense of \cite{kaplanEtAl2014}). In turn, it also generates an average quarterly MPC of around 15.8\%.

 \begin{table}
    \centering
     \small
   \begin{tabular}{ c | c c c} 
    \hline \hline
   Parameter &  Description & Value & Target  \\ 
   \hline
   $\beta$ & Time discounting & $0.9836$ & $K/Y = 11.22$ \\
   $\zeta$ & Prob. entrepreneur state & 0.0005 & Wealth share top 10  \\
   $\lambda$ & Prob. illiquid asset adjustment & $0.0404$ & $B/Y = 1.8$ \\
   $\Bar{R}$ & Borrowing penalty & 0.0356 & 16\% borrower share \\
   $\underbar{a}$ & Borrowing limit & -1.4467 & 100 \% avg. quart. income \\
   $G_{ss}$ & Government consumption & $0.5668$ & Budget clearing \eqref{Gbr_eq} \\
   $\varphi$ & Return wedge & 0.0093 & $R^l_{ss}=1.0$; MPCs \\
   $\Psi$ & Liquidity Supply &  & See Section \ref{subsec:discipline}\\
   \hline \hline
   \end{tabular}
   \caption{Internally calibrated parameters}
   \label{tab:int_parameters}
   \end{table}

\section{Monetary-fiscal interactions in the 2-asset HANK}\label{sec:fisc_imps}

 Armed with the suitably calibrated 2-asset HANK, we are now ready to explore how the asset market shapes monetary-fiscal interactions in this model. For this purpose, I first study the aggregate response to an expansionary fiscal policy shock before turning to the self-financing result and model determinacy properties. Additional robustness checks on the main insights in this section are discussed in Appendix \ref{app:robustness}.

 \subsection{Aggregate effects of fiscal stimulus}\label{subsec:fisc_shock}

 To keep the analysis in this subsection transparent, I again consider a simple fiscal policy shock: The only exogenous disturbance will be a uniform one-time shock to government transfers $T_{it}$ without any persistence, which may be viewed as the government sending out ``stimulus checks''. 
 To aid this interpretation, I choose the size of the shock to amount to 2\% of annualized SS GDP. In terms of the USA's 2019 GDP, this would amount to circa USD 1,300 per capita, roughly the size of the one-time payments distributed under the CARES act in 2020.

\begin{figure}
    \centering
   \caption{Model IRFs to transfer shock}
    \includegraphics[scale = 0.55]{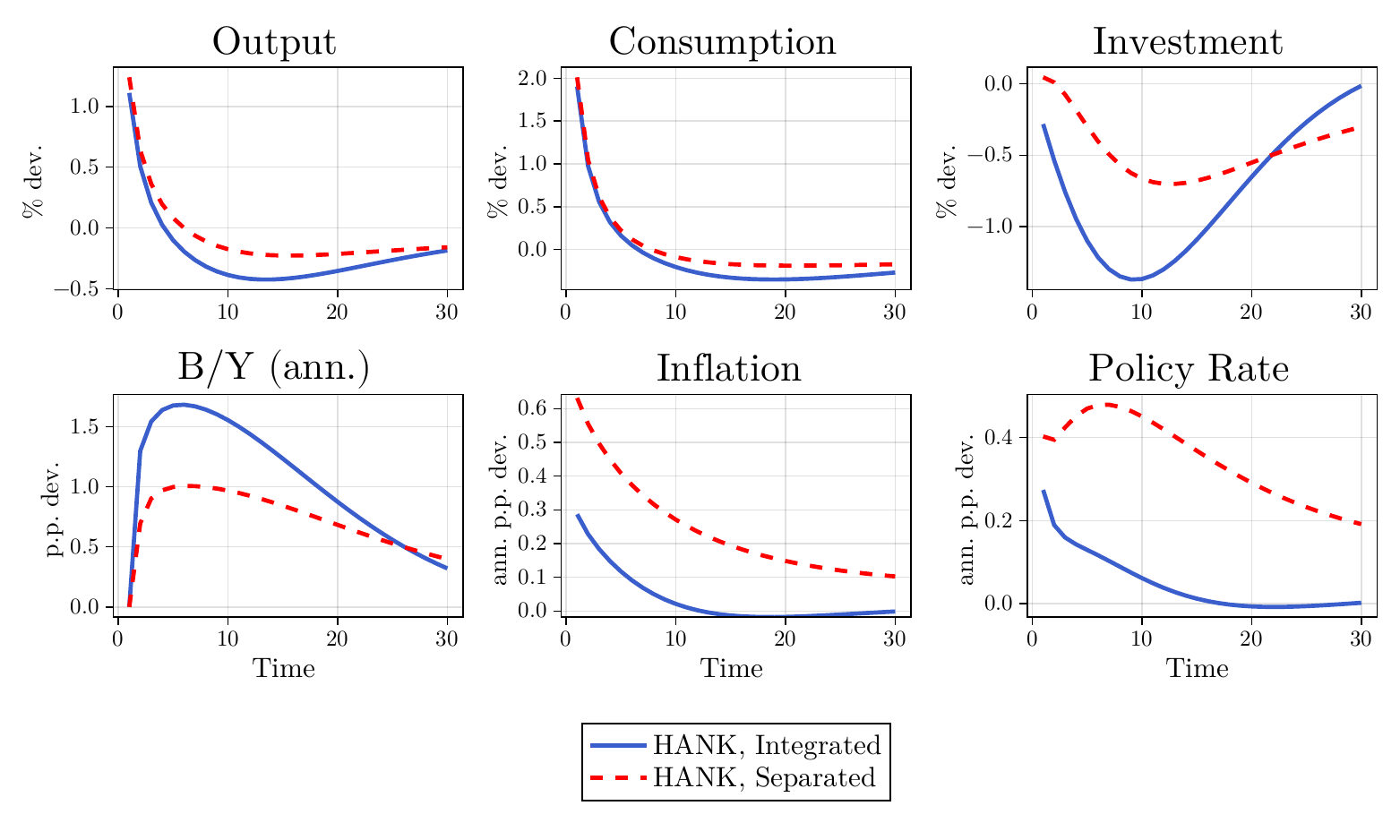}
    \fignote{$B/Y$ represents the real market value of public debt $B^g$ over annualized GDP. Figures display relative (in \%) or percentage point (p.p.) deviations from Steady State. An expanded figure with more IRFs is provided as Appendix Figure \ref{fig:T_shock9} in Appendix \ref{app:figures}.}
    \label{fig:T_shock}
 \end{figure}

 The response of key aggregate variables to the fiscal stimulus is depicted in Figure \ref{fig:T_shock}, both for the cases with integrated and separated asset markets.
Qualitatively, the responses under either model version align: The response of consumption looks as we would expect in a model featuring many HtM-households with high MPCs, increasing substantially at impact and fading out after a few quarters. 
Output and inflation similarly increase at impact, causing the central bank to raise its nominal rate. At the same time, investment decreases, which reflects crowd-out amid higher real interest rates after the shock.\footnote{The result of investment importantly shaping the transmission of the shock is reminiscent of \cite{rupertEtAl2019}.} This, in combination with slowly increasing taxes to combat the now-higher public debt, causes the output- and consumption-responses to turn negative after a few quarters.

Quantitatively, however, we see stark differences: As in the THANK model, inflation and the policy rate rise much more in the separated $\Psi\rightarrow\infty $ scenario. This also causes the market value of public debt to increase less due to the devaluation of the nominal long-term debt. Additionally, investment declines much less, which, together with the public debt-driven increase in aggregate demand, stabilizes the output responses after a few quarters. This contrasts with the consumption response, which is very similar in either case and again reminiscent of the THANK findings, where the consumption responses to the stimulus were similar despite very different inflation dynamics.
Thus, in my overall conventional 2-asset HANK model, the asset market structure remains a key driver of fiscal policy's impact on inflation and equilibrium interest rates. It does not seem to work through the consumption behavior usually emphasized by the HANK literature, but now also through the extent to which investment is crowded out by higher public debt. This feature was absent in the THANK analysis.

\subsection{Self-financing}

In Section \ref{subsec:fisc_shock}, we saw that the HANK model's asset market structure again matters significantly for the real fiscal costs of the policy: unlike in the THANK model without SS government debt, inflation-driven debt devaluations now play an important role.
\begin{figure}
    \centering
   \caption{Model IRFs: Self-financing}
    \includegraphics[scale = 0.55]{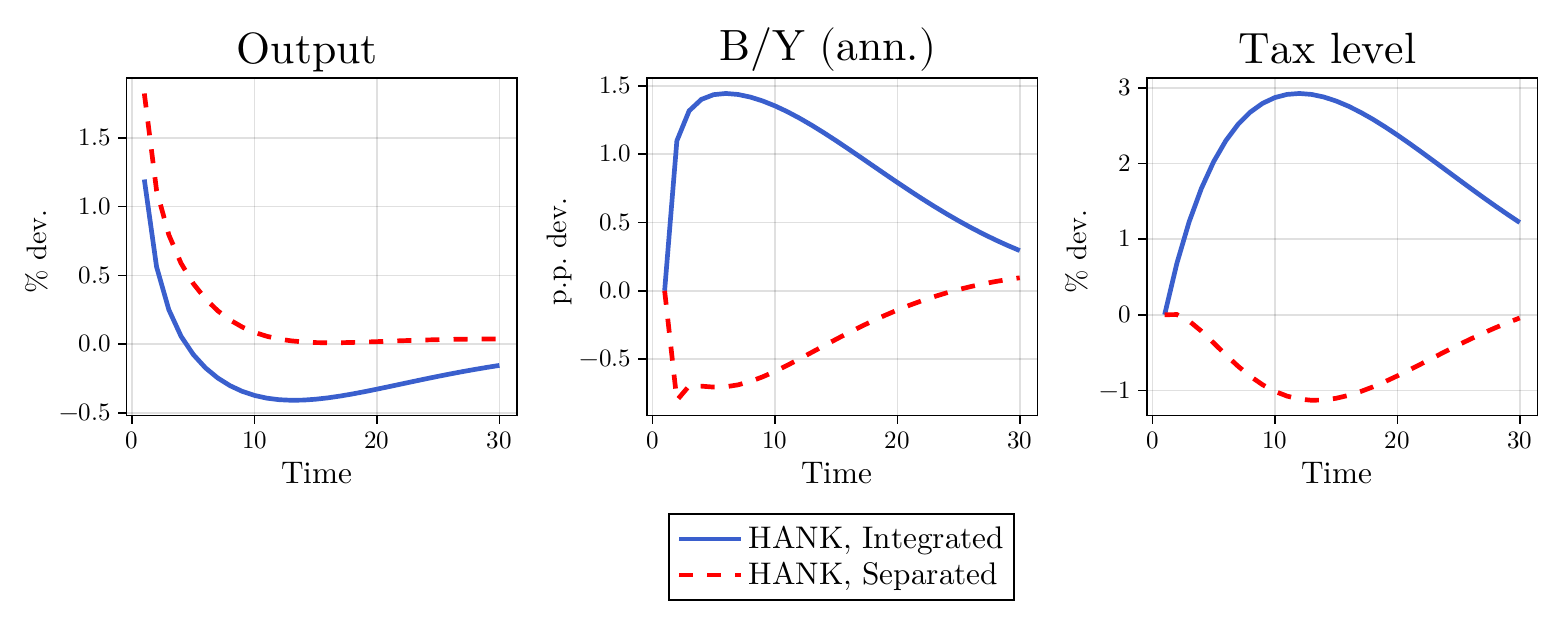}
    \fignote{The figure displays IRFs to a one-time transfer shock under $\theta_\pi = 1.05$ and $\theta_y = 0$. $B/Y$ represents the real market value of public debt $B^g$ over annualized GDP. Figures display relative (in \%) or percentage point (p.p.) deviations from Steady State.}
    \label{fig:T_selffinance}
 \end{figure}
To make this even clearer, Figure \ref{fig:T_selffinance} again follows \cite{angeletos_selffinance} by considering a scenario with a smaller but still ``active'' monetary response to inflation, $\theta_\pi = 1.05$. For comparability with other work, I remove the output response as well, so $\theta_y=0$.

Under separated markets, the response of output is noticeably stronger compared to the baseline, and we notice self-financing features: The real value of public debt even declines at impact, which the fiscal authority's instrument rule translates into lower net tax rates. This is, however, accompanied by high and persistent inflation. It also contrasts clearly with the response under integrated asset markets: Output eventually declines below its original level due to the capital crowd-out, and we still need noticeable tax increases to stabilize public debt. 
Clearly, the remarkable possibility of a self-financing fiscal stimulus seems to again depend on macro-level assumptions on liquid asset supply and not just the (again identical) micro-level patterns of consumer behavior.

\subsection{Policy rules and determinacy}

\begin{figure}
    \centering
    \caption{Determinacy analyses}
    \label{fig:determinacy}
    \begin{subfigure}{.4\textwidth}
        \caption{Integrated markets $\Psi=0$}\label{fig:det_int}
        \includegraphics[width=\linewidth]{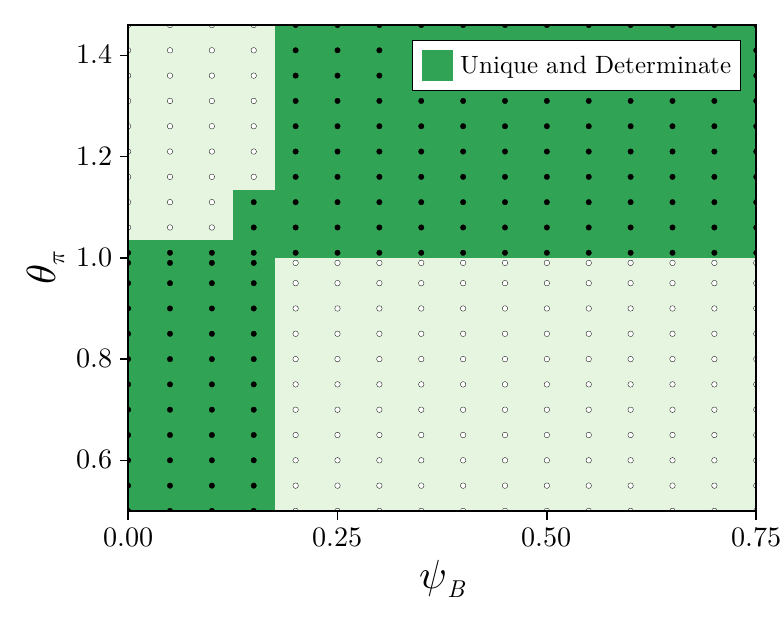}
    \end{subfigure}
    \begin{subfigure}{.4\textwidth}
        \caption{Separated markets $\Psi\rightarrow \infty$}\label{fig:det_seg}
        \includegraphics[width=\linewidth]{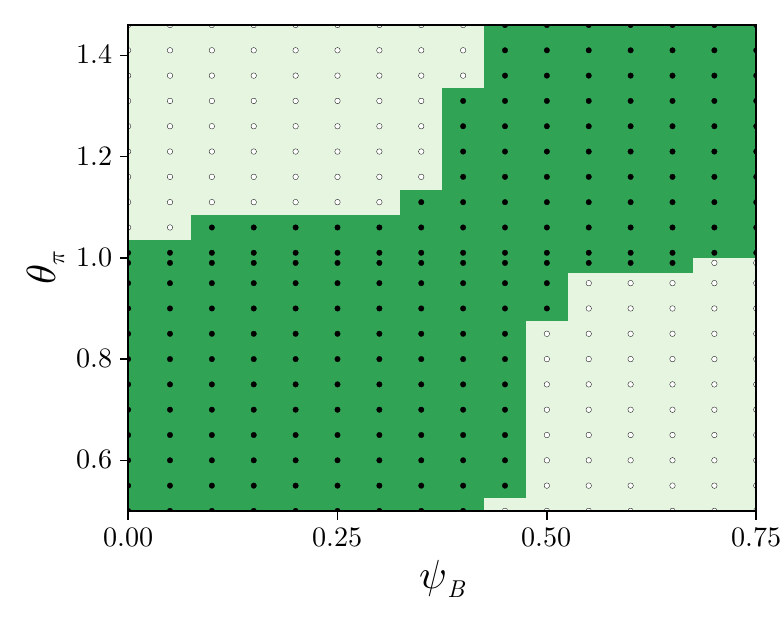}
    \end{subfigure}
    \begin{subfigure}{.4\textwidth}
        \centering
        \caption{RANK}\label{fig:det_rank}
        \includegraphics[width=\linewidth]{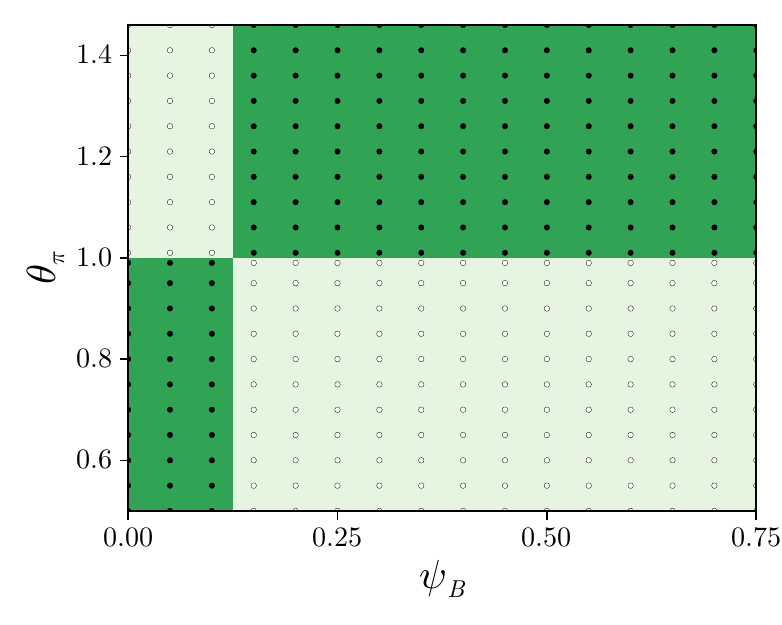}
    \end{subfigure}
    \fignote{Dots indicate the evaluated parameter combinations.}
\end{figure}

We can also do an equivalent to the THANK determinacy analysis in \ref{sec:thank_det}.
The results of that exercise are displayed in Figure \ref{fig:determinacy}, where the outcomes for the alternative scenarios are contrasted with the ones of a RA version of the model.\footnote{In the RA version, all parameters are unchanged, while the household block is replaced with a representative agent with discount factor $1/(1+r_{ss}^k)$. Since a no-arbitrage condition between capital and public debt has to hold in an RA framework, it furthermore assumes integrated asset markets with $\varphi=0$.} The RA model naturally provides for the classical \cite{leeper1991}-dichotomy discussed above. 
In contrast to the THANK model, however, the assumption of integrated asset markets doesn't result in full equivalence with the RA case. In the $\psi_B$-range consistent with active fiscal policy in RANK, macroeconomic stability can be achieved with an inflation reaction slightly above one. Moreover, some values in the RANK's ``passive fiscal'' region work with a $\theta_\pi$ slightly below one. Overall, though, the model's determinacy patterns are rather close to the RANK economy when contrasted with the separated $\Psi\rightarrow \infty$ case in Figure \ref{fig:det_seg}.

In this latter case, we don't only see a much wider ``bridge'' between the  ``active fiscal'' and ``active monetary'' regions, but also that active monetary policy now requires a significantly higher $\psi_B$: Since there is a positive SS stock of public debt in the 2-asset HANK, the stronger effects of fiscal expansions on interest rates under separated asset markets raise the governments refinancing costs and thus requires stronger adjustment under active fiscal policy.

\section{Liquidity Supply and Fiscal Inflation in HANK}\label{sec:liq_inflation}

In the previous sections, we found that the HANK model's asset market structure crucially shaped its aggregate response to fiscal shocks as well as its determinacy properties. Why is that? As I will elaborate more in this section, the key is that the differing scope for public debt to crowd-out other assets strongly affects the impact of public debt on the ``natural'' and ``neutral'' rates of interest. 

\subsection{Public debt and the natural rate}

\begin{figure}
    \centering
    \caption{Model: Long-run effects of gov't debt on bond returns}
    \includegraphics[scale = 0.425]{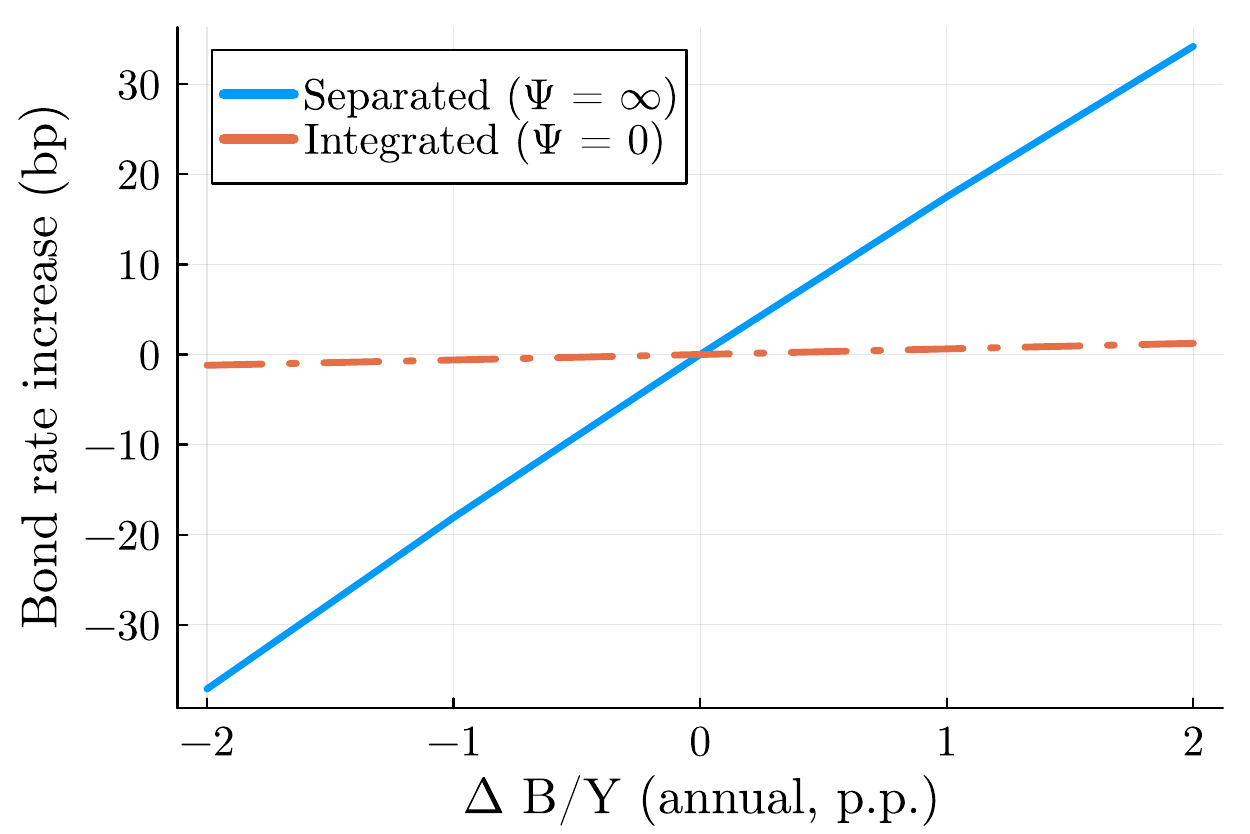}
    \label{fig:BY_graph1}
    \fignote{$\Delta B/Y$ denotes the change in the gov't debt-to-GDP ratio compared to the Baseline Steady State.}
 \end{figure}

To illustrate the effect of public debt supply on these rates in a way that abstracts from the model's real and nominal adjustment frictions, I analyze how government bond returns change with public debt in the long run. In either model version, this is equivalent to the change in the long-run natural rate of interest. The exercise involves computing new steady states for higher and lower Debt-to-GDP ratios under the parameterization specified in Section \ref{sec:calib} and yields Figure \ref{fig:BY_graph1}.\footnote{For the exercise, the following assumptions on government policy are made: The central bank adjusts its nominal rate target so that $\pi_{ss}=1$ is also achieved in the new steady state. At the same time, the fiscal authority keeps $G_{ss}$ at the same value as in the Baseline SS and adapts the tax level $\tau_t$ to clear its budget. For the pure long-run effects, it seems to make little difference whether taxes or $G_{ss}$ are adjusted.}

We see very different impacts of the government's debt supply on the return to its bonds. Under separated asset markets, a 1 p.p. higher annual Debt-to-GDP ratio causes the SS real treasury return $1/Q^B+(1-\delta^B)$ to increase strongly, by approx. 15 basis points (bp) in annualized terms. In case of the integrated asset market $\Psi\rightarrow 0$, we have the opposite: The response of the real liquid rate is hardly noticeable, reminiscent of the THANK model in which there was none at all. 

Considering empirical attempts to estimate the effects of public debt supply, does either of these effects appear plausible? According to a summary in \cite{rachelEtAl2019}, such estimates indicate medium- to long-term effects of 3 and 6 bp per percentage point increase in the Debt-to-GDP ratio. \cite{furceriEtAl2025} find a similar magnitude in a more recent study. For the purpose of this paper, I take this to be indicative about reasonable SS effects. But under the separated market assumption ($\Psi \rightarrow \infty$), the effect is much stronger, almost 3 times \emph{more} than the upper end of the empirical estimates. In the integrated asset market part, it is much \emph{smaller}, amounting to not even a third of the empirical estimates' lower range (dashed line).\footnote{In the THANK model, this is even more extreme: The SS effects of public debt supply on the interest rate on public debt is even larger under separated asset markets and exactly 0 in the integrated case.}

These results encapsulate insights that should be of interest beyond this paper's focus on monetary-fiscal interactions: Firstly, it also indicates that asset markets assumptions are of first-order importance for model-based analyses of secular interest rate changes \citep[such as][]{platzerEtAl2022}. Secondly, a stronger response of interest rates to public debt supply is often taken to imply a stronger crowd-out of capital \citep[see, e.g., the discussion in][]{laubach2009}. The comparison above shows that this does not need to be the case in 2-asset models, as the separated asset market version provides for less crowd-out than the integrated one.

\subsection{The mechanism}

Having established that the asset market structure is key for public debt's effect on the long-run natural rate, why should that matter for the differences in aggregate responses and determinacy?
The different investment responses in the 2-asset HANK are easy to rationalize, as public debt can freely crowd out capital under integrated but not under separated markets. But why the differences for inflation and monetary-fiscal interactions? While the previous sections already provided related insights, the analysis below aims to improve understanding with the help of a simple analytical exercise.

In that regard, it is useful to consider a simplified version of the model's central bank reaction function, a textbook Taylor rule of the form 
\begin{align}
    1+i_t =  \pi^* R^* + \theta_\pi(\pi_t-\pi^*), \label{eq:taylor_1}
\end{align}
allowing for a positive net inflation target $\pi^*$. $R^*$ denotes the natural (gross) rate consistent with the SS level of government debt $B^g_{ss}$. Now, assume the amount of government debt in circulation rises temporarily to $B_1^g > B_{ss}^g$.
Notice that we can add and subtract a term $\tilde{R}(B^g_1)$ to \eqref{eq:taylor_1} and re-write it as
\begin{align}
    1+i_t = \Tilde{\pi}_1 \tilde{R}(B_1^g) + \theta_\pi(\pi_t - \Tilde{\pi}_1)~~\text{with}~~\Tilde{\pi}_1 := \pi^*\frac{\theta_{\pi}-R^*}{\theta_\pi - \tilde{R}(B_1^g)} .\label{eq:pi_tilde}
\end{align}
If $\tilde{R}(B_1^g)>R^*$, then $\Tilde{\pi}_1 > \pi^*$ as long as $\theta_\pi > \tilde{R}(B_1^g)$. So, if public debt rises and the central bank sticks to rule \eqref{eq:taylor_1} while a current debt-dependent neutral rate $\Tilde{R}$ is de facto higher than $R^*$, the central bank would seem to operate as if having a higher ``implicit'' inflation target.
\begin{figure}
    \centering
    \caption{Debt and Inflation under alternative fiscal policy}\label{fig:infl_parameters}
    \includegraphics[scale = 0.55]{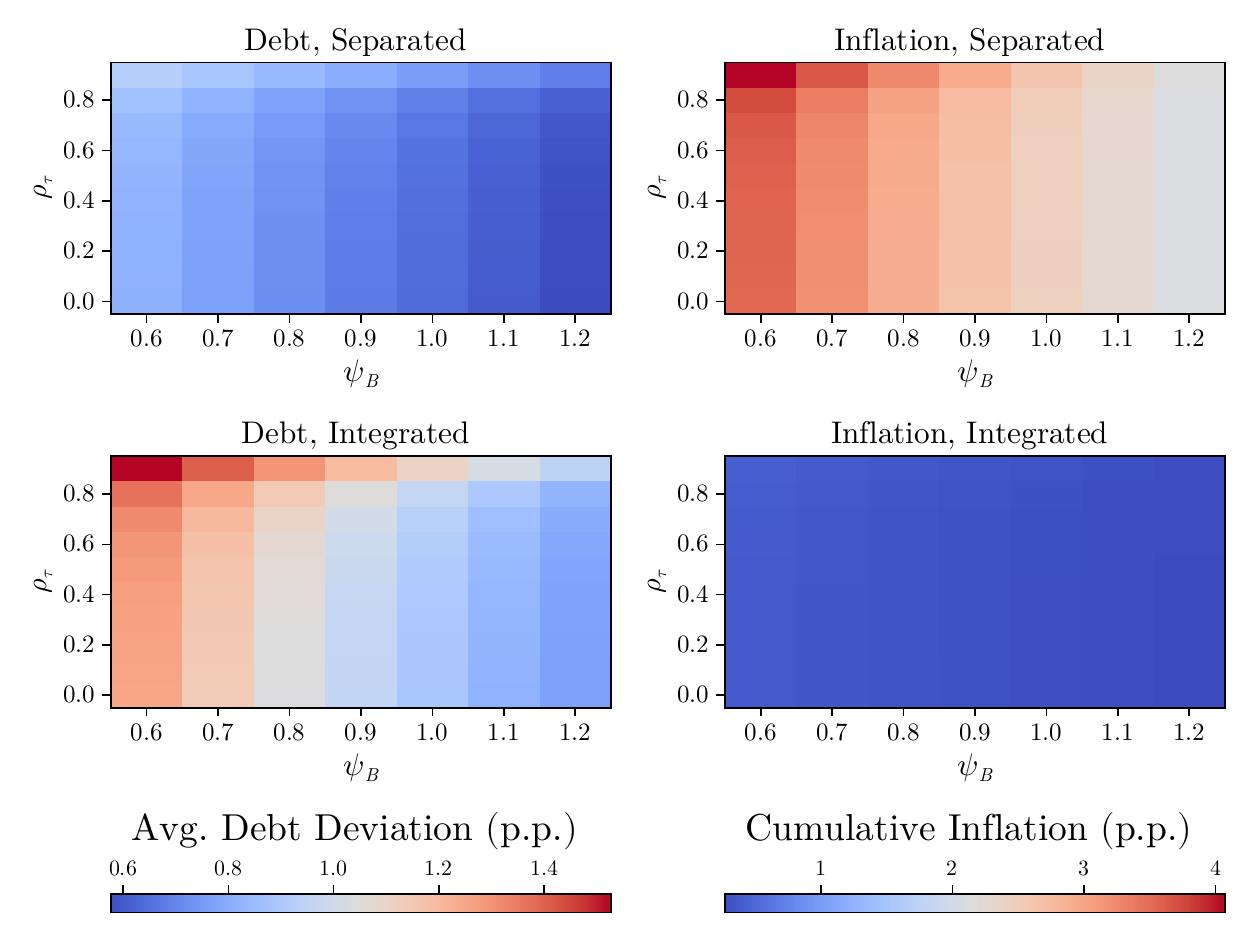}
    \fignote{The panels on the left-hand side indicate the average percentage-point (p.p.) deviation of public debt (market value) relative to annualized GDP during the first 40 quarters after a one-time transfer shock as in Section \ref{sec:fisc_imps}. The panels on the right-hand side indicate cumulative inflation during the same period. 
    }
 \end{figure}
This effect is the stronger, the more equilibrium interest rates depend on public debt supply, and is active even if the public debt level eventually returns to its original value (and the long-run natural rate doesn't actually change).

To back up that argument, I re-solve the model under various combinations of the parameters $\rho_\tau$ and $\psi_B$ that determine the persistence of the tax level and its responsiveness to the value of public debt, respectively.\footnote{All considered combination result in stable and determinate equilibria.}  For the same transfer shock as in the previous Section \ref{sec:fisc_imps}, Figure \ref{fig:infl_parameters} indicates for each of these combinations the resulting average deviation of public debt during the first 40 quarters (10 years) as well as cumulative inflation under that period: The results for the separated HANK model are displayed in the upper panels and the ones for the integrated $\Psi=0$ version in the lower ones.

Unsurprisingly, combinations of high $\rho_\tau$ and low responsiveness $\psi_B$ result in higher public debt. In the separated asset market HANK, this is systematically associated with stronger increases in the price level. In comparison, in the $\Psi=0$ case with very small effects of public debt supply on liquid interest rates, we always see lower inflation and a much less pronounced association with public debt, mirroring the THANK case where there was no effect at all. This affirms that the combination of a higher value of public debt and a high responsiveness of the neutral and natural rates to debt supply is crucial for the model's inflation response.

Equation \eqref{eq:pi_tilde} is also useful to understand the differences in model determinacy properties. We have
\begin{equation}
    \frac{d \Tilde{\pi}_1}{d \pi_t} = \pi^*\frac{\theta_\pi - R^*}{(\theta_\pi - \tilde{R}(B_1^g))^2}\frac{d\tilde{R}(B_1^g)}{dB_1^g}\frac{dB_1^g}{d\pi_t} \label{eq:dpi_tilde}
\end{equation}
so if the neutral rate increases in public debt, inflation-induced debt devaluations $\frac{dB_1^g}{d\pi_t}$ decrease (increase) the ``implicit'' monetary policy stance $\Tilde{\pi}_1$ if $\theta_\pi>R^*$ ($\theta_\pi<R^*$). This changes the extent to which the central bank induces higher (lower) real interest rates in response to inflation and thus results in a fuzzy separation between ``active'' and ``passive'' monetary policy. \eqref{eq:dpi_tilde} furthermore suggests this to become more pronounced the stronger the effect of public debt on the neutral rate, i.e. the higher $\frac{d\tilde{R}(B_1^g)}{dB_1^g}$ is. This rationalizes the stronger deviation from RANK under separated markets. Additionally, consider that if $\frac{d\tilde{R}(B_1^g)}{dB_1^g}$ is high, positive deviations of public debt raise fiscal authority's real refinancing costs more strongly and a higher $\psi_B$ is necessary to stabilize debt in the absence of a favorable monetary policy stance.

\subsection{The role of household heterogeneity}\label{subsec:heterogeneity_role}

Above, we saw that different asset market structures in HANK models importantly shape their implications for monetary-fiscal interactions. Yet, they are unrelated to household heterogeneity in that the different assumptions are consistent with the same SS micro moments. How do these effects not present in typical RA models come about then?

We already saw in Section \ref{sec:thank_ss} that in the THANK model, a higher probability of becoming constrained increases the sensitivity of the natural rate of interest with respect to public debt, \emph{conditional} on asset markets being separated.
Indeed, there is a similar link with distributional moments in the 2-asset HANK model: As is well known, such frameworks need to feature a sufficiently high gap between the return on liquid and illiquid assets to give rise to relatively high MPCs \citep[c.f.][]{kaplanEtAl2022}.
With that, the model can generate a substantial number of Wealthy HtM agents as households are incentivized to forego holding large amounts of liquidity in order to reap the illiquid assets' higher returns.
In that case, however, it also seems intuitive that if agents are to hold larger quantities of liquid government bonds, they will have to be compensated with substantially higher returns.
This argument, in turn, predicts the interest rate effects of public debt in 2-asset HANK models with \emph{separated} asset markets to be closely linked to their initial return gap and MPCs.

To analyze the link between the initial return gap and interest rate effects of public debt supply in the baseline separated case more formally, I re-calibrate my baseline framework under lower SS return gaps. In particular, I aim for the new parameterization to remain consistent with the aggregate moments targeted in \ref{sec:calib} but under a lower rate gap, using the wedge $\varphi$ to vary it between 1 and 3.75 annualized p.p. (baseline calibration).
A summary of the respective parameter results is provided in Appendix Table \ref{tab:alt_calib}: Overall, if the model is to remain compatible with the same moments under a lower returns gap, it requires a higher $\lambda$ (i.e., capital to be less illiquid) and the borrowing penalty $\bar{R}$ to be lower. The former is because the lower rate gap makes capital relatively less attractive, so it needs to be less illiquid for the aggregate household portfolio to remain the same. The latter is necessary as a higher liquid return makes it more costly to borrow by itself.\footnote{In an earlier version of the paper, the rate gap was varied by decreasing the capital return instead of increasing the liquid return through $\varphi$. In that case, the recalibrated discount factors $\beta$ varied more strongly, but the key conclusions were essentially the same.}

\begin{figure}
\centering
\caption{Implications of steady state return gaps}\label{fig:tradeoffs}
    \begin{subfigure}{.5\textwidth}
    \centering
    \caption{Liquid rate sensitivity}
    \includegraphics[width=\linewidth]{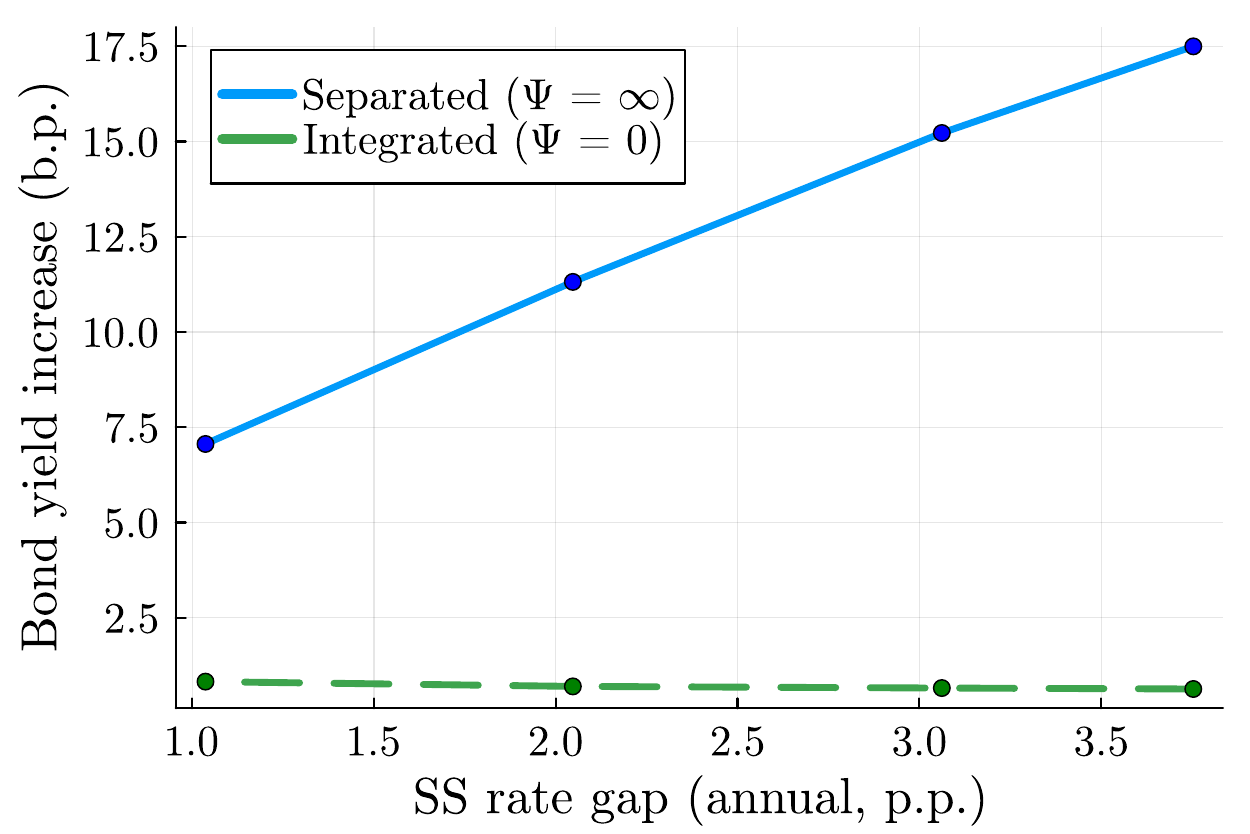}
    \label{fig:tradeoffs_1}
    \end{subfigure}%
    \begin{subfigure}{.5\textwidth}
    \centering
    \caption{Micro-moments}
    \includegraphics[width=\linewidth]{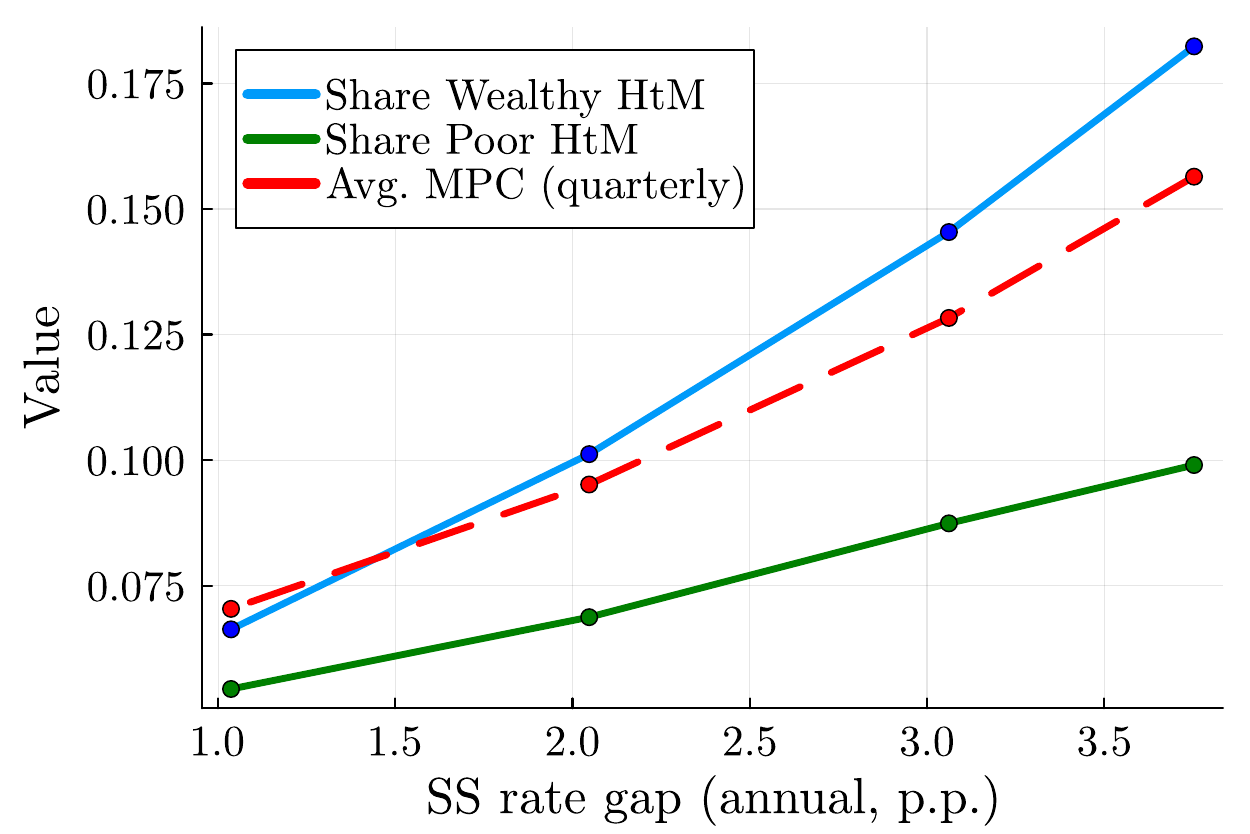}
    \label{fig:tradeoffs_2}
    \end{subfigure}
    \fignote{``Bond Yield Increase'' refers to the difference of the bond yield $\frac{1}{Q^b}+(1-\delta^B)$, expressed in annualized terms, from the calibrated Steady State after solving for a new stationary equilibrium with 1 p.p. higher annual Debt-to-GDP ratio. ``SS rate gap'' denotes the difference between annualized $r^k$ and $r^l$ in the calibrated steady state.}
\end{figure}

Figure \ref{fig:tradeoffs} then visualizes the implications of different initial return gaps: In Panel \ref{fig:tradeoffs_1}, we see that calibrating the model to be consistent with a lower initial return gap indeed decreases the long-run sensitivity of government bond returns with respect to bond supply under a separated asset market (solid line). In fact, it can even generate a response close to the 3-6 b.p. range if the return gap is low enough. 
But as Panel \ref{fig:tradeoffs_2} illustrates, this comes at the cost of having substantially lower average MPCs and few HtM households.
In line with the THANK model, under \emph{integrated} asset markets, the response of interest rates to public debt supply doesn't change much with the calibration and is always low (dashed line in Panel \ref{fig:tradeoffs_1}).

The above results suggest a tension in the HA(NK) literature on fiscal policy: Under the common assumption of separated asset markets, if one is to get the micro moments right, one may end up severely overstating the aggregate effects of liquidity supply.
But is that necessarily so? While the ``random adjustment'' setup used in my baseline model is appealing due to its relative simplicity, a popular alternative in the literature is to specify costs for illiquid asset adjustment. Since my proposal to alleviate the tension will effectively be a portfolio cost for the LAF in the aggregate, could additional \emph{micro-level} adjustment costs as already used in the literature resolve the issue as well? 

I expand on this issue in Appendix \ref{app:alt_illiquid}, in which I extend my 2-asset HANK model with the popular adjustment cost specification by \cite{auclertssj}. With these, it is still the case that generating a high average MPC requires higher return gaps, which in turn raises public debt's liquidity value. However, if the adjustment costs are high enough, it may be possible to generate high average MPCs without an excessive sensitivity of the natural rate to public debt supply, while also matching the baseline targets. But such calibrations provide a poor fit with other micro moments, such as the almost complete absence of HtM households without illiquid wealth (the ``poor'' HtM). 
I also find that under such high adjustment costs, the average MPC in the HANK model is no longer strongly connected to the fraction of HtM households as defined by their liquid asset holdings.
I conclude that richer illiquid asset adjustment frictions don't obviously resolve the tension, although I cannot exclude this possibility categorically.

\subsection{Disciplining the asset market}\label{subsec:discipline}

Overall, the above results for the off-the-shelf assumptions on asset market structure are somewhat unsatisfying.
This is not only because they fail to generate certain results, but also indicate a drawback of heterogeneous agent business cycle theory:
While related frameworks can relate to rich cross-sectional evidence, they may also require subtle modelling choices that matter for aggregate outcomes but cannot be disciplined by micro evidence alone. Instead, further macro-discipline is required.

\begin{figure}
    \centering
    \caption{Treasury return responses by $\Psi$}
    \includegraphics[scale = 0.45]{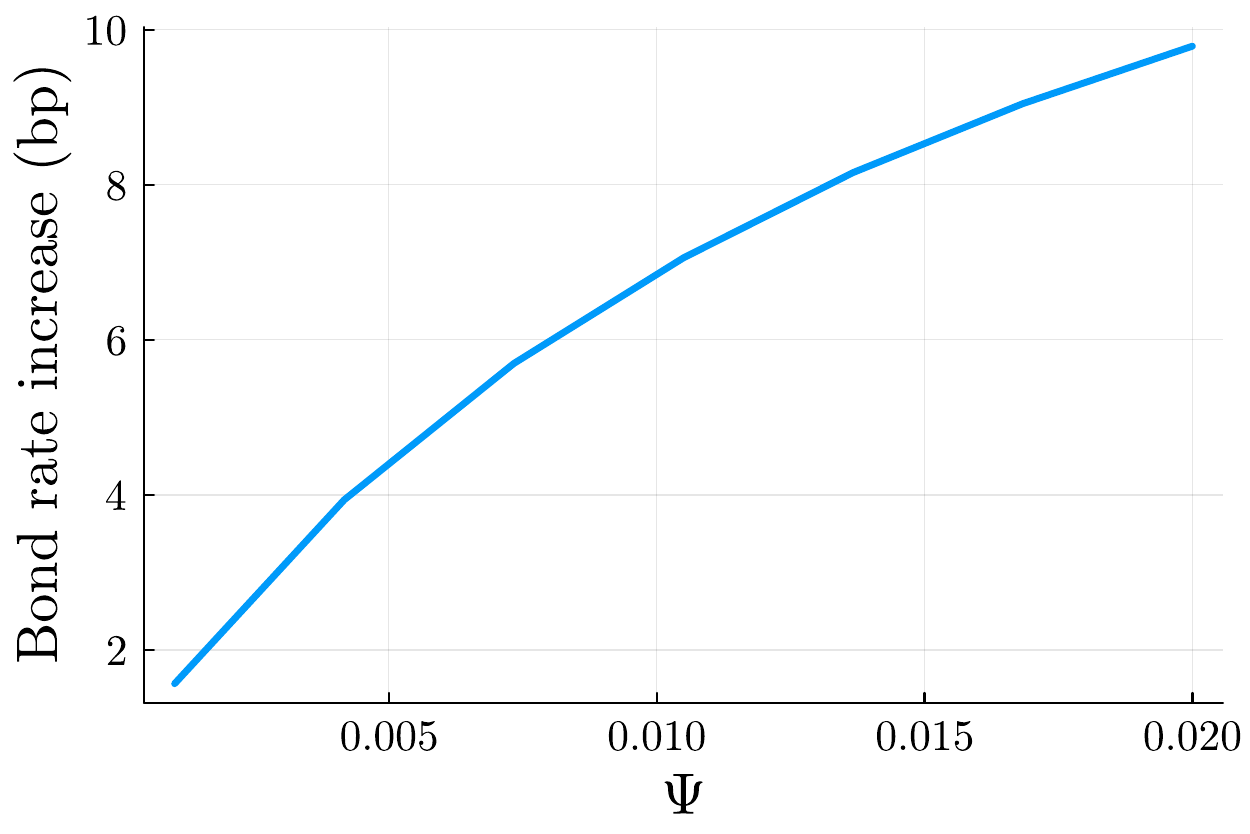}
    \label{fig:BY_graph2}
    \fignote{``Bond rate increase'' refers to the difference of the bond yield $\frac{1}{Q^b}+(1-\delta^B)$, expressed in annualized terms, from the calibrated Steady State after solving for a new stationary equilibrium with 1 p.p. higher annual Debt-to-GDP ratio.}
 \end{figure}

Of course, my set-up casting the model's asset market structure in terms of a single parameter $\Psi$ immediately suggests a resolution:
If the effects of public debt supply on its returns are overly strong for $\Psi\rightarrow\infty$ and overly weak for $\Psi \rightarrow 0$, then a value in between will presumably result in a reasonable magnitude.
To explore this possibility, Figure \ref{fig:BY_graph2} displays, for different values of $\Psi$, how much the steady state return of treasury securities changes after a long-run 1 p.p. increase in the Debt-to-GDP ratio.
The mapping between $\Psi$ and the long-run return effect is clearly monotone, which reassures us that targeting the latter is well-suited to pin down $\Psi$.
We see that for values between $0.003$ and $0.0075$, the liquid return reaction in the model economy is indeed in line with the range from \cite{rachelEtAl2019}. As baseline calibration, I will adopt $\Psi = 0.005$, which results in a response close to \cite{laubach2009}'s estimate of 4 basis points.

Disciplining the asset market by targeting the long-run interest rate effect of public debt has the appeal of being simple, mapping a simple moment to a single target. While not a business cycle outcome, the THANK analysis already illustrated that, for given household moments, it is informative to distinguish between different asset market structures with very different business cycle implications (recall Section \ref{sec:thank_ss}).
However, being able to vary the asset market structure independently of SS household heterogeneity also enables one to estimate this quantity jointly with other aggregate parameters by relating to business cycle variation, e.g., using IRF matching or full-information methods.\footnote{Dynamically estimating parameters affecting steady state household heterogeneity is typically very costly in rich HANK models. Compare the discussion in \cite{bayerEtAlaer}.}

A potential issue with the proposed ``long-run'' calibration strategy is that while it imposes a reasonable long-run sensitivity, it is not clear that this must result in reasonable short-run dynamics of liquid returns or other measures of public debt's liquidity value such as its spreads with less liquid assets. It is also interesting to ask whether targeting short-run evidence such as empirical IRFs can easily discriminate between different asset market structures, as this would support alternative calibration approaches.
I provide a discussion of these issues in Appendix \ref{app:irf_comp}, where I relate the HANK model to empirical fiscal policy IRFs estimated by \cite{bayerEtAl2023b}.

These auxiliary exercises yield the following conclusions: Disciplining the asset market structure by matching empirical IRFs for macroeconomic outcomes alone may be challenging, as such outcomes are influenced by many unrelated economic mechanisms. For this reason, generating a reasonable fit for liquid returns is possible regardless of the asset market structure. 

Regarding the spread between public debt and less liquid assets, none of the considered HANK versions can replicate the strong empirical response of a spread between a composite of capital assets and treasuries. Likely, the measured capital returns are driven by economic forces beyond standard HANK models, such as time-varying risk premia.

However, the calibrated $\Psi = 0.005$ HANK model matches the empirical response of a spread between high-quality corporate bonds and treasuries well, while the separated asset market version substantially overshoots it. As this spread is a standard measure of public debt's pure liquidity premium in the finance literature \cite[c.f.][]{krishnamurthyEtAl2012}, it is arguably a more suitable comparison for certainty-equivalent linearized HANK models. 
Indeed, in standard 2-asset HANK models, one can typically apply Modigliani-Miller logic to the firms' capital structure, which is separate from the household-level frictions, to replace (illiquid) claims to capital with (illiquid) equity or (illiquid) corporate bonds. This motivates a focus on this empirical liquidity premium of public debt directly.

\section{How relevant is public debt's liquidity value?}\label{sec:mon_imps}

Using the calibration for the HANK model's asset market separation in Section \ref{sec:liq_inflation}, we can now examine whether the liquidity margin of Ricardian non-equivalence still meaningfully affects monetary-fiscal interactions once reasonably disciplined. In that regard, I start with the simple fiscal shock and determinacy exercises familiar from Section \ref{sec:fisc_imps}, before briefly studying a richer scenario inspired by the US post-2020 episode.

\subsection{Simple fiscal shock and determinacy}\label{subsec:calib_simple}

 \begin{figure}

\caption{Model Results for $\Psi=0.005$}\label{fig:fisc_calib}
\centering
    \begin{subfigure}{1\textwidth}
    \centering
    \caption{Model IRFs to transfer shock}
     \includegraphics[width = \linewidth]{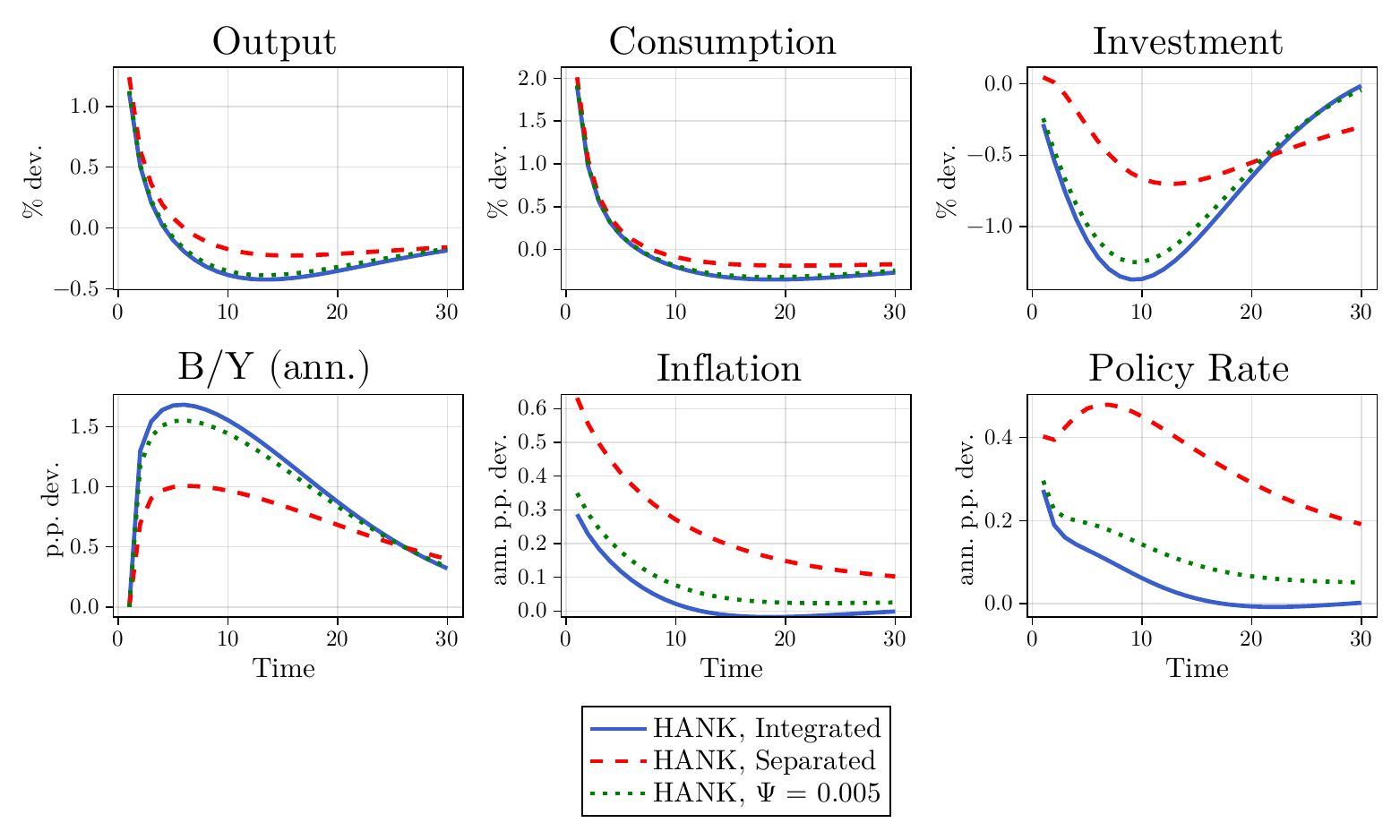}
     \label{fig:T_wcalib}
    \end{subfigure}

     \begin{subfigure}{1\textwidth}
    \centering
    \caption{Model IRFs: Self-financing}\label{fig:T_selffinance_all}
     \includegraphics[width = \linewidth]{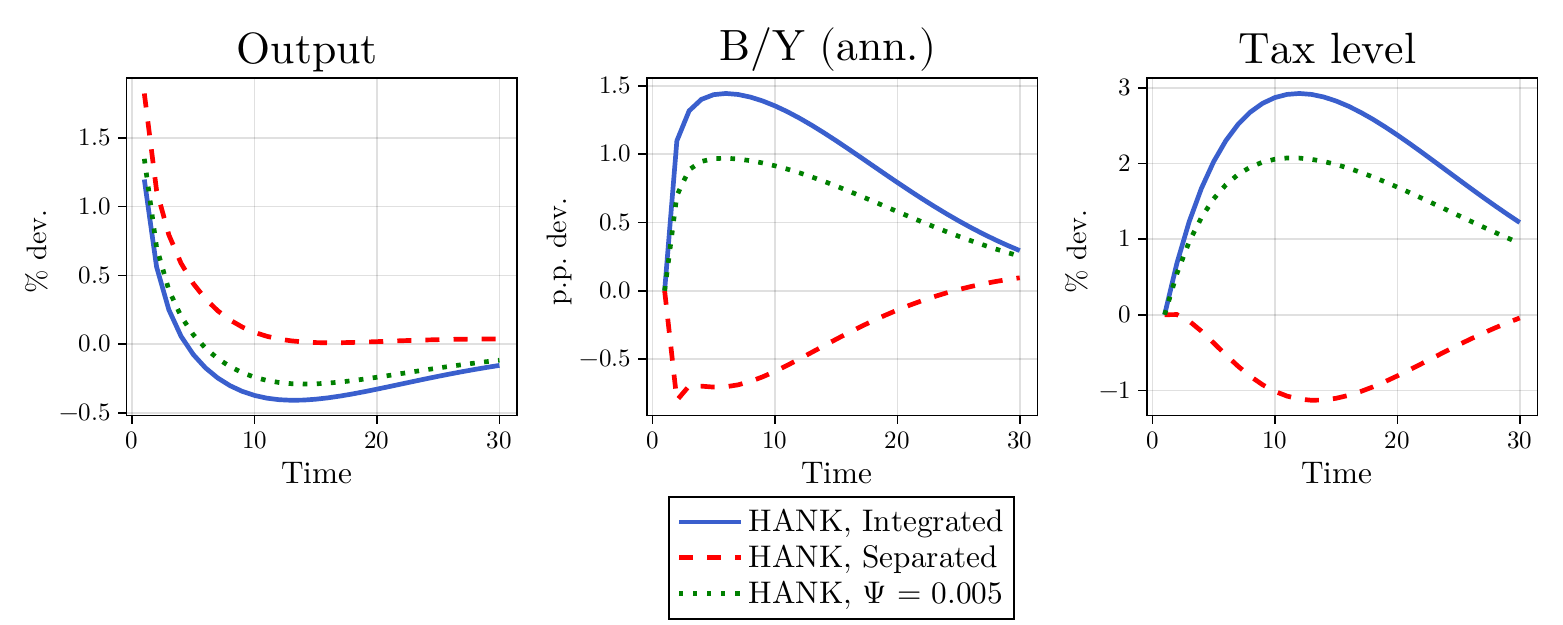}
    \end{subfigure}

    \begin{subfigure}{.4\textwidth}
    \centering
     \caption{Determinacy analysis}
     \includegraphics[width = \linewidth]{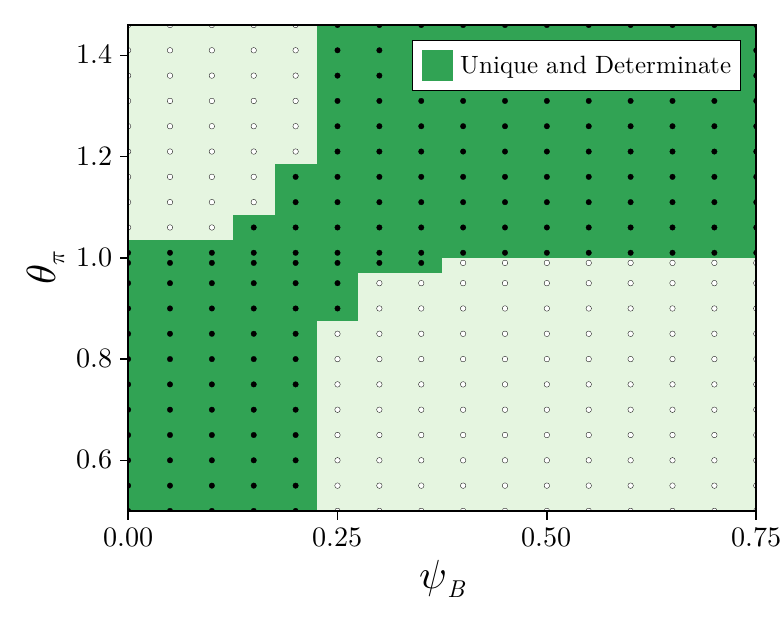}
     \label{fig:det_calib}
    \end{subfigure}

    \fignote{In Subfigures \ref{fig:T_wcalib} and $\ref{fig:T_selffinance_all}$,  $B/Y$ represents the real market value of public debt $B^g$ over annualized GDP. All lines display relative (in \%) or percentage point (p.p.) deviations from Steady State. In Subfigure \ref{fig:det_calib}, dots indicate the evaluated parameter combinations.}
\end{figure}

Figure \ref{fig:T_wcalib} provides the model responses analyzed in Section \ref{sec:fisc_imps} and contrasts them with the calibrated model featuring $\Psi=0.005$. The calibrated economy's responses lie between the polar cases, consistent with the interpolation property of the LAF cost in THANK (see Appendix \ref{app:thank_LAF}).
Overall, the responses resemble the integrated $\Psi=0$ case much more closely than the separated one. Compared to the integrated case, investment turns out slightly higher while inflation and the policy rate remain modestly elevated following the shock. Both differences reflect the still-elevated public debt exerting upward pressure on the neutral rate. While the calibrated version retains the fiscal non-neutrality stemming from the presence of constrained households, the additional relevance of government-provided liquidity appears modest ---  though this is for a scenario in which the overall debt increase is moderate and expected to be reined in within a few years. \\
For the self-financing exercise with less hawkish monetary policy, the $\Psi=0.005$ economy again appears qualitatively more similar to the integrated HANK. The quantitative differences for public debt- and tax dynamics, however, appear more meaningful. Compare Figure \ref{fig:T_selffinance_all}.

Turning to determinacy, the related results are displayed in Figure \ref{fig:det_calib}. As for the fiscal IRFs, the result is ``in between'' the ones for the separated and integrated cases. Nevertheless, the parameter region providing for a unique stable equilibrium doesn't resemble the four RANK quadrants as in Figure \ref{fig:det_rank} closely. So, even the disciplined liquidity value of public debt provides for a noticeable deviation from the RA case. 

Overall, the above suggests that regarding the responses to simple shocks under conventional monetary policy, the $\Psi=0.005$ economy can be approximated reasonably well by an integrated $\Psi=0$ economy. However, for policy reactions close to the boundaries of the classic ``active monetary/passive fiscal'' regime, as in the self-financing exercise or the determinacy analysis, differentiating between these cases is more important.

\subsection{An application to the post-2020 US}\label{sec:filtering}

The above scenarios might suggest that once properly calibrated, the liquidity effect of public debt is of limited relevance for conventional policy analyses. But is that still the case when public debt expansions are large and longer-lived?  
To provide an answer with real-world relevance, I apply my model to the US between 2020 and 2024 to assess whether distinguishing between the calibrated $\Psi=0.005$ model and the integrated $\Psi=0$ benchmark matters for understanding inflation dynamics in such a case. The aim of this exercise is not to provide a comprehensive macroeconomic assessment of this episode, but rather to gauge whether the implications of these 2-asset HANK models can differ by policy-relevant magnitudes. 

To analyze and conduct counterfactuals for this episode, we need the model to generate a situation in line with the economic dynamics of that time. Hence, I employ a filtering algorithm proposed by \cite{mckayEtAl2021} to obtain sequences of 5 aggregate shocks that make the framework match the evolution of 5 aggregate variables during the period 2020:Q1-2024:Q2.\footnote{This period was chosen to sidestep the challenge of having to relate the model to the numerous economic governance changes around and after the re-election of US President Trump.}
The various details and caveats of this exercise are presented in Appendix \ref{app:experiment}, but note that my model doesn't explicitly contain Covid-specific features and instead uses standard business cycle shocks as well as a transfer shock. It abstracts from non-transfer government spending during this period and thus generates a smaller increase in public debt compared to the data, which is conservative regarding the aggregate importance of its demand effect.
 \begin{figure}
 \centering
 \caption{Model-implied drivers of inflation}\label{fig:postcov_inflation}
     \begin{subfigure}{.5\textwidth}
    \centering
    \caption{$\Psi=0.005$}
    \includegraphics[width=\linewidth]{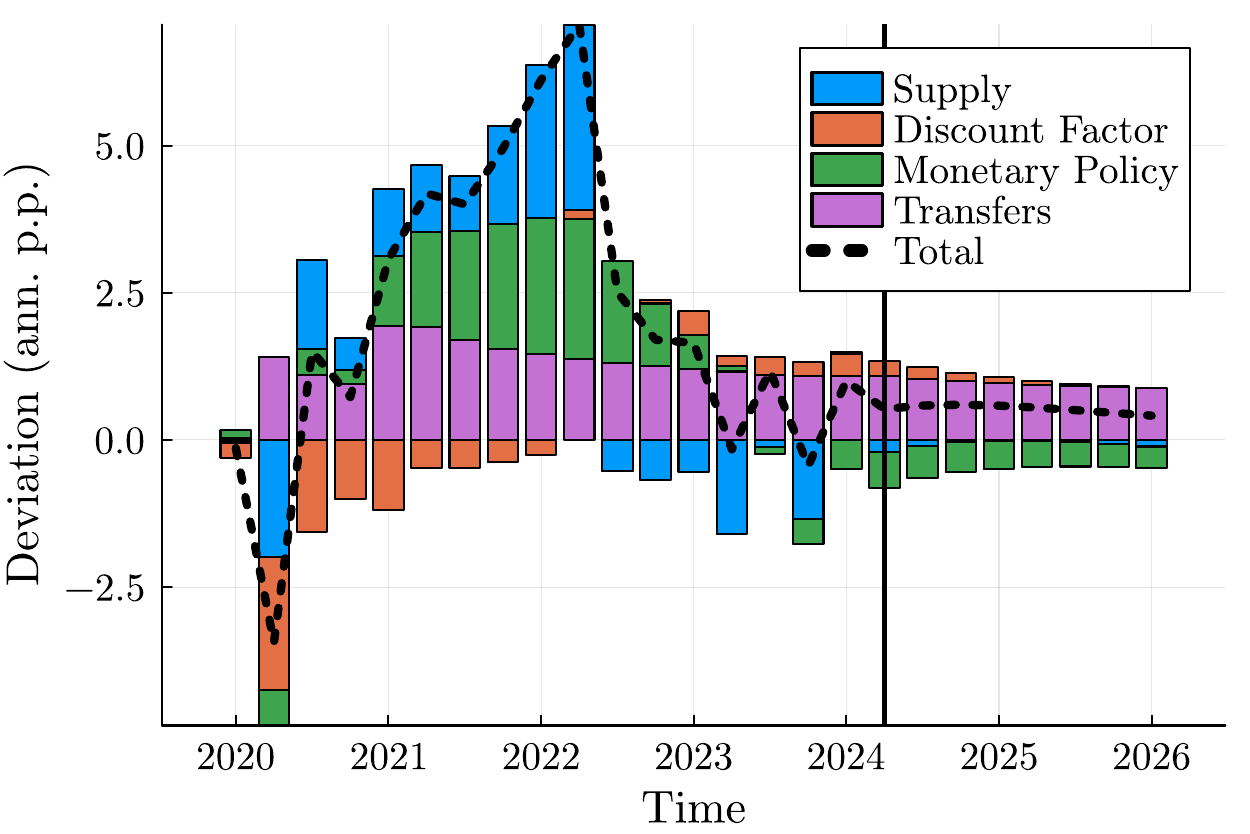}
    \label{fig:imp_decomp}
    \end{subfigure}%
    \begin{subfigure}{.5\textwidth}
    \centering
    \caption{Integrated asset markets ($\Psi=0$)}
    \includegraphics[width=\linewidth]{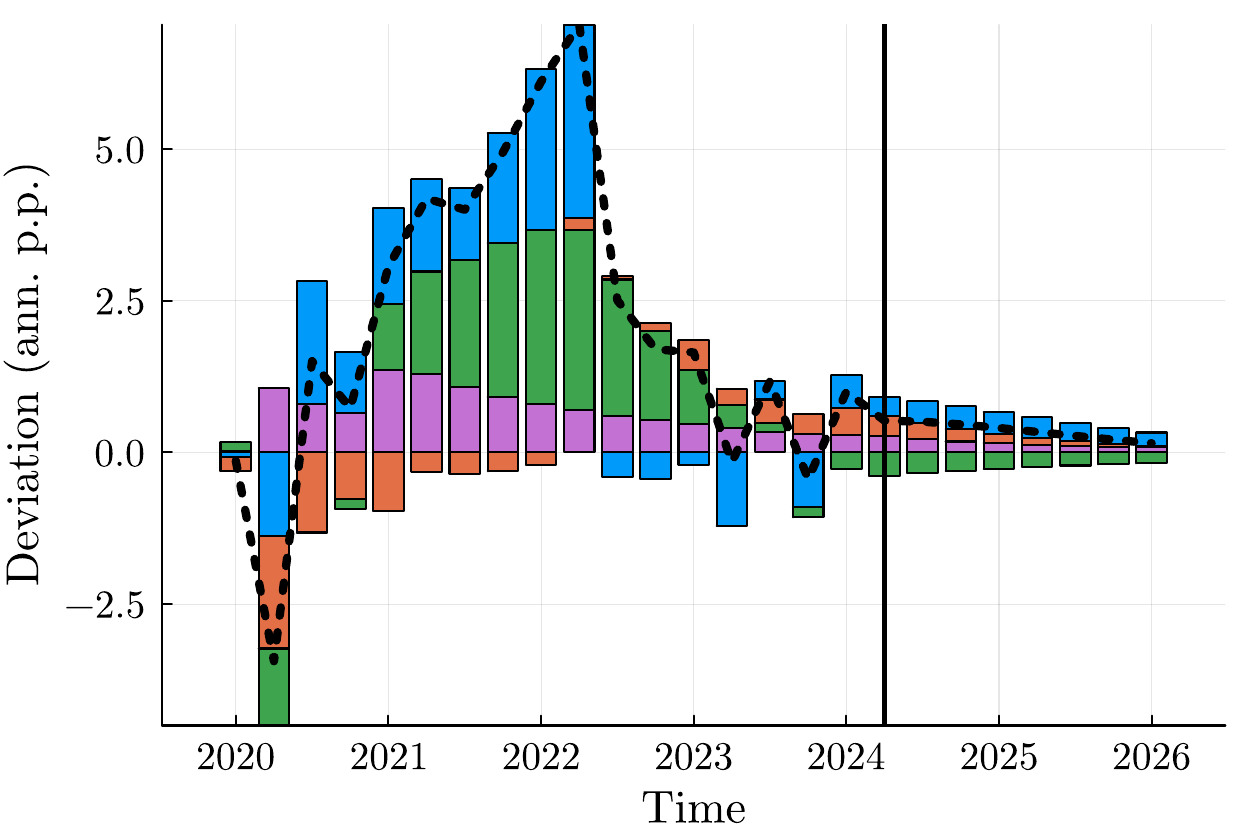}
    \label{fig:debt_contrib}
    \end{subfigure}
    \fignote{The figures display a decomposition of inflation between 2020:Q1 and 2024:Q2 into the contributions of filtered business cycle shocks. See Appendix \ref{app:experiment} for details on the HANK filtering exercise. Beyond 2024:Q2 (black vertical line), the model is simulated forward without additional shocks hitting. "Supply" collects the impact of ``cost-push''- and the investment technology shocks in the model.}
\end{figure}

Model-based decompositions of inflation are then presented in Figure \ref{fig:postcov_inflation}. The models rationalize the deflation at the beginning of the Covid pandemic with a combination of discount factor- and supply shocks, partly counteracted by the strong increase in transfers.\footnote{The negative contribution of Monetary Policy reflects the binding nominal interest ELB.}
While the filtering method assigns the 2022 peak in inflation to adverse supply shocks, it also suggests the combination of government transfers and accommodative monetary policy to be the key drivers of inflation during 2021-2022. This is in line with the findings of \cite{giannoneEtAl2024} who argued such ``demand side'' factors to be the most important determinants of the post-Covid price pressures.
Incidentally, the strong decline of inflation in 2022:Q3 is interpreted to be due to an unexpected easing (negative innovations) of ``cost push''-shocks, which aligns with a decrease of the oil price and an easing of supply chain bottlenecks at the time.

From the perspective of this paper's topic though, the most interesting observation are the different implied impacts of the transfer shocks. Their contribution encapsulates the liquidity effects of the large amount of public debt that financed them initially. In particular, we see the contribution of transfers to be noticeably larger and more persistent in the calibrated $\Psi=0.005$ economy. Indeed, the model suggests they were the reason for inflation staying above target after 2023 and would continue to keep inflation noticeably elevated (by about 0.5-0.7 p.p.), as it subsequently turned out to be.
In contrast, in the integrated $\Psi = 0$ economy, the filtering exercise assigns a noticeably larger share of inflation to loose monetary policy instead of the previous transfer expansions, which play hardly any role in 2023 and after.
Thus, an analysis with the integrated model would have predicted noticeably less ``last mile'' inflation during and after 2024.

All in all, the distinction between the calibrated $\Psi=0.005$ and $\Psi = 0$ mattered for analyzing post-2020 inflation dynamics through the lens of the 2-asset HANK. As such, disciplining this model aspect properly seems desirable to analyze previous and future macroeconomic shocks with this type of model, increasingly used by policy institutions \citep[e.g.,][]{acharyaEtAl2023}.

\section{Concluding Remarks}\label{sec:conclusion}

This paper studied monetary-fiscal interactions in HANK models. While previous works emphasized the implications of household consumption and savings behavior in such models, I highlight that several key outcomes depend substantially on less prominent assumptions on the asset market structure. As I demonstrate both in an analytical THANK model and a quantitative 2-asset HANK model, this is the case for the inflationary effects of fiscal policy, model determinacy properties and the fiscal costs of stimulus policies. Effectively, it does not just matter whether households have high MPCs and hold little liquid assets, but also how fundamentally scarce these liquid assets are.

While the potential importance of liquid asset supply depends on the micro moments a model generates, it cannot be determined from them alone and thus needs to be disciplined separately. Richer household-level frictions, such as convex costs for illiquid asset adjustment, cannot easily substitute for explicit discipline of the asset market structure, as they affect other aspects of the model's micro-level fit. Instead, I propose a parsimonious model extension calibrated to match the long-run effect of public debt on treasury returns.

In the calibrated model, the liquidity supply channel of Ricardian non-equivalence is restricted but not irrelevant. For standard fiscal shocks under conventional monetary policy, its effects are moderate, and the extended model behaves similarly to one with a simpler integrated asset market structure. However, this is not the case more generally:  First, the model's determinacy properties differ noticeably from the integrated case. Second, the channel matters more for shock responses either when monetary policy is less hawkish --- for example, in the self-financing exercise, the dynamics of fiscal outcomes differ by quantitatively meaningful magnitudes --- or when public debt expansions are large and persistent. In a stylized model experiment on the US post-Covid episode, the calibrated model attributes a noticeably larger and more persistent role to fiscal shocks for inflation than a model with integrated asset markets, of similar magnitude as the observed "last mile" inflation in 2023 and afterwards.

Regarding future work, my paper pinpoints a need to thoroughly assess and discipline the asset market structure of rich heterogeneous agent models. While the simple reduced-form approach taken can discipline the aggregate liquidity channel, more detailed micro-foundations of financial intermediation could provide richer policy implications. Additionally, it remains an open question how optimal central bank policy should account for the liquidity effects of public debt in settings with rich household heterogeneity.

\newpage

\subsubsection*{Declaration of generative AI use in the manuscript preparation process}

During the preparation of the manuscript, I used various generative AI tools. The AI chatbot Claude Opus/Claude Code was used to assist and review coding, writing and analytical derivations. The AI chatbot ChatGPT was used to assist coding and review some sections of the manuscript. The AI-based online tool \texttt{refine.ink} was used to review the manuscript. After using the tools, I checked and edited the content as needed and take full responsibility for the content of the article.

\bibliography{fisc_bib.bib}
\newpage

\appendix

\renewcommand\thefigure{\thesection.\arabic{figure}}
\renewcommand{\thetable}{\thesection.\arabic{table}}

\setcounter{figure}{0}    
\setcounter{table}{0}

\section*{Appendix}

\section{Derivations for the Tractable HANK model}

\subsection{Steady State derivations} \label{app:SS_derivations}

With time-invariant $C^H$ and $C^S$, we can re-arrange the Euler equation \eqref{teq:EU_A} to obtain 
\begin{align*}
    \frac{C^S}{C^H} = \left( \frac{1-\beta s R^A}{\beta (1-s) R^A}\right)^{1/\sigma}.
\end{align*}
Given that in the separated steady state (SS) $C^S = ((1-(1-\varphi)R^B \mathcal{B}-\mathcal{G})Y_{ss}-\lambda C^H )/(1-\lambda)$ and $C^H = \omega (1-\tau^y) w_{ss}N_{ss}+\bar{\tau}/\lambda+\frac{(1-\lambda)(1-s)}{\lambda}R^A A_{ss} $, we obtain a non-linear equation that can be solved for $R^A$. If log-utility $\sigma \rightarrow 1$ is assumed, we can simplify it to
\begin{align}
    \frac{(1-(1/\varphi-1) R^A\mathcal{B}-\mathcal{G})/(1-\lambda)}{\frac{\omega (1-\mu (\mathcal{T}+\mathcal{G}))}{1-\lambda + \lambda \omega} + \mathcal{T}/\lambda +  \frac{(1-s)}{\lambda}R^A \mathcal{B}}- \frac{\lambda}{1-\lambda} = \frac{1-\beta s R^A}{\beta (1-s) R^A} \label{appeq:RASS_1}
\end{align}
which uses that $(1-\lambda)A_{ss} = B_{ss} = \mathcal{B} Y_{ss}$, $ \varphi R^B = R^A$ and $\bar{\tau}=\mathcal{T}Y_{ss}$ by assumption as well as $N_{ss} = \frac{Y_{ss}}{(1-\lambda + \lambda \omega)}$.
This can be re-arranged to obtain a quadratic equation in $R^A$ which has a closed-form solution in principle, but the resulting expression is very tedious and in turn not very enlightening. However, analytically convenient expressions can be obtained if $\omega \rightarrow 0$ and $\mathcal{T}\rightarrow 0$, i.e. if the $H$ household has no labor income. In that case, we can arrange \eqref{appeq:RASS_1} to obtain
\begin{align*}
    \frac{(1-\mathcal{G})\lambda}{(1-\lambda)\mathcal{B}} - \frac{\lambda }{1-\lambda}(1-s)\left(1+\frac{1/\varphi-1}{1-s} \right)R^A = \frac{1}{\beta} - s R^A \\
     \Leftrightarrow R^A = \frac{1}{\beta}\frac{1-\lambda}{s-\lambda/\varphi} - \frac{\lambda(1-\mathcal{G})}{(s-\lambda/\varphi)\mathcal{B}}
\end{align*}
i.e., an expression that cleanly links the SS liquid return to the debt-to-GDP ratio $\mathcal{B}$. Focussing on the practically relevant case with $s>\lambda$, we have 
\begin{align*}
    \frac{\partial R^A}{\partial \mathcal{B}} = \frac{(1-\mathcal{G})\lambda}{(s-\lambda/\varphi)\mathcal{B}^2}>0~~\text{and}~~\frac{\partial^2 R^A}{\partial \mathcal{B}^2} = -\frac{2(1-\mathcal{G})\lambda}{(s-\lambda/\varphi)\mathcal{B}^3}<0, 
\end{align*}
i.e., the interest rate on liquid assets is increasing and concave in the debt-to-GDP ratio. 
Additionally, note that 
\begin{align*}
    \frac{\partial \lambda}{\partial s} = \frac{h-1}{(2-s-h)^2}<0,
\end{align*}
so 
\begin{align*}
    \frac{\partial^2 R^A}{\partial s \partial \mathcal{B}} = \frac{1-\mathcal{G}}{\mathcal{B}^2}\frac{\frac{\partial \lambda}{\partial s}(s-\lambda/\varphi) - \lambda(1-\frac{\partial \lambda}{\partial s}\frac{1}{\varphi})}{(s-\lambda/\varphi)^2} = \frac{1-\mathcal{G}}{\mathcal{B}^2}\frac{\frac{\partial \lambda}{\partial s} s - \lambda}{(s-\lambda/\varphi)^2} < 0
\end{align*}
meaning that the sensitivity of the liquid return to the debt-to-GDP ratio is decreasing in $s$.

\subsection{Log-Linearization} \label{app:loglin}

\subsubsection{The log-linearized model conditions}
We begin with log-linearizing the equations that are common to all model versions around $B_{ss}= A_{ss}=0$:
\begin{itemize}
    \item Aggregate resource constraint: 
    \begin{align*}
        C_t = Y_t - (1/\varphi-1)R_t^A (1-\lambda)A_t - G \implies \hat{c}_t = \frac{1}{1-\mathcal{G}}\hat{y}_t - \frac{(1-\varphi)(1-\lambda)}{\beta(1-\mathcal{G})}\tilde{a}_t.
    \end{align*}
    Since $A_{ss} = 0$ around the steady state, linearize w.r.t. $\tilde{a}_t = A_t/Y_{ss}$ instead. Recall that $R^A_{ss}=\varphi/\beta$.
    \item Production: 
    \begin{align*}
          Y_t = L_t = (1-\lambda + \lambda \omega)N_t \implies \hat{y}_t = \hat{n}_t  
    \end{align*}
    \item Labor supply:
    \begin{align*}
        C_t^{-\sigma} (1-\tau^y) w_t = \varsigma N_t^\phi \implies \hat{w}_t - \sigma \hat{c}_t = \phi \hat{n}_t \implies \hat{w}_t = \underbrace{\left( \phi + \frac{\sigma}{1-\mathcal{G}}\right)}_{:=\Phi}\hat{y_t} - \frac{\sigma  \tilde{\varphi}}{(1-\mathcal{G})}\tilde{a}_t
    \end{align*}
    where $\tilde{\varphi} := \frac{(1-\varphi)(1-\lambda)}{\beta}$.
    \item Consumption of $H$ households: Since $A_{ss} = T_{ss} = 0$ in the steady state, linearize w.r.t. $\tilde{a}_t = A_t/Y_{ss}$ and $\tilde{t} = T_t/Y_{ss}$ instead.
    \begin{align*}
        C_t^H &= (1-\tau^y)w_t N_t\omega + \frac{(1-\lambda) (1-s)}{\lambda}R_t^A A_{t} + \bar{\tau} + T_t \\ 
         \implies \hat{c}_t^H &= \delta(\hat{w}_t + \hat{n}_t)  + \underbrace{\frac{(1-\lambda)(1-s)}{\lambda}R^A_{ss}\frac{Y_{ss}}{C^H_{ss}}}_{:= \alpha}\tilde{a}_t + \frac{Y_{ss}}{C^H_{ss}}\tilde{t}_t \\
         &= \delta\left[\left(\Phi + 1\right)\hat{y}_t \right]+ \underbrace{(\alpha - \delta_a)}_{:=\tilde{\alpha}}\tilde{a}_t + \frac{Y_{ss}}{C^H_{ss}}\tilde{t}_t
    \end{align*}

    \begin{align*}
        \delta := \frac{\omega (1-\tau^y) w_{ss} N_{ss}}{C^H_{ss}} = \frac{\omega (1-\tau^y) w_{ss} N_{ss}}{\omega (1-\tau^y) w_{ss} N_{ss} + \bar{\tau}} \in (0,1)
    \end{align*}
    and $\delta_a := \delta \frac{\sigma(1-\varphi)(1-\lambda)}{\beta(1-\mathcal{G})}$.
    \item Consumption of $S$ households:
    \begin{align}  
        C_t^S &= (Y_t - (1/\varphi-1)R_t^A (1-\lambda) A_t - G - \lambda C_t^H)/(1-\lambda) \nonumber \\
        \implies \hat{c}_t^S &= \underbrace{\frac{1}{1-\lambda}\frac{Y_{ss}}{C^S_{ss}}}_{:=\theta_y}\left[\hat{y}_t - \tilde{\varphi}\tilde{a}_t\right] - \underbrace{\frac{\lambda}{1-\lambda}\frac{C^H_{ss}}{C^S_{ss}}}_{:=\theta_h}\hat{c}_t^H \nonumber \\
        &= \underbrace{\left(\theta_y - \theta_h \delta (\Phi + 1)\right)}_{:=\Theta}\hat{y}_t - \underbrace{\left[\theta_y\tilde{\varphi} + \theta_h\tilde{\alpha}\right]}_{:=\Theta_a}\tilde{a}_t - \theta_h \frac{Y_{ss}}{C^H_{ss}}\tilde{t}_t \label{appeq:CS_lin}
    \end{align}
    Note that $ \theta_h \frac{Y_{ss}}{C^H_{ss}} = \lambda \theta_y$.
    \item Euler equation for liquid assets:
    \begin{align}
        (C_t^S)^{-\sigma} &= \beta \mathbb{E}_t R_{t+1}^A\left[s(C_{t+1}^S)^{-\sigma} +(1-s)(C_{t+1}^H)^{-\sigma} \right] \nonumber \\
        \implies  \hat{c}_t^S &= -\frac{1}{\sigma}\mathbb{E}_t \hat{r}^A_{t+1} + \chi \mathbb{E}_tc^S_{t+1} + (1-\chi) \mathbb{E}_t c^H_{t+1} \label{appeq:CH_lin} \\
        \text{where } \chi &:= \frac{s (C^S_{ss})^{-\sigma}}{s (C^S_{ss})^{-\sigma} +(1-s)(C^H_{ss})^{-\sigma}} \in (0,1) \nonumber
    \end{align}
    Note that since $\varphi$ is assumed to fulfil \eqref{teq:varphi} to have SS equivalence, we have $\chi = s\varphi$.
    \item Euler equation for illiquid assets: 
    \begin{align*}
        (C_t^S)^{-\sigma} &= \beta  \mathbb{E}_t R_{t+1}^I(C_{t+1}^S)^{-\sigma} 
        \implies \hat{c}_t^S = -\frac{1}{\sigma}\mathbb{E}_t \hat{r}^I_{t+1} + \mathbb{E}_t \hat{c}^S_{t+1}
    \end{align*}
    This is the standard log-linear Euler equation.
    \item Budget constraint of the government: Recall that we are linearizing around a SS with $B_{ss}=0$ and $R^B_{ss}=1/\beta$.
    \begin{align}
        B_{t+1} &= R_t^B B_t + T_t + \bar{\tau} + G - \tau^y w_t L_t + (1-\lambda)\tau_t  \nonumber\\
        \implies \tilde{b}_{t+1} &= \frac{1}{\beta}\tilde{b}_t + \tilde{t}_t -  (1-\lambda)\tilde{\tau}_t - \frac{\tau^y w_{ss}L_{ss}}{Y_{ss}} (\hat{w}_t + \hat{n}_t) \nonumber\\
         &= \frac{1}{\beta}\tilde{b}_t + \tilde{t}_t -  (1-\lambda)\tilde{\tau}_t - \underbrace{\tau^y w_{ss}}_{:=\phi_\tau} (\hat{w}_t + \hat{n}_t) \nonumber\\
         &= \frac{1}{\beta}\tilde{b}_t + \tilde{t}_t -  (1-\lambda)\tilde{\tau}_t -  \phi_\tau \left[\left( \Phi + 1\right)\hat{y_t} - \frac{\sigma \tilde{\varphi}}{(1-\mathcal{G})}\tilde{a}_t\right] \label{app:lin_budget}
    \end{align}
    where the last expression substitutes $\hat{w}$ and $\hat{n}$.
    \item No-profit condition for Liquid Asset Funds:
    \begin{align*}
        \mathbb{E}_t R^A_{t+1} = \varphi\mathbb{E}_t R^B_{t+1}  = \varphi \mathbb{E}_t  \frac{1+i_{t}}{1+\pi_{t+1}} 
        \implies \mathbb{E}_t \hat{r}^A_{t+1} = \mathbb{E}_t \hat{r}^B_{t+1} =  \left(\hat{i}_{t} - \mathbb{E}_t \pi_{t+1}\right)
    \end{align*}
    since $\varphi$ is assumed to be constructed so that $R^B = 1/\beta$.
\end{itemize}

\subsubsection{Deriving the aggregate demand condition}

To derive the IS condition for the baseline THANK model with separated asset markets, we can plug the expression for $\hat{c}_t^H$ and $\hat{c}_t^S$ into the Euler equation for liquid assets to obtain
\begin{align}
  \Theta \hat{y}_t = \underbrace{\left(\chi \Theta + (1-\chi)\delta(1+\Phi) \right)}_{:=\Gamma}\mathbb{E}_t \hat{y}_{t+1} - \frac{1}{\sigma}\left(\hat{i_t}-\mathbb{E}_t\pi_{t+1}\right) + \underbrace{((1-\chi)\tilde{\alpha} -\chi \Theta_a )}_{:=\Gamma_a}\tilde{a}_{t+1} + \Theta_a \tilde{a}_t \label{appeg:liq_IS}
\end{align}
where the last terms are zero if fiscal policy implements $\tilde{b}_t=(1-\lambda)\tilde{a}_t = \tilde{t}_t = 0$. For the alternative model with integrated asset markets, we can make use its additional no-arbitrage condition $\mathbb{E}_t R^I_{t+1} = \mathbb{E}_t R^B_{t+1} \implies \mathbb{E}_t \hat{r}^I_{t+1} = \hat{i}_t - \mathbb{E}_t \pi_{t+1}$. Using that in the illiquid asset Euler equation, we obtain 
\begin{align}
    \hat{c}_t^S = -\frac{1}{\sigma}\left(\hat{i}_t - \mathbb{E}_t \hat{\pi}_{t+1}\right) + \mathbb{E}_t \hat{c}^S_{t+1} \label{app:illiq_cs}\\
    \implies \Theta \hat{y}_t = \Theta \mathbb{E}_t \hat{y}_{t+1} - \Theta_a(\tilde{a}_{t+1} - \tilde{a}_t) - \frac{1}{\sigma}\left(\hat{i}_t - \mathbb{E}_t \pi_{t+1}\right) \nonumber
\end{align}
It is further useful to note that under integrated asset markets, if we substitute $\hat{c}_t^S$ using \eqref{app:illiq_cs} in the liquid asset Euler equation, we get 
\begin{align*}
0 = (\chi - 1)\mathbb{E}_t \hat{c}^S_{t+1} + (1-\chi)\mathbb{E}_t \hat{c}^H_{t+1} - \frac{1}{\sigma}(\hat{i_t}-\mathbb{E}_t \pi_{t+1} - \mathbb{E}_t \hat{r}^I_{t+1})
\end{align*}
which, given that the no-arbitrage condition between liquid and illiquid assets requires $\hat{i_t}-\mathbb{E}_t\pi_{t+1} - \mathbb{E}_t \hat{r}^I_{t+1} = 0$ under integrated markets, implies $\mathbb{E}_t \hat{c}^S_{t+1} = \mathbb{E}_t \hat{c}^H_{t+1}$. This means the expected deviations (but not levels) of consumption of $S$ and $H$ households are equalized in this case. Given this result, we can recover the liquid asset from dynamics as 
\begin{align*}
  \mathbb{E}_t \hat{c}^S_{t+1} = \mathbb{E}_t \hat{c}^H_{t+1} &\implies  \mathbb{E}_t\left(\Theta \hat{y}_{t+1} - \Theta_a \tilde{a}_{t+1} \right) = \mathbb{E}_t \left( \delta\left[\left(\Phi + 1\right)\hat{y}_{t+1} \right]+ \tilde{\alpha}\tilde{a}_{t+1} \right) \\
    &\implies \tilde{a}_{t+1} = \underbrace{\frac{\Theta - \delta(1+\Phi)}{\Theta_a + \tilde{\alpha}}}_{:=\Lambda}\mathbb{E}_t \hat{y}_{t+1} 
\end{align*}
(recall that $\tilde{a}_{t+1}$ is determined at time $t$). While the denominator of $\Lambda$ is positive under any reasonable parameterization, the numerator must be negative if $\Gamma>\Theta$, as
\begin{align*}
    \Gamma>\Theta \Leftrightarrow \frac{\chi \Theta + (1-\chi)\delta(1+\Phi)}{\Theta} > 1 \Leftrightarrow (1-\chi)\frac{\delta(1+\Phi)}{\Theta} > 1 - \chi \Leftrightarrow \frac{\delta(1+\Phi)}{\Theta} > 1
\end{align*}
and $\geq 0$ if not.
If no shock moving $\hat{y}_t$ occurs in period $t$, we have $(\tilde{a}_{t+1} - \tilde{a}_t) = \Lambda(\mathbb{E}_t \hat{y}_{t+1} - \hat{y}_t)$ and thus
\begin{align*}
    \hat{y}_t = \mathbb{E}_t \hat{y}_{t+1} - \frac{1}{\sigma \left( \Theta - \Theta_a \Lambda\right)}\left(\hat{i}_t - \mathbb{E}_t \pi_{t+1}\right).
\end{align*}
Note that 
\begin{align*}
    \Omega := \Theta - \Theta_a \Lambda = \Theta - \underbrace{\frac{\Theta_a}{\Theta_a + \tilde{\alpha}}}_{\in(0,1)}\left(\Theta - \delta(1+\Phi)\right) > 0.
\end{align*}
and that $\Omega < \Theta$ if $\Gamma <  \Theta $ and $\Omega > \Theta$ if $\Gamma > \Theta$.

\subsubsection{The NK Phillips Curve}

Since  $\hat{w}_t = \Phi \hat{y_t} - \sigma \frac{\tilde{\varphi}}{1-\mathcal{G}}\tilde{a}_t$, the New Keynesian Phillips curve $\pi_t = \kappa_w \hat{w}_t + \beta \mathbb{E}_t \pi_{t+1}$ becomes
\begin{align}
    \pi_t = \kappa \hat{y}_t - \kappa_a \tilde{a}_t + \beta \mathbb{E}_t \pi_{t+1} \label{teq:NKPC}
\end{align}
with $\kappa_a := \kappa_w\sigma\tilde{\varphi}/(1-\mathcal{G})$. The additional term appears since the cost of liquidity provision affects the level of consumption and thus the labor supply condition. While this effect should be very small for reasonable parameterizations, it is not 0 at first order. If $\tilde{b}_t = 0$ in the baseline model version, the conventional NKPC is recovered. For the integrated asset market version, we have $\tilde{a}_t = \Lambda \hat{y}_t$ without any shocks hitting, in which case \eqref{teq:NKPC} is isomorphic to the standard case with adjusted slope $\tilde{\kappa} = \kappa - \kappa_a\Lambda$.

\subsection{Analysis of the simple fiscal shock}\label{app:fisc_shock}

We start by deriving the responses of output and inflation to the baseline model in which necessarily $(1-\lambda)\tilde{a}_t = \tilde{b}_t$. Since $\tilde{t}_t=0$ for $t\geq 1$, the IS relationship in these periods is given by 
\begin{align*}
    \Theta \hat{y}_t = \Gamma \mathbb{E}_t \hat{y}_{t+1} - \frac{1}{\sigma}\left(\hat{i}_t - \mathbb{E}_t \pi_{t+1}\right) + \Theta_b \tilde{b}_t + \Gamma_b \tilde{b}_{t+1}
\end{align*}
where $\Theta_b:= \Theta_a/(1-\lambda)$ and $\Gamma_b := \Gamma_a/(1-\lambda)$.
Given the exogenous process $\tilde{b}_{t+1}=\rho_b \tilde{b}_t$, we can use the methods of undetermined coefficients to solve for the responses of output and inflation for $t\geq 1$. Conjecture that $\hat{y}_t = \psi_y b_t$ and $\pi_t = \psi_\pi b_t$. Plugging this into the above equation and the Phillips curve, we get the relationship
\begin{align*}
    \psi^{BL}_\pi = \frac{\kappa \psi_y - \kappa_b}{1-\rho_b \beta}
\end{align*}
where similarly $\kappa_b := \kappa_a/(1-\lambda)$. We can subsequently use this in the IS condition to obtain 
\begin{align}
    \psi^{BL}_y = \frac{\Theta_b(1-\rho_b\chi) +\rho_b(1-\chi)\frac{\tilde{\alpha}}{1-\lambda} + \frac{\kappa_b}{\sigma}\frac{\phi_\pi - \rho_b }{1-\rho_b\beta}}{\Theta - \Gamma \rho_b + \frac{\kappa}{\sigma}\frac{\phi_\pi - \rho_b }{1-\rho_b\beta}}. \label{app:psi_y_BL}
\end{align}
The numerator of this expression is generically positive, while $\Gamma<\Theta$ ensures that the denominator is as well. In that case, if public debt is elevated in the aftermath of a fiscal shock, this will be accompanied by higher output, and, given that $\kappa_b$ should be small under any reasonable calibration, also higher inflation. What about period $t=0$? Assuming that the one-time transfer $\tilde{t}_0$ is financed by issuing an equal amount of public debt $\tilde{b}_1 = \tilde{t}_0$, we will have 
\begin{align}
    \pi_0 = \kappa \hat{y}_0 + \beta \mathbb{E}_0 \pi_1 = \kappa \hat{y}_0 + \beta \psi_\pi \tilde{b}_1 \label{teq:pi0_seg}
\end{align}
(if the economy starts in steady state, the $\kappa_b \tilde{b}_0$ in \eqref{teq:NKPC} is 0). 
We can now substitute all endogenous terms in the IS condition for $t=0$
\begin{align}
    \Theta\hat{y}_0 = \Gamma \psi^{BL}_y \tilde{b}_1 + \Gamma_b \tilde{b}_1 + \theta_h \frac{Y_{ss}}{C^H_{ss}}\tilde{t}_0 -\frac{1}{\sigma}\left(\phi_\pi(\kappa y_0 + \beta \psi^{BL}_{\pi}\tilde{b_1}) - \psi^{BL}_{\pi}\tilde{b_1}\right) \label{app:IS_0_seg}
\end{align}
we can obtain 
\begin{align}
    \hat{y}_0^{BL} = \frac{\theta_h \frac{Y_{ss}}{C^H_{ss}}\tilde{t}_0 + \left(\Gamma \psi^{BL}_y + \Gamma_b + \frac{(1-\phi_\pi\beta)}{\sigma}\psi_{\pi}^{BL}  \right)\tilde{b}_1 }{\Theta + \frac{\kappa \phi_\pi}{\sigma}} . \label{app:y0_seg}
\end{align}
Now, a slight complication arises in that $\tilde{t}$ is endogenous, in the sense that in period $t=0$, the real response to fiscal expansion provides the government with extra resources that it can distribute as well. To back out what amount of $\tilde{t}_0$ is consistent with the exogenously given we can turn to \eqref{app:lin_budget}, which implies under $\tilde{\tau}_t = 0$ and $\tilde{a}=\tilde{b}_0=0$ that 
\begin{align*}
\tilde{t}_0 = \tilde{b}_1 + \phi_\tau (1+\Phi) \hat{y}_0.
\end{align*}
Substituting this into \eqref{app:y0_seg}, we can solve out $\hat{y}_0$ to obtain
\begin{align}      
\hat{y}_0 = \frac{\theta_h \frac{Y_{ss}}{C^H_{ss}} + \left(\Gamma \psi^{BL}_y + \Gamma_b + \frac{(1-\phi_\pi\beta)}{\sigma}\psi_{\pi}^{BL}  \right)}{\Theta + \frac{\kappa \phi_\pi}{\sigma} - \theta_h \frac{Y_{ss}}{C^H_{ss}}\phi_\tau(1+\Phi)}\tilde{b}_1. \label{app:y0_seg_end}
\end{align}
We note that since $(1-\phi_\pi\beta)<0$, $\left(\Gamma \psi^{BL}_y + \Gamma_b + \frac{(1-\phi_\pi\beta)}{\sigma}\psi_{\pi}^{BL}  \right)$ and in turn $\hat{y}_0$ cannot be guaranteed to be positive for arbitrary calibrations.\footnote{While $\Gamma_b$ can be negative, we will typically have $\Gamma \psi^{BL}_y + \Gamma_b >0$ as I explain below} However, for typical calibrations without overly strong central bank reaction $\phi_\pi$ and a NKPC slope $\kappa$ of moderate size, it seems to be the case. 

Let us now turn to the model version with integrated asset markets. Abstracting from further shocks in $t\geq 1$, the inflation coefficient will have to satisfy
\begin{align*}
    \pi_t = \tilde{\kappa} \hat{y}_t + \beta \mathbb{E}_t \pi_{t+1}  \implies \psi^{INT}_{\pi} = \frac{\tilde{\kappa} \psi^{INT}_y}{1-\rho_b \beta}
\end{align*}
in the aftermath of the shock. Turning to the IS condition, we note that for $t>1$, one can substitute any instance of $\tilde{a}_t$ by $\Lambda \hat{y}_t$ (no additional shocks), implying we will end up with system \eqref{teq:int_IS} which is \emph{independent} of $\tilde{b}$. Hence, output and inflation in $t\geq 1$ must evolve independently of public debt $\tilde{b}$ and we have $\hat{y}_t = \pi_t = 0$ for all $t\geq 1$.
This makes solving for the initial responses in $t=0$ straightforward, as we will have $\pi_0 = \kappa \hat{y}_0$ and thus
\begin{align}
    \Theta \hat{y}_0 &= (\Theta - \Theta_a \Lambda)\underbrace{\mathbb{E}_t\hat{y_1}}_{=0} + \theta_h \frac{Y_{ss}}{C^H_{ss}}\tilde{t}_0 - \frac{1}{\sigma}\left(\phi_\pi \kappa \hat{y}_0 - \underbrace{\pi_1}_{=0}\right) \nonumber \\
    \implies \hat{y}^{INT}_0 &= \frac{\theta_h \frac{Y_{ss}}{C^H_{ss}}\tilde{t}_0}{\Theta + \frac{\kappa \phi_\pi}{\sigma}} \implies   \hat{y}_0^{INT} = \frac{\theta_h \frac{Y_{ss}}{C^H_{ss}}\tilde{b}_1}{\Theta + \frac{\kappa \phi_\pi}{\sigma} - \theta_h \frac{Y_{ss}}{C^H_{ss}}\phi_\tau(1+\Phi)}.\label{app:y0_int_end}
\end{align}
Comparing this to \eqref{app:y0_seg_end}, we can see that $y_0^{BL} > y_0^{INT}$ if $\left(\Gamma \psi^{BL}_y + \Gamma_b + \frac{(1-\phi_\pi\beta)}{\sigma}\psi_{\pi}^{BL}  \right)$ is positive. This need not be the case for all reasonable calibrations, but is more likely to be the case if $\phi_\pi$ is not too large. Intuitively, since inflation is higher for $t\geq 1$ in the baseline model, it is also in $t=0$, causing the central bank to react more strongly which can potentially dampen the output response. \\
Nevertheless, we will typically have initial inflation $\pi_0$ being higher in the baseline model. Denoting as $\pi_0^{BL}-\pi_0^{INT}:=\Delta \pi_0$ the difference between initial inflation responses, the NK Phillips curve implies 
\begin{align*}
    \Delta \pi_0 = \kappa \underbrace{(\hat{y}_0^{BL} - \hat{y}_0^{INT})}_{:=\Delta \hat{y}_0} + \beta (\mathbb{E}_0 \pi_1^{BL} - \mathbb{E}_0 \pi_1^{INT}) = \kappa \Delta \hat{y}_0 + \beta \psi_\pi^{BL} \tilde{b}_1
\end{align*}
given the previous results. Since $\psi_\pi^{BL} >0$, clearly $\Delta \pi_0 >0$ if $\Delta \hat{y}_0 >0$. However, even if $\Delta \hat{y}_0 \leq 0$, the respective IS relations require that
\begin{align}
    \Delta \hat{y}_0 =  \frac{1}{\Theta}\left(\Gamma \psi^{BL}_y + \Gamma_b \right)\tilde{b}_1 - \frac{1}{\sigma \Theta}\left( \phi_\pi \Delta \pi_0 - \pi_1^{BL} \right) + \frac{\theta_h}{\Theta} \frac{Y_{ss}}{C^H_{ss}}\Delta \tilde{t}_0  \nonumber\\
    \Leftrightarrow 
    \Delta \hat{y}_0 \left[1-\phi_\tau(1+\Phi)\frac{\theta_h}{\Theta} \frac{Y_{ss}}{C^H_{ss}} \right] = \left[ \frac{1}{\Theta}\left(\Gamma \psi_y^{BL}+\Gamma_b \right)+\frac{1}{\sigma \Theta}\psi^{BL}_\pi\right]\tilde{b}_1 - \frac{1}{\sigma \Theta}\phi_{\pi} \Delta \pi_0 \label{app:Deltapi_0}
\end{align}
Following the arguments above, we should have $\psi^{BL}_{\pi} > 0$, and below, I further argue that we should usually have $\Gamma \psi^{BL}_y + \Gamma_b>0$ as well. If $\left[1-\phi_\tau(1+\Phi)\frac{\theta_h}{\Theta} \frac{Y_{ss}}{C^H_{ss}} \right]>0$, then by \eqref{app:Deltapi_0} we can only have $\hat{y}_0<0$ if $\Delta \pi_0 > 0$. The potential cause of initially lower output in the separated baseline model is precisely a higher interest which the central bank only engineers if inflation is indeed higher.\\
Note that conventional calibrations should fulfill
\begin{align}
    \left[\Theta-\phi_\tau(1+\Phi)\theta_h \frac{Y_{ss}}{C^H_{ss}} \right] = \theta_y [1-\lambda \phi_{\tau} (1+\Phi)]-\theta_h \delta (1+\Phi) >0 \label{app:Deltapi0_cond}
\end{align} 
given that $\theta_y -\theta_h \delta (1+\Phi)>0$ is equivalent to $\Theta >0$ and $\phi_\tau$ usually $<0.5$. In effect, condition \eqref{app:Deltapi0_cond} holds whenever the model is not close to the $\Theta = 0$ boundary.\\
Regarding the issue whether we should typically have $(\Gamma \psi^{BL}_y + \Gamma_b) > 0$: Considering \eqref{app:psi_y_BL} as well as the definition of $\Gamma_b:= ((1-\chi)\tilde{\alpha} - \chi \Theta_a)/(1-\lambda) $, this will be the case if 
\begin{align*}
    \frac{\Gamma(\Theta_a(1-\rho_b\chi) +\rho_b(1-\chi)\tilde{\alpha} )}{\Theta - \Gamma \rho_b + \bar{\kappa}} + (1-\chi)\tilde{\alpha} - \chi \Theta_a >0  \\
    \Leftrightarrow \Theta_a\left(\Gamma - \chi\Theta - \chi \bar{\kappa} \right) + (1-\chi)\tilde{\alpha}(\Theta + \bar{\kappa}) >0 , 
\end{align*}
where $\bar{\kappa} := \frac{\kappa}{\sigma}\frac{\phi_\pi - \rho_b }{1-\rho_b\beta}$. Recall that $\Theta_a$ equals $ \theta_h\tilde{\alpha}$ plus a small adjustment term depending on the liquidity cost $(1-\varphi)$, where $\theta_h := \frac{\lambda}{1-\lambda}\frac{C^H_{ss}}{C^S_{ss}}$. So, the condition will typically be satisfied if 
\begin{align*}
    \theta_h\left(\Gamma - \chi\Theta - \chi \bar{\kappa} \right) + (1-\chi)(\Theta + \bar{\kappa}) >0 \\
    \Leftrightarrow (1-\chi)\Theta + \theta_h(\Gamma - \chi \Theta) > \left[\chi(1+\theta_h)-1 \right]\bar{\kappa} \\
    (1-\chi)\theta_y > \left[\chi(1+\theta_h)-1 \right]\bar{\kappa}
\end{align*}
where the last step uses the definitions of $\Theta$ and $\Gamma$. If $\left[\chi(1+\theta_h)-1 \right]<0$, this holds trivially, but for many calibrations $\chi(1+\theta_h)>1$. Since $\theta_y := \frac{1}{1-\lambda}\frac{Y_{ss}}{C^S_{ss}}>1$ and $\left[\chi(1+\theta_h)-1 \right]/(1-\chi)$ should be of moderate size since $\theta_h<0.5$, $(\Gamma \psi^{BL}_y + \Gamma_a) > 0$ typically holds for moderately-sized $\bar{\kappa}$ (e.g., when the central bank reaction is not overly strong or the Phillips curve not very steep).

\subsection{Model determinacy properties}\label{app:determinacy}

Consider the baseline economy: After the inclusion of the feedback rules $(1-\lambda)\tilde{\tau}_t = \phi_b \tilde{b}_t$ and $\hat{i}_t = \phi_\pi \pi_t$, the system of equations characterizing the endogenous dynamics of the economy becomes  
\begin{align*} 
    \Theta \hat{y}_t &= \Gamma \mathbb{E}_t \hat{y}_{t+1} - \frac{1}{\sigma}\left(\phi_\pi \pi_t - \mathbb{E}_t \pi_{t+1}\right) + \Theta_b \tilde{b}_t + \Gamma_b \tilde{b}_{t+1} \\
    \pi_t &= \kappa \hat{y}_t - \kappa_b \tilde{b}_t + \beta \mathbb{E}_t \pi_{t+1} \\
    \tilde{b}_{t+1} &= \frac{1}{\beta}\tilde{b}_t - \phi_b \tilde{b}_t  - \phi_\tau \left[\left( \Phi + 1\right)\hat{y_t} - \frac{\sigma \tilde{\varphi}}{(1-\mathcal{G})}\tilde{b}_t\right]
\end{align*}
This is a system with two ``jump'' variables $\hat{y}_t$ and $\pi_t$ and one predetermined variable $\tilde{b}_t$. After substituting the $\mathbb{E}_t\pi_{t+1}$ using the NKPC, we can re-write it in the form $\mathbb{E}_t z_{t+1} = \mathbf{A} z_t $ with 
\begin{align*}
    \mathbf{A} = \begin{pmatrix}   \frac{\Theta^* + \frac{\kappa}{\sigma \beta}}{\Gamma} & - \frac{(1-\beta\phi_\pi)}{\Gamma \sigma \beta} & -\frac{(\Theta^*_b + \frac{\kappa_b}{\sigma \beta})}{\Gamma} \\ -\frac{\kappa}{\beta} & \frac{1}{\beta} & \kappa_b/\beta \\ -(\phi_\tau)(1+\Phi) & 0 & \delta_b\end{pmatrix}
\end{align*}
where
\begin{align*}
    \Theta^* &:= \Theta + \Gamma_b\phi_\tau(1+\Phi) \\
    \delta_b &:= \frac{1}{\beta} - \phi_b + \phi_\tau \frac{\sigma \tilde{\varphi}}{(1-\mathcal{G})} \\
    \Theta^*_b &:= \Theta_b + \Gamma_b\delta_b.
\end{align*}
We note that this system nests the standard RANK model as a special case with $\Theta^* = \Gamma = 1$ and $\Theta_b = \kappa_b = 0$. For the purpose of studying determinacy, it also nests the model with integrated assets markets if $\sigma$ and $\kappa$ are replaced with the appropriate adjusted values $\tilde{\sigma}= \sigma \Omega$ and $\tilde{\kappa}=\kappa-\kappa_a \Lambda$.\footnote{Due to certainty equivalence, studying the deterministic case providing for $\tilde{a}_t = \Lambda \hat{y}_t$ is sufficient for this purpose.}\\
To proceed, I build on Proposition C.2 in \cite{woodford2003}, which provides necessary and sufficient conditions for determinacy in a system of the form $\mathbb{E}_t z_{t+1} = \mathbf{A} z_t$ with $2$ jump variables and one predetermined variable. 
Specifically, said Proposition states that determinacy can arise in three mutually exclusive cases, each stated in terms of the characteristic polynomial of $\mathbf{A}$
\begin{align*}
    \mathcal{P}(\lambda) = \lambda^3 + A_2 \lambda^2 + A_1 \lambda + A_0
\end{align*}
evaluated at $\lambda = 1$ and $\lambda = -1$, as well as additional conditions on the coefficients $A_0$, $A_1$ and $A_2$. For the analysis below, I shall use $\kappa_b \approx 0$, noting that the model's determinacy properties are unlikely to change substantially due to this term given that $\kappa_w \tilde{\varphi}$ should be small under any reasonable calibration.\footnote{The only need for the liquidity cost term $\tilde{\varphi}$ in the baseline model with separated asset market is to allow for steady state equivalence with the integrated alternative. Hence, the determinacy properties of a model version without it is arguably the economically interesting object anyway.}
We have 
\begin{align*}
    A_2 &= -tr(\mathbf{A}) = -\left(\frac{\Theta^* + \frac{\kappa}{\sigma \beta}}{\Gamma} + \frac{1}{\beta} + \delta_b\right) \\
    A_1 &= M_{12} + M_{13} + M_{23} = \frac{\Theta^* + \frac{\kappa \phi_\pi}{\sigma}}{\Gamma \beta} + \frac{1}{\Gamma}\left(\left(\Theta^* + \frac{\kappa}{\sigma\beta}\right)\delta_b - \Theta_b^*\phi_\tau(1+\Phi) \right) + \frac{\delta_b }{\beta} \\
    A_0 &= -\det(\mathbf{A}) = -\frac{1}{\Gamma \beta}\left[\left(\Theta^*+\frac{\kappa \phi_\pi}{\sigma}\right)\delta_b  - \Theta_b^*\phi_\tau(1+\Phi) \right]
\end{align*}
where the $M$ terms denote the principal minors of $\mathbf{A}$. \\
Evaluating $\mathcal{P}(-1) = -1 + A_2 - A_1 + A_0$, we obtain after some re-arranging
\begin{align}
    \mathcal{P}(-1) = \frac{1}{\Gamma \beta}\left[\Theta_b^*\phi_\tau(1+\Phi)(1+\beta) - (1+\delta_b)\left(\left(\Theta^*+\Gamma \right)(1+\beta)+\frac{\kappa (1+\phi_\pi)}{\sigma}  \right)\right]. \label{apeq:P_minus_1}
\end{align} 
In the case of a RANK model or the THANK with integrated asset markets, we have $\Theta_b^* = 0$, so clearly $\mathcal{P}(-1)<0$. In the baseline model, this does not have to be the case for arbitrary calibrations, which already highlights the more complex determinacy properties of this model version. However, for reasonable calibrations, we should have $\mathcal{P}(-1)<0$ as well.
In this regard, note that $\mathcal{P}(-1)<0$ if 
\begin{align*}
(1+\delta_b)\left(\left(\Theta^*+\Gamma \right) \right) &> \Theta_b^*\phi_\tau(1+\Phi) \\
\Leftrightarrow (1+\delta_b)\left(\left(\Theta+\Gamma \right)\right) &> (\Theta_b - \Gamma_b)\phi_\tau(1+\Phi)
\end{align*}
which uses the definition of $\Theta^*$, $\Theta_a^*$ and is already weaker than necessary due to the omission of the $\frac{\kappa(1+\phi_\pi)}{\sigma}$ term. We further note that $\Theta+\Gamma \geq (1+\chi)\Theta$, and $\Theta_b-\Gamma_b \approx [((1+\chi)\theta_h - (1-\chi))\alpha]/(1-\lambda) $ if we omit the small adjustment terms stemming from the liquidity cost. Since $\alpha > 0$, we can thus expect $\mathcal{P}(-1)<0$ under the even weaker condition 
\begin{align}
    \Theta > \frac{\theta_h \alpha \phi_\tau}{(1+\delta_b)(1-\lambda)}(1+\Phi) = \frac{(1-s)\varphi}{(1+\delta_b)\beta}\frac{Y_{ss}}{C^S_{ss}}\phi_\tau\frac{(1+\Phi)}{1-\lambda}  \label{app:Pmin1_cond} \\
    \Leftrightarrow 1 > \frac{(1-s)\varphi}{(1+\delta_b)\beta}\frac{\tau_y}{\mu}(1+\Phi)  + \lambda\delta\frac{C_{ss}^H}{Y_{ss}} (1+\Phi) \nonumber,
\end{align}
where the second step uses the definitions of the various composite parameters. Note that the $\Theta>0$ condition is equivalent to $1> \lambda\delta\frac{C_{ss}^H}{Y_{ss}} (1+\Phi)$.\footnote{$\theta_y - \theta_h \delta(1+\Phi) > 0 \Leftrightarrow \frac{Y_{ss}}{(1-\lambda)C^S_{ss}} > \frac{\lambda}{1-\lambda}\frac{C^H_{ss}}{C^S_{ss}}\delta (1+\Phi)\Leftrightarrow 1 > \lambda C^H_{ss}\delta(1+\Phi)/Y_{ss}$.} Additionally, we notice that $\frac{(1-s)\varphi}{(1+\delta_b)\beta}\frac{\tau_y}{\mu}$ will typically be small.
All calibrations considered by \cite{bilbiie2025} have $s>0.8$. If $\delta_b>0$, $\tau_y < 0.5$ and $R^{A}_{SS}=\varphi/\beta \approx 1$, that means $\frac{(1-s)\varphi}{(1+\delta_b)\beta}\frac{\tau_y}{\mu}$ should be at most around $0.08$.
Considering that typical calibrations of $\Phi = (\phi + \sigma/(1-\mathcal{G}))$ are around 2 or 3 and condition \eqref{app:Pmin1_cond} was weaker than necessary, we can expect $\mathcal{P}(-1)<0$ for any reasonable calibration that is not too close to the $\Theta \leq 0$ boundary.
\\
Now, if $\mathcal{P}(-1)<0$, \cite{woodford2003}'s Proposition C.2 implies that a necessary condition for determinacy is $\mathcal{P}(1) > 0$. Evaluating $\mathcal{P}(1) = 1 + A_2 + A_1 + A_0$, we obtain after some re-arranging
\begin{align}
    \mathcal{P}(1) &= 1 + A_2 + A_1 + A_0 \nonumber\\
    &= \frac{1}{\Gamma \beta}\left[(1-\delta_b)\left((\Theta-\Gamma)(1-\beta) + \frac{\kappa(\phi_\pi-1)}{\sigma} \right) + (1-\beta)(\Theta_b +\Gamma_b)\phi_\tau (1+\Phi) \right] \label{apeq:P_plus_1} 
\end{align}
We note that in the case of RANK and the integrated asset market version of the THANK, $\Theta^* = \Gamma=1$ and $\Theta_a^*=0$, the condition $\mathcal{P}(1)>0$ simplifies to 
\begin{align*}
    (1-\delta_b)(\phi_\pi - 1) > 0
\end{align*} 
which is this models' version of \cite{leeper1991}'s result that determinacy can only obtain if either a) the monetary authority reacts more than 1-to-1 to inflation $\phi_\pi > 1$ while the fiscal authority stabilises debt $\delta_b < 1$ (``active monetary, passive fiscal'') or b) the monetary authority reacts less than 1-to-1 to inflation $\phi_\pi < 1$ while the fiscal authority doesn't stabilise debt $\delta_b > 1$ (``passive monetary, active fiscal'').
In contrast, the full condition \eqref{apeq:P_plus_1} doesn't factor cleanly, so even conditionally on $\phi_\pi > 1$ or $\delta_b > 1$, the possible values of the respective other parameters that can ensure $\mathcal{P}>0$ will have to fulfill
\begin{align*}
 (1-\delta_b)(\phi_\pi - 1) > -\frac{\sigma(1-\beta)}{\kappa}\left[(1-\delta_b)(\Theta-\Gamma)+(\Theta_b +\Gamma_b)\phi_\tau (1+\Phi) \right]   ,
\end{align*}
and thus, unlike in Leeper's condition, depend generically on the respective other value. 
\\ 
Ultimately, note that given $\mathcal{P}(-1)<0$, $\mathcal{P}(1)>0$ is only a necessary but not sufficient condition for determinacy. For sufficiency, \cite{woodford2003}'s Proposition requires either $\vert A_2 \vert > 3$ or $A^2_0 - A_0 A_2 + A_1 - 1>0$. Note that the first condition will typically be satisfied if $\delta_b$ is close to 1 or above, i.e., around or beyond the boundary between passive and active fiscal policy in standard models. For lower $\delta_b$, condition $A^2_0 - A_0 A_2 + A_1 - 1 >0$ needs to be checked. While analyzing the expression directly is very tedious, there are arguments that $\mathcal{P}(1)>0$ will be typically sufficient (and not only necessary) for determinacy in case $\delta_b\in(0,1)$, at least under IS equation discounting (above $\vert A_2 \vert > 3$ can be expected to apply).\\
Drawing on similar arguments as above, we should have $\Theta^*/\Gamma > \Theta_b \phi_\tau (1+\Phi)$ under reasonable calibrations providing for IS equation discounting $\Gamma/\Theta < 1$. That means there is a $\delta_{b0} \in (0,1)$ for which $A_0 = 0$ and $A_0>0$ for $\delta_b<\delta_{b0}$. Since $A_2$ is clearly negative for any $\delta_b \in (0,1)$, it follows that all products in $A^2_0 - A_0 A_2$ are positive for $\delta_b \in (0,\delta_{b0})$ and given the size of the $\Theta^*/\Gamma$ term in $A_1$, one can expect that $A^2_0 - A_0 A_2 + A_1 - 1 >0$ will hold for all $\delta_b \in (0,\delta_{b0})$. For the same reason, we should have $A^2_0 - A_0 A_2 + A_1 - 1= A_1 - 1>0$ at $\delta_{b0}$ itself.\\
What about the case $\delta_{b}\in (\delta_{b0},1)$? We note that $-\det(\mathbf{A})=A_0$ is $<0$ and monotonously decreasing in $\delta_b$ in that case. In turn, as $\det(\mathbf{A})=\lambda_1 \lambda_2 \lambda_3$ (where $\lambda$'s denote the eigenvalues of $\mathbf{A}$), the eigenvalue product must be $>0$ and increasing in $\delta_b$. If we have determinacy at $\delta_{b0}$ and thus two eigenvalues outside the unit circle, the increase should be driven by the eigenvalue inside the unit circle as long as $\mathcal{P}(1)>0$ and $\mathcal{P}(-1)<0$ hold so that there is no unit circle crossing at $\pm 1$. Given that determinacy is again ensured for high $\delta_b$, one would expect determinacy properties to be preserved for $\delta_b \in (\delta_{b0},1)$ as well. However, this is a heuristic argument and doesn't strictly rule out crossings by complex conjugate pairs in this range.

\subsection{Extended LAF structure in THANK}\label{app:thank_LAF}

A version of the extended LAF as described in Section \ref{sec:lafs} can also be introduced into the THANK setup. However, as the THANK model was linearized around $B_{ss}=A_{ss}=0$ for tractability, it requires a slight reformulation of the LAF setup, given that the equivalent to the ratio $B^l_{ss}/A^l_{ss}$ is no longer well-defined. \\
As a consequence, I now assume it faces the problem 
\begin{align}
    \max_{B^l_t}~~ & \mathbb{E}_t \left[\varphi (R^B_{t+1}B^l_{t} + R^I_{t+1}(A^l_{t+1}-B_t^l))- \frac{\Psi}{2 Y_{ss}}(B_{t+1}^l - A^l_{t+1})^2\right]  \label{eq:thank_laf} 
\end{align}
featuring the quadratic cost in levels instead of portfolio shares. The constant liquidity cost again reverts to the ``iceberg'' specification previously used in the THANK. Since relationship between the central bank rate and government bond rate is trivial in the absence of long-term bonds, I also omit the reserve asset from the formulation.\\
The LAF FOC and its linearized counterpart are 
\begin{align}
    \varphi(\mathbb{E}_tR^B_{t+1}- E_t R_{t+1}^I) = \frac{\Psi}{Y_{ss}}(A^l_{t+1}-B_{t+1}^l) \nonumber \\
    \implies \mathbb{E}_t \hat{r}^I_{t+1} - \mathbb{E}_t \hat{r}^B_{t+1} = \bar{\Psi}\left[\tilde{b}_{t+1} - (1-\lambda)\tilde{a}_{t+1}\right],\label{appeq:thank_foc}
\end{align}
which uses liquid asset market clearing $A_t^l = (1-\lambda)A_t$ and defines $\bar{\Psi} := \Psi/R^A_{ss} = \Psi \beta/\varphi$. For the liquid asset return, we have 
\begin{align}
   R_t^A &= \varphi \left[R_t^B\frac{B_t^l}{A_t^l}+R_t^I \frac{A_t^l-B_t^l}{A_t^l}\right] - \frac{\Psi}{2Y_{ss}A_t ^l }(B_t^l - A_t^l)^2  \nonumber \\
   \implies \hat{r}_t^A &= \hat{r}_t^B
\end{align}
around $A^l_{ss}=B^l_{ss}$.\footnote{The expression for $R_t^A$ divides by $A_t^l$ which is $A_{ss}^l=0$ in the considered SS. However, it has the limit $R_{ss}^A = \varphi R_{ss}^B$ in the case $A_{ss}^l = B_{ss}^l \rightarrow 0$, and $A^l_{ss}= B^l_{ss}$ is required by SS invariance. Similarly, its linearized version holds for any $A_{ss}^l = B_{ss}^l > 0$ and thus also applies at the limit $A_{ss}^l = B_{ss}^l\rightarrow 0$.}

\subsubsection{Liquid savings dynamics}

Recall that if we substitute $\hat{c}_t^S$ in the linearized liquid asset Euler equation using the illiquid Euler, we obtain 
\begin{align*}
    0= \frac{1}{\sigma}(\mathbb{E}_t \hat{r}^I_{t+1} - \mathbb{E}_t \hat{r}^B_{t+1}) + (1-\chi)(\mathbb{E}_t \hat{c}^H_{t+1}- \mathbb{E}_t \hat{c}^S_{t+1}).
\end{align*}
Substituting the return difference using \eqref{appeq:thank_foc} and the consumption terms using \eqref{appeq:CS_lin} and \eqref{appeq:CH_lin}, we obtain
after some re-arranging
\begin{align}
   \tilde{a}_{t+1} = \upsilon(\Psi) \frac{\tilde{b}_{t+1}}{1-\lambda} + (1-\upsilon(\Psi))\Lambda \mathbb{E}_t \hat{y}_{t+1} ~~\text{with}~~ \upsilon(\Psi) := \frac{\frac{1}{\sigma}(1-\lambda)\bar{\Psi}}{\frac{1}{\sigma}(1-\lambda)\bar{\Psi} + (1-\chi)(\tilde{\alpha}+\Theta_a)} \in (0,1) \label{appeq:LAF_atilde}
\end{align}
(taking the expected transfer shock to be $0$). The liquid savings dynamics turn out to be a convex combination of the ones under separated and integrated asset markets, with the weights determined by the LAF cost $\Psi$.
If $\Psi \rightarrow \infty$, we have $\upsilon \rightarrow 1$ and the separated dynamics $ (1-\lambda)\tilde{a_t}= \tilde{b}_t$ are recovered. If $\Psi \rightarrow 0$, we have $\upsilon \rightarrow 0$ and the integrated dynamics $ \tilde{a}_{t+1}= \Lambda \mathbb{E}_t\hat{y}_{t+1}$ obtain.

\subsubsection{Aggregate IS relation}

Building on the derived liquid asset dynamics, we can now characterize the aggregate demand IS relation for our ``in-between'' economy. Note that Equation \eqref{appeg:liq_IS} derived earlier does not change compared to the baseline --- the LAF extension only enters the IS curve through equation \eqref{appeq:LAF_atilde}. Focussing on the case with no shocks hitting in $t$ so that $\mathbb{E}_{t-1}\hat{y}_t = \hat{y}_t$, we can combine these equations to obtain the aggregate IS condition 
\begin{align}
    \Theta_\Psi \hat{y}_t = \Gamma_\Psi \mathbb{E}_t\hat{y}_{t+1} - \frac{1}{\sigma}(\hat{i}_t - \mathbb{E}_t\pi_{t+1}) + \upsilon(\Psi)(\Gamma_b \tilde{b}_{t+1} + \Theta_b \tilde{b}_{t}) \label{appeq:thank_laf_IS}
\end{align}
where 
\begin{align*}
    \Theta_\Psi &:= \Theta - \Theta_a (1-\upsilon(\Psi))\Lambda  \\
    \Gamma_\Psi &:= \Gamma + \Gamma_a(1-\upsilon(\Psi))\Lambda.
\end{align*}
If $\Psi\rightarrow \infty$ and $\upsilon(\Psi)\rightarrow 1$, \eqref{appeq:thank_laf_IS} reduces to the IS relation for the separated Baseline economy \eqref{teq:IS_seg} as stated in the main text. If $\Psi\rightarrow 0$ and $\upsilon(\Psi)\rightarrow 0$, the integrated IS condition \eqref{teq:int_IS} obtains.\footnote{Recall that $(\Theta_a + \tilde{\alpha})\Lambda = \Theta - \delta(1+\Phi)$. In turn, using the definition of $\Gamma_a := (1-\chi)\tilde{\alpha} - \chi \Theta_a$, we have
\begin{align*}
\Gamma + \Gamma_a \Lambda = \chi \Theta + (1-\chi)\delta(1+\Phi)+(1-\chi)(\tilde{\alpha}+\Theta_a)\Lambda - \Theta_a \Lambda = \Theta - \Theta_a \Lambda.
\end{align*}
}\\
As for the liquid asset dynamics, introducing the LAF allows one to effectively interpolate between the separated and integrated model IS curves, with the interpolation weight determined by the LAF cost $\Psi$.

\subsubsection{New Keynesian Phillips curve}

In the baseline model, liquid asset dynamics also introduced a small term into the NKPC, as the LAF costs affect aggregate consumption and thus the labor supply condition $\tilde{w}_t = \Phi \hat{y}_t - \sigma\frac{\tilde{\varphi}}{1-\mathcal{G}}\tilde{a_t}$.
In turn, the NKPC in the ``in-between economy'' will be of the form 
\begin{align*}
    \pi_t = \kappa_\Psi \hat{y}_t - \kappa_{b,\Psi} \tilde{b}_t + \beta \mathbb{E}_t \pi_{t+1}
\end{align*} 
where $\kappa_\Psi := \kappa - (1-\upsilon(\Psi))\kappa_a \Lambda$ and $\kappa_{b,\Psi}:= \kappa_a\upsilon (\Psi)/(1-\lambda)$. Evidently, the ``interpolation'' property also extends to this part of the model.

\subsubsection{Structural similarity to the separated baseline}

Above, it became clear that by extending the LAF with the adjustment cost set-up, one can effectively interpolate the key model conditions between the separated and integrated case, with the interpolation weights determined by the cost $\Psi$. But what about the subsequent model results?

Note that \eqref{appeq:thank_laf_IS} has the same structure as the baseline IS condition \eqref{teq:IS_seg}. The same holds for the NKPC. In turn, by substituting $\Theta \rightarrow \Theta_\Psi$, etc., the derivations for the out-of-steady-state results, such as the fiscal shocks or model determinacy carry over with parameter substitutions. Everything can proceed with same method as in the baseline, but with $\Psi$-dependent coefficients, coinciding with the integrated and separated cases at the limits.

This structural similarity also encapsulates insights about the usefulness of varying the asset market structure. In principle, the dynamics of the ``in-between'' economy could also be generated by a separated THANK model with different $\Theta, \Gamma$, and so on. But changing the size of these composites would require different micro-heterogeneity: For example, the composite parameters $\Theta_b$ and $\Gamma_b$ depend practically only on heterogeneity-related moments (and a steady-state interest rate). Thus, generating the LAF-driven dynamics through recalibration would require shifting these moments.

\section{The quantitative 2-asset HANK model}\label{app:two_asset_hank}

\subsection{The household problem in recursive form}\label{app:ha_recursive}

Letting $\Gamma_t$ denote a set containing the economy's \emph{aggregate} state at period $t$, we are now ready to state the Bellman equations corresponding to the households' dynamic utility maximization problem, which are 
\begin{align}
 V^{a}(a_{it}, k_{it}, s_{it}, \Xi_{it}; \Gamma_t)    &= \max_{c_{it}, k_{it+1}, a_{it+1}} \left\lbrace
 \frac{c_{it}^{1-\xi}-1}{1-\xi} - \varsigma\frac{N_t^{1+\gamma}}{1+\gamma}  +\beta \mathbb{E}_t A_{t+1} V(a_{it+1}, k_{it+1}, s_{it+1}, \Xi_{it+1}; \Gamma_{t+1}) \right\rbrace \nonumber\\
 &\text{s.t. to \eqref{BC_adjust}, \eqref{inc_eq}, $k_{it}\geq 0$ and $a_{it}\geq \underbar{a}$ } \label{Vadjust_eq}
\end{align}
for a household able to adjust its capital stock and
\begin{align}
 V^{na}(a_{it}, k_{it}, s_{it}, \Xi_{it}; \Gamma_t)    &= \max_{c_{it}, a_{it+1}} \left\lbrace
 \frac{c_{it}^{1-\xi}-1}{1-\xi}  - \varsigma\frac{N_t^{1+\gamma}}{1+\gamma} +\beta \mathbb{E}_t A_{t+1} V(a_{it+1}, k_{it},s_{it+1}, \Xi_{it+1}; \Gamma_{t+1}) \right\rbrace \nonumber\\
 &\text{s.t. to \eqref{BC_nonadjust}, \eqref{inc_eq}, $k_{it}\geq 0$ and $a_{it}\geq \underbar{a}$ } \label{Vnonadjust_eq}
\end{align}
for a household that is unable to do so. The ex-ante value function $V(\cdot)$ is given by
\begin{align*}
V(a_{it+1}, k_{it+1}, s_{it+1}, \Xi_{it+1}; \Gamma_{t+1}) &= \lambda  V^{a}(a_{it+1}, k_{it+1}, s_{it+1}, \Xi_{it+1}; \Gamma_{t+1}) \\
&+ (1-\lambda) V^{na}(a_{it+1}, k_{it+1}, s_{it+1}, \Xi_{it+1}; \Gamma_{t+1})  .
\end{align*}

\subsection{Production}

As \cite{bayerEtAlaer}, I make the simplifying assumption that firms solving forward-looking problems (such as the retailers' price setting problem) discount the future at the households' discount parameter $\beta$.\footnote{Since I will linearize the model with respect to aggregate shocks, only the steady-state value of the discount factor in the firms' dynamic problems will matter for the dynamic model responses. \cite{bayerEtAl2019} and \cite{lee2021b} report that using different specifications does not significantly affect results in their 2-asset HANK models with many similar features.}
\subsubsection{Final goods production}\label{subsec:final_goods}

The economy's final good is produced by a representative firm that assembles a set of differentiated inputs $j\in[0,1]$ according to the CES technology
\begin{align}
    Y_t = \left(\int^1_0 y_{jt}^{\frac{\epsilon_t-1}{\epsilon_t}}dj\right)^{\frac{\epsilon_t}{\epsilon_t-1}}. \label{AggY_eq}
\end{align}
Denoting the intermediate prices as $p_{jt}$, this gives rise to the familiar input demand schedule 
\begin{align}
    y_{jt} = \left(\frac{p_{jt}}{P_t}\right)^{-\epsilon_t} Y_t \label{ydemand_eq}
\end{align}
with optimal price index $P_{t} = (\int^1_0 p_{jt}^{1-\epsilon_t})^{1/(1-\epsilon_t)}$.
Note that I allow the elasticity of substitution between different varieties to exogenously vary over time, i.e., I allow for what is commonly referred to as ``cost-push'' shocks in the literature. For notational convenience, I define $\mu_t:= \frac{\epsilon_t}{\epsilon_t-1}$ to denote the target mark-up of the retailers presented in the next section.

\subsubsection{Retailers}

There is a unit mass of retailers, each of which produces a given variety of the differentiated input as a monopolist, taking into account demand schedule \eqref{ydemand_eq}. Their only input are intermediate goods, which they purchase at real price $mc_t$ (also referred to as ``marginal costs'') from the competitive intermediate goods producers. However, they are subject to nominal rigidities \`a la \cite{calvo1983}, i.e., in any given period their nominal price remains fixed with probability $\lambda_Y$. 

If receiving a re-set opportunity, a retailer will choose a price to maximize the corresponding expected net present value of real profits
\begin{align*}
   \mathbb{E}_0 \sum^{\infty}_{t=0}\beta^t \lambda_Y^t \left( \frac{p^*_{jt}}{P_t}-mc_t\right)\left(\frac{p^*_{jt} }{P_t}\right)^{\frac{-\mu_t}{\mu_t-1}}Y_t.
\end{align*}
In equilibrium, log-linearizing the first order conditions of the resulting price setting problem and combining them with the (also linearized) equations governing price dynamics will give rise to the standard log-linear Phillips curve 
\begin{align}
  \log\left(\pi_t\right) = \kappa_Y \left(mc_t - \frac{1}{\mu_t}\right) + \beta \mathbb{E}_t  \log\left(\pi_{t+1}\right)  \label{phillips_eq}
\end{align}
with $\kappa_Y := \frac{(1-\lambda_Y)(1-\lambda_Y\beta)}{\lambda_Y}$. 

\subsubsection{Intermediate goods producers}

The homogeneous intermediate good is produced by a continuum of firms that use a constant-returns-to-scale technology represented by production function
\begin{align}
 F(u_tK_t, H_t) =  (u_t K_t)^\alpha H_t^{1-\alpha}.\label{F_eq}
\end{align}
$K_t$ and $H_t$ denote the input of capital and labor services. $u_t$ is the degree of capital utilization that determines capital depreciation according to 
\begin{align*}
    \delta(u_t) = \delta_0 + \delta_1(u_t-1) + \frac{\delta_2}{2}(u_t-1)^2.
\end{align*}
Taking the price $h_t$ for labor services as well as the capital rental rate $r_t$ and its output price $mc_t$ as given, an intermediate goods producer solves the static profit maximization problem
\begin{align*}
    \max_{K_t, H_t, u_t} mc_t F(u_tK_t, H_t)  - h_tH_t - (r_t + q_t\delta(u_t))K_t,
\end{align*}
the solution of which can be characterized using the following first order conditions:
\begin{align}
    h_t &= (1-\alpha)mc_t (u_t K_t)^\alpha H_t^{-\alpha}\label{foc_H_eq}  \\
    r_t + q_t\delta(u_t) &= \alpha mc_t  u_t (u_t K_t)^{\alpha-1} H_t^{1-\alpha}   \label{foc_K_eq} \\
    q_t(\delta_1+\delta_2(u_t-1)) &= \alpha mc_t  (u_t K_t)^{\alpha-1} H_t^{1-\alpha}.\label{foc_u_eq}
\end{align}
\subsubsection{Capital goods producer}

Capital goods producers use the final good as input and operate a technology subject to adjustment costs: Using $I_t$ units of the final good, they can produce
\begin{align*}
Z_t^I\left[1-\frac{\phi}{2}\left(\frac{I_t}{I_{t-1}}-1\right)^2\right]I_t 
\end{align*}
units of capital. Investment-specific productivity $Z^I_t$ is exogenous and potentially following a time-varying shock process. 
Taking the price of capital $q_t$ as given, the producers choose $I_t$ to maximize the net present value of real profits 
\begin{align*}
   \mathbb{E}_0 \sum^{\infty}_{t=0} \beta^t \left(q_t Z_t^I \left[1-\frac{\phi}{2}\left(\frac{I_t}{I_{t-1}}-1\right)^2\right]I_t - I_t \right)~~
\end{align*}
and their optimal interior solution will fulfill first-order condition
\begin{align}
    1 + q_t Z_t^I\left(\frac{\phi}{2}\left(\frac{I_t}{I_{t-1}}-1\right)^2 - 1 + \phi\left(\frac{I_t}{I_{t-1}}-1 \right)\frac{I_t}{I_{t-1}}\right) = \beta \mathbb{E}_t q_{t+1} Z_{t+1}^I \phi\left(\frac{I_{t+1}}{I_{t}}-1 \right)\left(\frac{I_{t+1}}{I_{t}}\right)^2. \label{foc_I_eq}
\end{align}

\subsubsection{Labor market}\label{sec:Labour}

The labor market follows a set-up as in \cite{auclertEtAljpe}. 
Labor services are produced by a representative labor packer that aggregates a range of differentiated labor inputs $u\in(0,1)$ according to 
\begin{align*}
    N_t = \left(\int^1_0 N_{ut}^{\frac{\epsilon_w-1}{\epsilon_w}}du\right)^{\frac{\epsilon_w}{\epsilon_w-1}}
\end{align*}
and will thus demand 
\begin{align}
    N_{ut} = \left( \frac{W_{ut}}{W_t}\right)^{-\epsilon_w} N_t \label{eq:Nu_demand}
\end{align}
of each labor variety.
Each of the differentiated labor types is supplied by a union that sets the nominal wage $W_{ut}$ as a monopolist to maximize the utility of its members. The latter are required to work according to a uniform schedule, i.e., all $u$ workers have to supply the same amount of hours $N_{ut}$.
Unfortunately, in every period the leadership of a union $u$ suffers utility costs $\frac{\psi}{2}\left(\frac{W_{ut}}{W_{ut-1}}-\pi_{SS}\right)^2$ for changing the nominal wage, perhaps due the administrative burden of adjusting contracts.
In turn, every union solves
\begin{align}
    \max_{\lbrace W_{u_\tau} \rbrace^\infty_{\tau = t}} \mathbb{E}_t \sum^{\infty}_{\tau = t}\beta^{\tau-t} \left( \int \left(u(c_{i\tau})-\varsigma\frac{N_{i\tau}^{1+\gamma}}{1+\gamma}\right)  di -\frac{\psi}{2}\left(\frac{W_{u\tau}}{W_{u\tau-1}}-1\right)^2   \right), \label{eq:union_prob}
\end{align}
taking demand schedule \eqref{eq:Nu_demand} into account. 

Households are exogenously distributed over unions in a uniform manner: Note that the law of large numbers applies thus also \textit{within} unions so that the distribution of agents $i$ overall and within any union $u$ coincides.\footnote{Since all labor varieties will be symmetric in equilibrium, her labor type $u$ doesn't matter for an individual's consumption-saving problem.} 
Due to symmetry, the F.O.C.s of \eqref{eq:union_prob} then give rise to an aggregate \emph{Wage Phillips curve} of the form 
\begin{align}
    \pi^w_{t} (\pi^w_t-1) &= \kappa_w\left(\varsigma N_t^{1+\gamma} - \frac{\epsilon_w-1}{\epsilon_w}(1-\tau_p)(1-\tau^w_t) \int \left(u'(c_{it})\left(s_{it}\frac{W_{t}}{P_t}N_{t}\right)^{1-\tau_p}\right)di \right) \nonumber \\
    &+ \beta \mathbb{E}_t\pi^w_{t+1} (\pi^w_{t+1}-1), \label{eq:wage_phillips}
\end{align}
with $w_t := W_t/P_t$ and $\pi_t^w := \frac{W_t}{W_{t-1}}$. For convenience, I define $\kappa_w := \frac{\epsilon_w}{\psi}$ to denote the slope of the Wage Phillips curve. 
\\~\\
\textbf{Derivation:}\\
Given the uniform hours $N_{it} = N_{ut}$ for all union members and demand schedule \eqref{eq:Nu_demand}, we have
\begin{align*}
    \frac{\partial}{\partial W_{ut}} N_{ut} = -\epsilon_w \frac{N_{ut}}{W_{ut}}
\end{align*}
and 
\begin{align*}
    \frac{\partial}{\partial W_{ut}} u(c_{it}) = u'(c_{it})(1-\tau_p)(1-\tau_w)\left(s_{it}\frac{W_{ut}}{P_t}N_{ut}\right)^{-\tau_p}s_{it}\frac{N_{ut}}{P_t}(1-\epsilon_w) ~~,
\end{align*}
the latter reflecting that due to the envelope theorem, the marginal utility of additional resources should equal the marginal utility of consumption.
In turn, the F.O.C. corresponding to \eqref{eq:union_prob} is 
\begin{align*}
    &(1-\tau_p)(1-\tau_w)\frac{1-\epsilon_w}{W_{ut}} \int \left(u'(c_{it})\left(s_{it}\frac{W_{ut}}{P_t}N_{ut}\right)^{1-\tau_p}\right)  di + \frac{\epsilon_w}{W_{ut}}\varsigma N_{ut}^{1+\gamma} \\
    -&\psi\left(\frac{W_{ut}}{W_{ut-1}}-1\right)\frac{1}{W_{ut-1}} + \beta\mathbb{E}_t\psi\left(\frac{W_{ut+1}}{W_{ut}}-1\right)\frac{W_{ut+1}}{(W_{ut})^2} = 0~~.
\end{align*}
If we now use that unions are symmetric and thus $N_{ut}=N_t$ and $W_{ut}=W_t$, re-arranging yields
\begin{align*}
   \pi^w_{t} (\pi^w_t-1) &= \frac{\epsilon_w}{\psi}\left(\varsigma N_t^{1+\gamma} -\frac{\epsilon_w-1}{\epsilon_w}(1-\tau_p)(1-\tau_w) \int \left( u'(c_{it})\left(s_{it}\frac{W_{t}}{P_t}N_{t}\right)^{1-\tau_p}\right)di \right) \\
   &+ \beta \mathbb{E}_t\pi^w_{t+1} (\pi^w_{t+1}-1)~~.
\end{align*}

\subsection{External calibration}\label{app:ext_calib}

The model's externally calibrated parameters are displayed in Table \ref{tab:ext_parameters}: I set the households' risk aversion parameter to $\xi=1.5$, within the range of standard values used in the literature.
Regarding technology, I use the conventional value of $\alpha=0.33$ for the Cobb-Douglas parameter for capital and set a quarterly depreciation rate for capital of $\delta_0=0.0175$, implying approx. 7\% annual depreciation.
 Similarly, I set $\mu_t$ to a conventional value of $1.1$, resulting in a SS markup of 10\%. The elasticity of substitution between different labor varieties is assumed to be the same as for goods and thus $\epsilon_w = 11$.\footnote{Since I set $\kappa_w$ independently of $\epsilon_w$ and calibrate $\varsigma$ to achieve $N_{ss}=1$, the value of this parameter is practically of limited importance.}
 The slope of the price Phillips curve is set to $\kappa_Y = 0.06$, in line with the recent evidence by \cite{lenzuEtAl2023}. In the HANK literature, wages are often assumed to be substantially stickier than prices, even though estimated DSGE models do not always support this. I set $\kappa_w = 0.015$ to be consistent with the former, a value based on the estimate of \cite{auclertEtAl2020a}.
 However, given the empirical controversies surrounding these parameters, related robustness checks will be discussed in Appendix \ref{app:robustness}.\\
 Several other parameters governing the economy are calibrated in line with \cite{bayerEtAlaer}: 
 First, I set the probability of exiting the $\Xi=1$ state within a given period to be $6.25$\%, which these authors choose based on \cite{guvenenEtAl2014}. The process for labor income $s$ follows a discretized AR(1) process in logs, with persistence $0.98$ and normal innovations with standard error $0.12$. This reflects evidence by \cite{storeslettenEtAl2004}.
 I also adopt \cite{bayerEtAlaer}'s estimated tax progressivity parameter $\tau_p = 0.12$. The investment adjustment cost is chosen to be $3.5$ and the ratio $\delta_2/\delta_1$ set to be $1$, reflecting the results of their model estimation. For a given $\delta_2/\delta_1$-ratio, I always set $\delta_1$ and $\delta_2$ to achieve $u_t = 1.0$ in steady state. Again, since estimates for the investment- and utilization adjustment parameters also vary in the literature, these values will be subjected to robustness tests as well. 

 Regarding monetary policy, I parameterize the model's interest rule with standard values also employed by \cite{bayerEtAl2023a} to study post-2020 macroeconomic dynamics in the US. In particular, this includes setting the Central Bank's inflation reaction parameter to the value $\theta_\pi = 1.5$ first proposed by \cite{taylor1993}, making it clear that monetary policy is ``active'' in my model. The ELB is set to be 2 percentage points (in annual terms) below the SS policy rate $r^R_{SS}$, given that an exercise in Section \ref{sec:mon_imps} effectively assumes the model to be in SS before 2020. In that year, the nominal rate was around 2\% and the ELB thus 2 p.p. below it.
 
 On the fiscal side, the SS labor tax level $\tau$ is chosen to be consistent with $G/Y\approx 17.5\%$ given the model's other targets, in line with the average ratio of government consumption and investment to GDP in 2014-2019.
 The tax rate on the ``entrepreneurs'''s profit incomes is set to 24\%, reflecting the top tax bracket for qualified dividends in the US.
 \cite{bianchiEtAl2023} report the average maturity of US treasury debt to typically vary between 4.5-5.5 years, so I choose $\delta^B=0.05$ to be consistent with an average 5-year (20 quarter) duration.\\
 Initially, I set $\rho_\tau = 0.85$ and $\psi_B = 1.25$, ad hoc values that result in a moderately drawn-out public debt response. 
 In particular, these parameters imply that for a simple transfer shock as studied in Section \ref{sec:fisc_imps}, public debt will start to clearly decline after approx. 2 years. 
 As will become clear, different combinations of parameters yield similar results as long as they generate public debt responses of comparable magnitude.
 
 \begin{table}
 \centering
  \small
\begin{tabular}{ c | c c c} 
\hline \hline 
Parameter &  Description & Value & Source  \\ 
\hline
 $\xi$ & Risk aversion  & 1.5 & Standard  \\
 $\iota$ & Exit prob. entrepreneurs & 1/16 & \cite{bayerEtAlaer} \\
 $\alpha$ & Cobb-Douglas parameter & 0.33 & Standard \\
 $\delta_0$ & Steady State depreciation &  0.0175 & Standard \\
 $\mu$ & SS Goods markup & 1.1 & Standard \\
 $\kappa_Y$ & Slope of price Phillips curve & 0.06 & \cite{lenzuEtAl2023} \\
 $\kappa_w$ & Slope of wage Phillips curve & 0.015 & \cite{auclertEtAl2020a}\\
 $\epsilon_w$ & EOS labor varieties & 11 & Standard \\
 $\phi$ & investment adjustment cost & 3.5 & \cite{bayerEtAlaer} \\
$\delta_2/\delta_1$ & utilization parameters & 1.0 & \cite{bayerEtAlaer}\\
$\gamma$ & Inverse Frisch & 1.0 & Standard \\
$\delta^B$ & Government debt duration & $0.05$ &  5 years avg. maturity \\
$\tau$ & Tax level & $0.2$ & $G/Y \approx 17.5\%$ \\
$\tau^p$ &Tax progressivity & 0.12 & \cite{bayerEtAlaer} \\
$\tau^\Xi$ & Profit Tax & 0.24 & US Tax Code \\
$(\rho_R, \theta_\pi, \theta_y)$ & Taylor rule parameters & (0.8,1.5,0.2) & \cite{bayerEtAl2023a} \\
$R^{LB}$ & Effective Lower Bound & $R^R_{SS} - 0.005$ &  2 p.p. below $R^R_{SS}$ \\
$(\rho_\tau, \psi_B)$ & Tax rule parameters & (0.85, 1.25) & See text \\
\hline \hline
\end{tabular}
\caption{Externally set parameters}
\label{tab:ext_parameters}
 \end{table}

 \subsection{Distributional Moments}\label{app:dist_moments}

\begin{figure}
   \centering
   \includegraphics[scale=0.5]{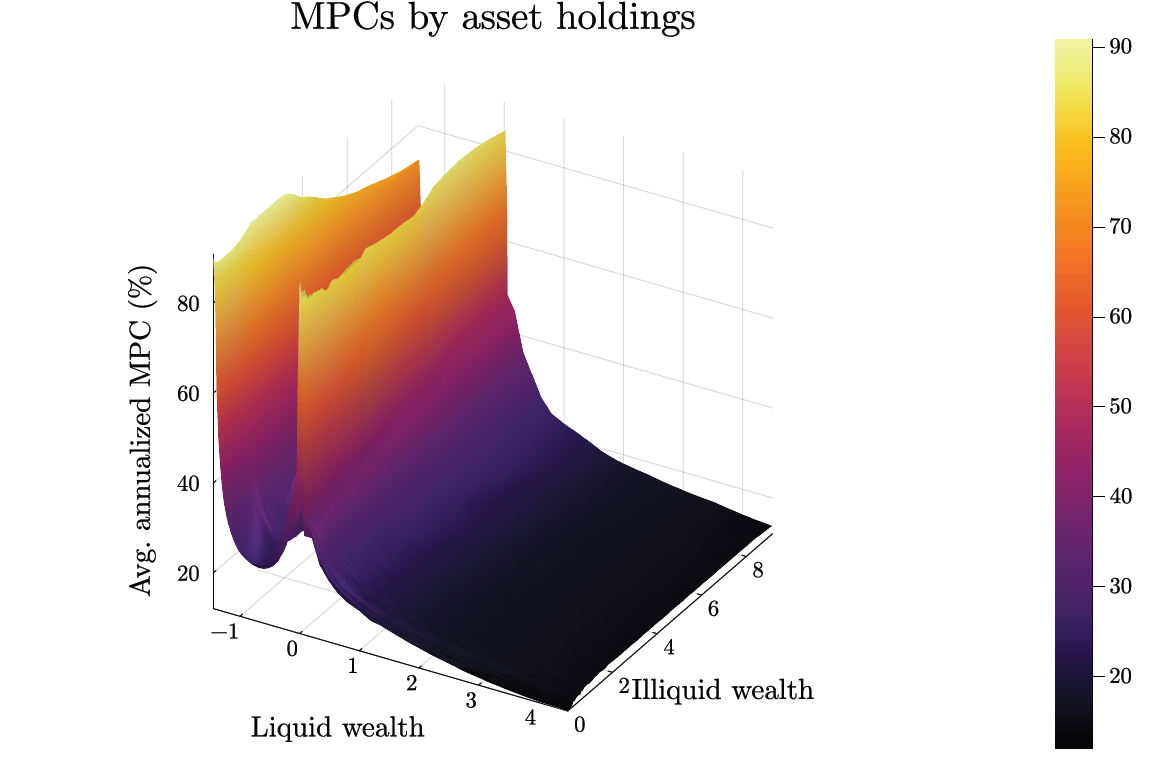}
   \caption{Model MPC distribution}
   \label{fig:MPCgraph}
 \end{figure}

 In this section, I validate the internal calibration by analyzing various model-generated moments that were not directly targeted.\\
 Table \ref{tab:distr_moments} compares various untargeted moments of the model's SS income- and wealth distributions with their empirical counterparts as reported by \cite{kruegerEtAl2016}. 
 The latter are based on the 2006 Panel Study of Income Dynamics (PSID) and the 2007 Survey of Consumer Finance (SCF), respectively.\footnote{In the data, disposable income is defined as the sum of after-tax earnings, income generated by assets held as well as unemployment benefits. In the model, it only comprises the first two as there is no unemployment.
 In both model and data, Net Worth relates to both liquid and illiquid assets.}
 Arguably, the model achieves a fairly good fit, in particular for Net Worth.\\
 Since I am employing a two-asset model, it is not only relevant to assess how closely the framework matches data moments related to the distribution of overall net worth, but also the different asset classes held by the households. I do so in Table \ref{tab:port_moments}:
 First, I am considering moments of the illiquid- and liquid wealth distribution separately. In particular, I compare them with statistics reported by \cite{kaplanEtAl2018}, who rely on the 2004 SCF. As in the data, the model generates a more unequal distribution of liquid assets and ownership of both asset classes is concentrated in their respective Top 10\%, with the bottom 50\% holding hardly any. 
 For both liquid and illiquid assets, I generate somewhat more equal asset distributions, with the share held by the Top 10\% not as high and the share of the Next 40\% substantially larger than in the SCF data. But, as noted by \cite{kaplanEtAl2018}, it is ``notoriously challenging'' to match the extreme right tail of wealth distributions with income risk alone. From that perspective, I view my model's performance as satisfactory.\\
    \begin{table}
         \centering
    \begin{threeparttable}
     \begin{tabular}{|c | c c | c c |}
     \hline 
          & \multicolumn{2}{c|}{Disposable Income} &  \multicolumn{2}{c|}{Net Worth}  \\
         & Model & Data & Model & Data \\\hline
         Quint. 1 & 6.7 & 4.5  & 0.0 & -0.2  \\
         Quint. 2 & 10.8 & 9.9 & 0.9 & 1.2 \\
         Quint. 3  & 14.8  & 15.3 & 4.1 & 4.6 \\
          Quint. 4  & 20.4 & 22.8 & 11.2 & 11.9  \\
           Quint. 5  & 47.2  & 47.5 & 83.8 & 82.5 \\\hline
           Gini & 0.40 & 0.42 & 0.80 & 0.78\\
         \hline 
     \end{tabular}
     \begin{tablenotes}
         \small
         \item Note: ``Data'' refers to moments computed by \cite{kruegerEtAl2016} using PSID and SCF.
     \end{tablenotes}
     \caption{Distributional moments comparison}
     \label{tab:distr_moments}
     \end{threeparttable}
 \end{table}
 \begin{table}[]
     \centering
     \begin{tabular}{|c | c  | c |}
     \hline 
       Moments  & Model & Data (incl. source)  \\\hline
    \textit{Illiquid asset shares} & & \citep[from][]{kaplanEtAl2018} \\
    Top 10\% & 63.0 & 70 \\
    Next 40\% & 34.9 & 27\\
    Bottom 50\% & 2.1 & 3 \\\hline
        \textit{Liquid asset shares} & &  \citep[from][]{kaplanEtAl2018}\\
        Top 10\% & 70.6 & 86  \\
        Next 40\% & 31.0 & 18  \\
        Bottom 50\% &  -1.6 & -4\\
        \hline
        \textit{Hand-to-Mouth (HtM) Status} & & \citep[from][]{kaplanEtAl2014} \\
        Share HtM & 28.1 & 31.2 \\
    Share Wealthy HtM & 18.2 & 19.2  \\
    Share Poor HtM & 9.9 & 12.1 \\
       \hline
     \end{tabular} 
     \caption{Portfolio moments: Model vs. Data}
     \label{tab:port_moments}
 \end{table}
 Finally, I analyze how many households are \textit{Hand-to-Mouth} (HtM) in the sense of \cite{kaplanEtAl2014}, i.e., whether their liquid asset holdings are less than 2 weeks ($\approx$ 1/6 of a model period) of current household income above either $0$ or the borrowing constraint.
 I also classify them as ``Wealthy HtM'' if they additionally hold illiquid assets and ``Poor HtM'' if they do not.  The model matches the empirical evidence on the size of either group of agents well.
 As visualized in Figure \ref{fig:MPCgraph}, these agents with low liquid wealth tend to have particularly high MPCs. In turn, my framework is able to generate an average quarterly MPC of 15.8\% and an average annualized MPC of 36.7\%.\footnote{I compute individuals' annualized MPCs $aMPC_i$ as $aMPC_i = 1 - (1-qMPC_i)^4$ following \cite{carrollEtAl2017}. Note that these \emph{annualized} MPCs will not exactly equal the individuals' \emph{annual} MPCs.}
 The former is of a similar magnitude as the corresponding value reported by \cite{kaplanEtAl2018}.

\subsection{Definition of equilibrium}\label{app:equilibrium}

\begin{definition}
    A \textbf{Recursive Equilibrium} of the model consists of 
    \begin{itemize}
        \item value functions $V^a(a_{it},k_{it},s_{it},\Xi_{it}; \Gamma_t)$, $V^{na}(a_{it},k_{it},s_{it},\Xi_{it}; \Gamma_t)$
        \item household policies $a^{a}(a_{it},k_{it},s_{it},\Xi_{it}; \Gamma_t)$, 
        $a^{na}(a_{it},k_{it},s_{it},\Xi_{it}; \Gamma_t)$, 
        $k(a_{it},k_{it},s_{it},\Xi_{it}; \Gamma_t)$ and $c^{a}(a_{it},k_{it},s_{it},\Xi_{it}; \Gamma_t)$,
        $c^{na}(a_{it},k_{it},s_{it},\Xi_{it}; \Gamma_t)$,
        \item firm- and union policies $I_t$, $K_t$, $H_t$, $Y_t$, $N_t$, $u_t$, $B_t^l$, $\Pi_t$, $y_{jt}\forall j \in [0,1]$, $w_t$
        \item prices $h_t$, $r^k_t$, $q_t$, $R_t^l$, $mc_t$
        \item inflation $\pi_t$
        \item government policies $G_t, B^g_{t+1}, \tau_t , r_{t+1}^R$,
    \end{itemize}
    so that 
    \begin{enumerate}
        \item Given prices $R_t^l$, $r^k_t$, $q_t$, wages $w_t$ and profits $\Pi_t$, the value functions $V^a(a_{it},k_{it},s_{it},\Xi_{it}; \Gamma_t)$, $V^{na}(a_{it},k_{it},s_{it},\Xi_{it}; \Gamma_t)$  solve the households' Bellman equations in \eqref{Vadjust_eq} and \eqref{Vnonadjust_eq} and $a(a_{it},k_{it},s_{it},\Xi_{it}; \Gamma_t)$, $k(a_{it},k_{it},s_{it},\Xi_{it}; \Gamma_t)$, $c(a_{it},k_{it},s_{it},\Xi_{it};  \Gamma_t)$ are the resulting optimal policy functions.
        \item Expectations are model-consistent.
        \item $ y_{jt}\in [0,1]$ are consistent with demand schedule \eqref{ydemand_eq} and final output $Y_t$ given by \eqref{AggY_eq}.
        \item Inflation $\pi_t$ is consistent with Phillips curve \eqref{phillips_eq}.
        \item Given prices  $h_t$, $r_t$, $q_t$, $mc_t$, 
        the intermediate goods producers choices $K_t,H_t,u_t$ are consistent with optimality conditions \eqref{foc_H_eq}-\eqref{foc_u_eq}.
        \item Given price $q_t$, the intermediate goods producers choices $I_t$ are consistent with optimality condition \eqref{foc_I_eq}.
        \item The labor packer's zero profit condition $h_t = w_t$ is fulfilled.
        \item The dynamics of the wage level $w_t$ are consistent with \eqref{eq:wage_phillips}.
        \item The Liquid Asset Funds' portfolio choice is consistent with \eqref{eq:LAF2}.
        \item The return of liquid assets is given by \eqref{eq:LAF3}.
        \item Given inflation $\pi_t$ and output growth $Y_t/Y_{t-1}$, the monetary authority set $R^R_{t+1}$ according to \eqref{Taylor_eq}.
        \item Taking the remaining values as given, the government sets taxes according to \eqref{eq:taxrule} and issues debt $B^g_{t+1}$ so that \eqref{Gbr_eq} holds.
        \item The market for liquid asset clears, i.e.,
        \begin{align*}
            A^l_t = \int^1_0 a_{it}di ~~.
        \end{align*}
        \item The government bond market clears, i.e.,
     \begin{align*}
         B^l_t = B^g_t ~~.
     \end{align*}
     \item The capital market clears, i.e.,
     \begin{align*}
        K_t =  \frac{A^l_t-B^l_t}{q_{t-1}} + \int^1_{0} k_{it} di ~~.
     \end{align*}
    \item  The market for investment good clears, i.e.,
     \begin{align*}
     K_{t+1} = (1-\delta(u_t))K_t + \left[1-\frac{\phi}{2}\left(\frac{I_t}{I_{t-1}}-1\right)^2\right]I_t 
    \end{align*}
     \item The market for labor services clears, i.e.,
     \begin{align*}
    H_t = N_t\int^1_{m^\Xi}s_{it} di.
    \end{align*}
    \item The market for intermediate goods clears, i.e.,
     \begin{align*}
        \int^1_0 y_t(j) dj = F_t(u_tK_t, H_t)
     \end{align*}
     \item The final good market clears, i.e.,
     \begin{align*}
    Y_t = &C_t + G_t + I_t + \Bar{R}\int^1_0 \mathfrak{1}(a_{it}\leq 0) a_{it}  di +\left[\varphi + \frac{\Psi}{2}\left(1-\frac{B_t^l}{A_t^l}\right)^2 \right]A_t^l~~
     \end{align*}
     where $\mathfrak{1}(\cdot)$ denotes the indicator function.
     \item The distributions of income and wealth evolve according to household's policy function and the exogenous transition probabilities $\pi^s(\cdot\vert \cdot)$, $\zeta$ and $\iota$.
    \end{enumerate}
\end{definition}

\subsection{Details on numerical implementation}\label{app:numerics}

\subsubsection{Details on Steady State Solution}

The household problem needs to be solved on a discretization of the state space: I choose $90$ grid points for both $a$ and $k$, either of which are non-linearly spaced as household decision functions tend to be more non-linear for lower levels of assets.
In particular, the grid points for both $a$ for $k$ are spaced according to the ``double exponential'' rule, i.e.
\begin{align*}
        \mathcal{X} = x_{min} + \exp(\exp(\mathbf{u}(\log(1 + \log(1+x_{max}-x_{min})),n_i)) - 1) - 1  
\end{align*}
where $x_{min}$ is the minimum value on the grid for variable $x$, $x_{max}$ the maximum value and $\mathbf{u}(x_{max}, n_i)$ a vector of $n_i$ equidistant points on the interval $[0,x_{max}]$. Since household value- and policy functions will feature an additional kink around $a=0$ when the borrowing penalty kicks in, I add 5 additional grid points in the immediate vicinity of that point. Given that individual labor productivity is discretized to $15$ points, this means that the household problem is solved on a tensor grid of $90\times 90\times (15+1) = 129600$ points (the ``entrepreneur'' status adds an ``income'' state to the $15$ ``skill'' states). The discretization of the individual labor productivity process follows an off-the-shelf method \`a la \cite{tauchen1986}. Whenever interpolation is needed off the grid, I use linear interpolation\\
For the implementation of the multidimensional EGM algorithm, I follow the replication codes for \cite{bayerEtAlaer} closely.\footnote{As of October 2025, these replication codes are available under \url{https://github.com/BASEforHANK/BASEtoolbox.jl}. For the analysis in Appendix \ref{app:alt_illiquid}, I combine their algorithm with the 2-asset EGM scheme proposed in the Appendix to \cite{auclertssj}.}
Given the random illiquid asset adjustment, the EGM scheme only needs to iterate over marginal value functions (i.e. the derivatives of $V$ with respect to $a$ and $k$) and does not compute $V$ directly.\\
For finding the SS, I iterate over $r^l_{ss}$ and $r^k_{ss}$: Given these values, the remaining SS variables can be backed out and the household-problem solved. I then use a heuristic updating procedure to search for $r^l_{ss}$ and $r^k_{ss}$ so that the asset markets clear.

\subsubsection{Details on State Space Perturbation}

As already indicated in the main text, the model's dynamic equilibrium is approximated using First-Order Perturbation around its non-stochastic steady state.
For the State Space perturbation \`a la \cite{bayerEtAlaer}, note that when using the discretized representations of the marginal value functions as well as the joint income/asset distribution, the equilibrium can be represented as the solution to a non-linear difference equation of the form
\begin{align}
  \mathbb{E}_t F(\mathbf{y_t},\mathbf{x_t},\mathbf{y_{t+1}},\mathbf{x_{t+1}})=0   \label{eq:sgu}
\end{align}
as, e.g., used by \cite{schmittgroheEtAl2004}. $\mathbf{y}$ denotes a vector of control variables, which includes the households' marginal value functions on the grid and $\mathbf{x}$ a vector of state variables, which includes the discretized distribution.

In theory, one could find the linearized equilibrium using the standard approach of computing the Jacobians of $F$ as in \eqref{eq:sgu} and then solve the resulting linear difference equation relying on methods such as \cite{klein2000}. In practice, however, such an approach would involve very high computational costs for the 2-asset HANK model, given that the full $\mathbf{y}$ and $\mathbf{x}$ have a combined length exceeding $200,000$.

To overcome this problem, \cite{bayerEtAlaer} propose a procedure which conducts dimension reduction in 2 steps, one before computing the Jacobians and one after.
Specifically, it first uses a Discrete Cosine Transform (DCT) to dimension-reduce the marginal value functions: The values resulting from such a DCT are coefficients of a fitted multidimensional Chebychev polynomial, of which only a subset are selected to be perturbed: \cite{bayerEtAlaer} propose to use the nodes that are most important for describing the derivatives of the SS marginal value functions to changes in the set of prices that households directly take into account (e.g., interest rates and the wage). The other coefficients are kept at their SS values.

For reducing the dimensionality of the joint distribution in the first step, the authors furthermore suggest splitting it into marginals and a copula, where the latter is in effect treated as an interpolator mapping the SS marginal distributions into the joint distribution. That ``interpolator'' can also be dimension-reduced through a DCT or just kept fixed, so one only perturbs the marginals as well as selected coefficients of the copula, which have substantially lower dimension. Overall, in my application the procedure manages to shrink the effective dimensionality of the system to a manageable number of approx. 1400, for which an initial perturbation solution is obtained.\\
The second step further reduces the set of DCT coefficients by using the first step solution to check which ones vary only very little with the aggregate shocks and are thus not important for explaining model dynamics. It is useful if the model has to be repeatedly solved for different parameters, such as for checking model stability for different parameters as e.g., in Section \ref{sec:fisc_imps}. For a more detailed exposition, see \cite{bayerEtAlaer}. 

\subsubsection{Details on Sequence Space Perturbation}\label{app:ssj}

As already mentioned in the main text, in addition to the State Space perturbation method described above, I also use a Sequence Space linearization method \`a la \cite{auclertssj}, as its allows flexibly exposing the economy to various news shocks. This proves useful particular for the analysis in Section \ref{sec:filtering}: Firstly, the analysis in that Section requires the model to be able to handle a binding lower bound on the nominal interest rate. 
As already pointed out by \cite{mckayEtAl2021}, this can be achieved (relatively) easily in a Sequence space setting using news shocks. The idea is that if an aggregate shock would cause the central bank (CB) to violate the ELB in a certain number of periods, one can solve for a sequence of pre-announced monetary policy (news) shocks that would keep the economy at the ELB instead. 
The CB then enforces the ELB by announcing exactly these shocks. Secondly, the same feature of a sequence space solution makes it easy to consider different interest rate rules. This is because in a linearized model, those can similarly be imposed by announcing a suitable set of news shocks to the policy rule in place \citep{mckayEtAl2023_count}.

I obtain the Sequence Space Jacobians (SSJs) of the model's heterogeneous agent block as described in \cite{auclertssj}, although I rely on automatic differentiation instead of finite differences to ensure accuracy of the derivatives. For obtaining the general equilibrium SSJ's, I then build on the model representation proposed by \cite{bhandariEtAl2023}, which, for given Heterogeneous Agents SSJs, yields the same GE SSJs as \cite{auclertssj}'s Directed Acyclical Graph (DAG) approach.\footnote{An example application for a simpler HANK model can be found under \url{https://mhaense1.github.io/SSJ_Julia_Notebook/SSJ_notebook_2.html}.} I use a truncation horizon of $T=500$ periods, as my 2-asset HANK model features relatively-long lived IRFs due to the presence of investment adjustment costs and shocks with quite high persistence.

\section{Robustness analyses}\label{app:robustness}

This appendix aims to address potential concerns regarding the generality and robustness of the paper's main insights in Section \ref{sec:fisc_imps}. In particular, the respective model exercises were conducted only in the context of a particular fiscal policy scenario and the model features several parameters whose values are subject to empirical controversies.

 \subsection{Alternative Fiscal Shocks}

\begin{figure}
    \centering
    \caption{Model IRFs to Government spending shock}
    \includegraphics[scale = 0.6]{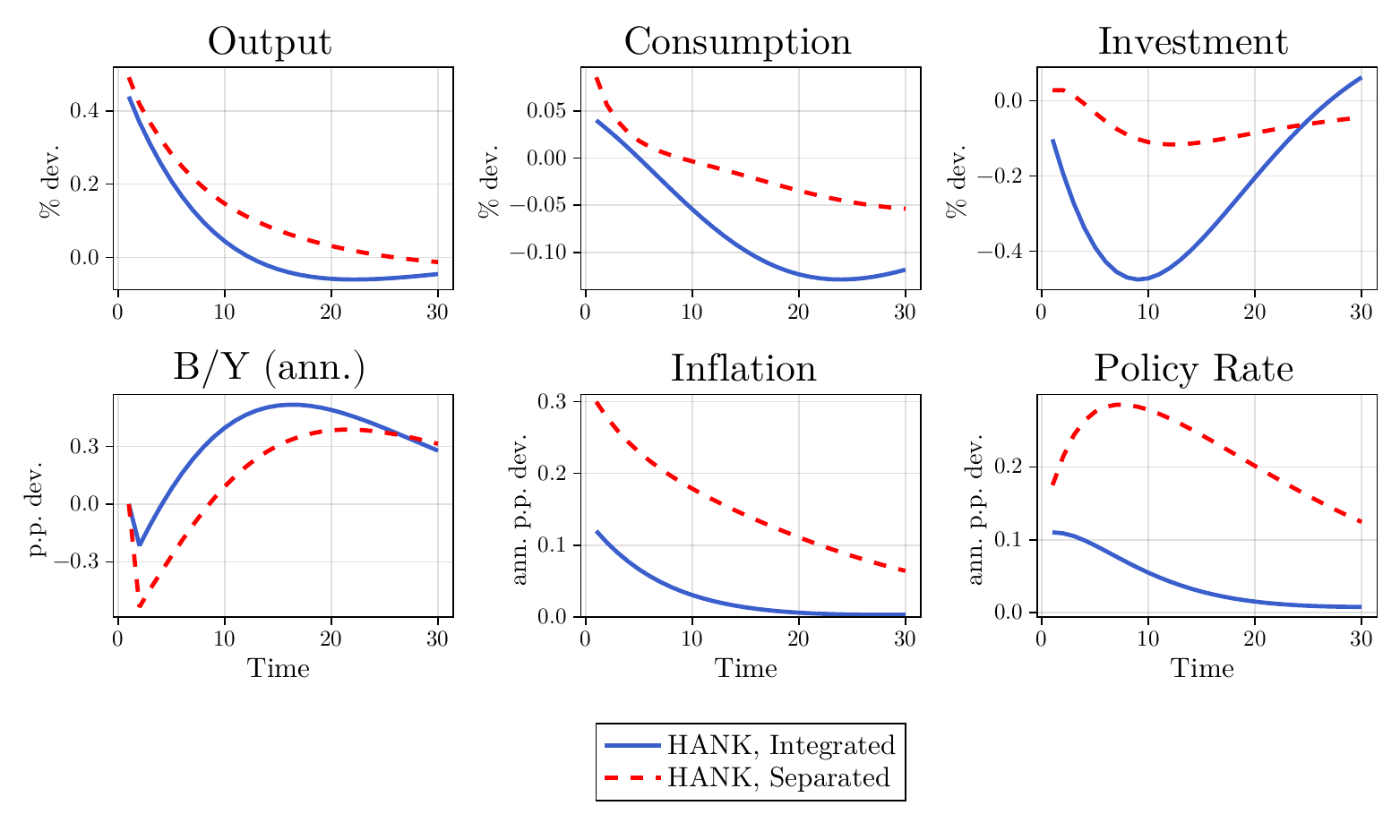}
    \fignote{$B/Y$ represents the real market value of public debt $B^g$ over annualized GDP. Figures display relative (in \%) or percentage point (p.p.) deviations from Steady State.}
    \label{fig:G_shock}
 \end{figure}

Firstly, to what extent do the results in Section \ref{sec:fisc_imps} depend on the used fiscal policy scenario? To answer this question, I firstly considered an alternative scenario with a government consumption shock instead of the transfer shock. In particular, Appendix Figure \ref{fig:G_shock} displays the macroeconomic response to a persistent increase in government consumption $G$, assumed to increase by $2.5\%$ on impact and afterwards following an AR(1)-decay with persistence $0.9$. While the aggregate responses naturally differ, considering different asset market structures has quantitatively comparable effects.\\
Secondly, I computed IRFs for the original transfer shock but assuming that the fiscal authority consolidates by reducing government consumption $G$ instead of changing taxes. The respective results are provided in Appendix Figure \ref{fig:G_adj}. For this alternative, the aggregate wealth effect due to less wasteful (in this model) government spending causes household consumption to increase slightly more and investment to decrease less. But as long as public debt and hence its liquidity effect evolve similarly, there are no major changes compared to the baseline fiscal rule.

\subsection{Alternative model parameterizations}

Like every New Keynesian model, the framework used for the above analysis contains several parameters subject to empirical controversies, in particularly the ``slopes'' of the New Keynesian Price- and Wage Phillips curves. Additionally, since investment demand was found to be important for the transmission of the debt-driven inflation, it is relevant to check whether the above results are due to specific values for the model's capital utilization- and adjustment costs. To address this, I varied the parameters one-by-one to assert robustness with respect to the Baseline calibration's parameter choices. The resulting IRFs to the transfer shock are displayed in Appendix \ref{app:figures} and briefly summarized below.

\begin{figure}
    \centering
    \caption{IRFs to transfer shock: Varying $\kappa$}
    \includegraphics[scale = 0.55]{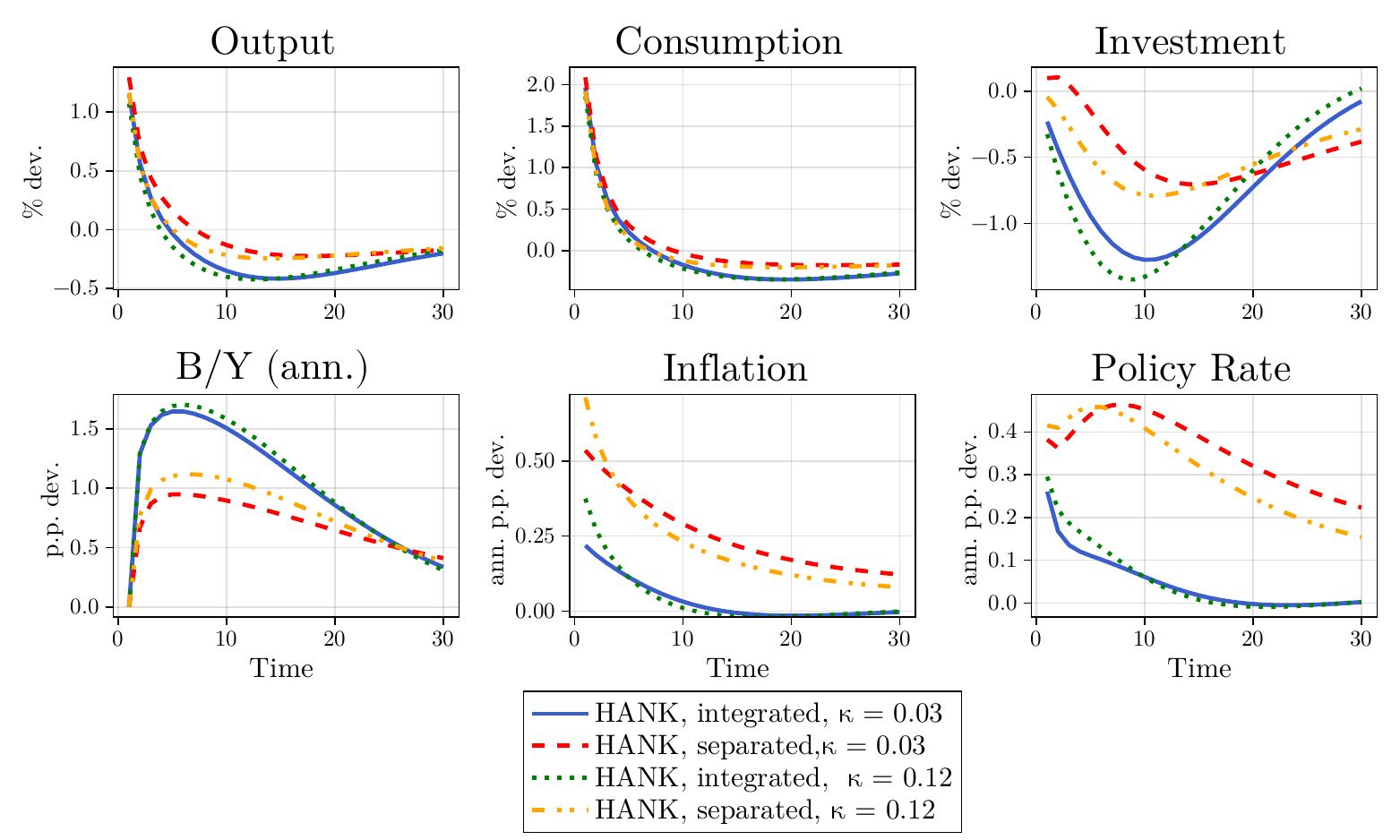}
    \fignote{ $B/Y$ represents the real market value of public debt $B^g$ over 4 times (=annualized) GDP.}
    \label{fig:T_kappay}
 \end{figure}

\begin{figure}
    \centering
    \caption{IRFs to transfer shock: Varying $\kappa_w$}
    \includegraphics[scale = 0.55]{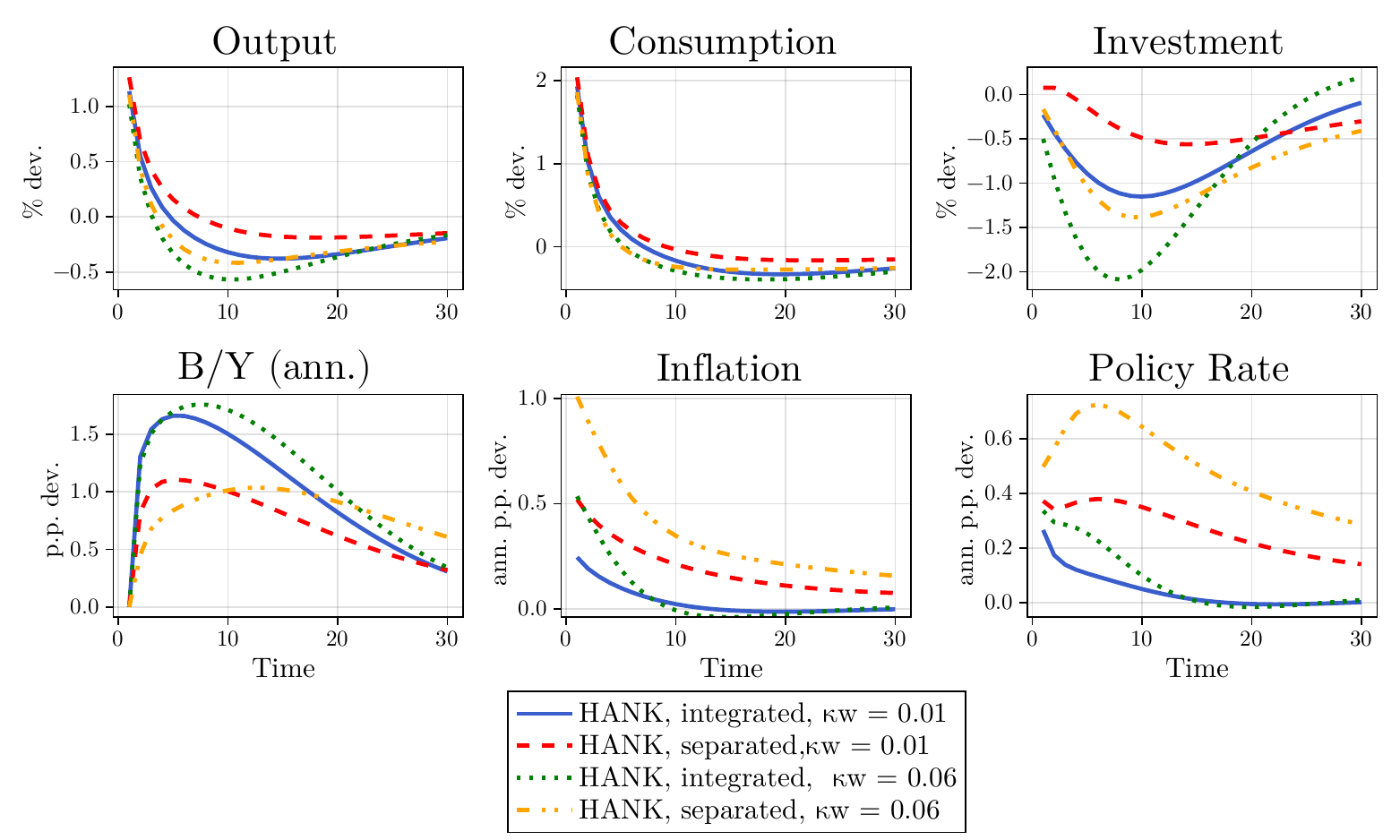}
    \fignote{ $B/Y$ represents the real market value of public debt $B^g$ over 4 times (=annualized) GDP.}
    \label{fig:T_kappaw}
 \end{figure}

\begin{figure}
    \centering
    \caption{IRFs to transfer shock: Varying $\phi$}
    \includegraphics[scale = 0.55]{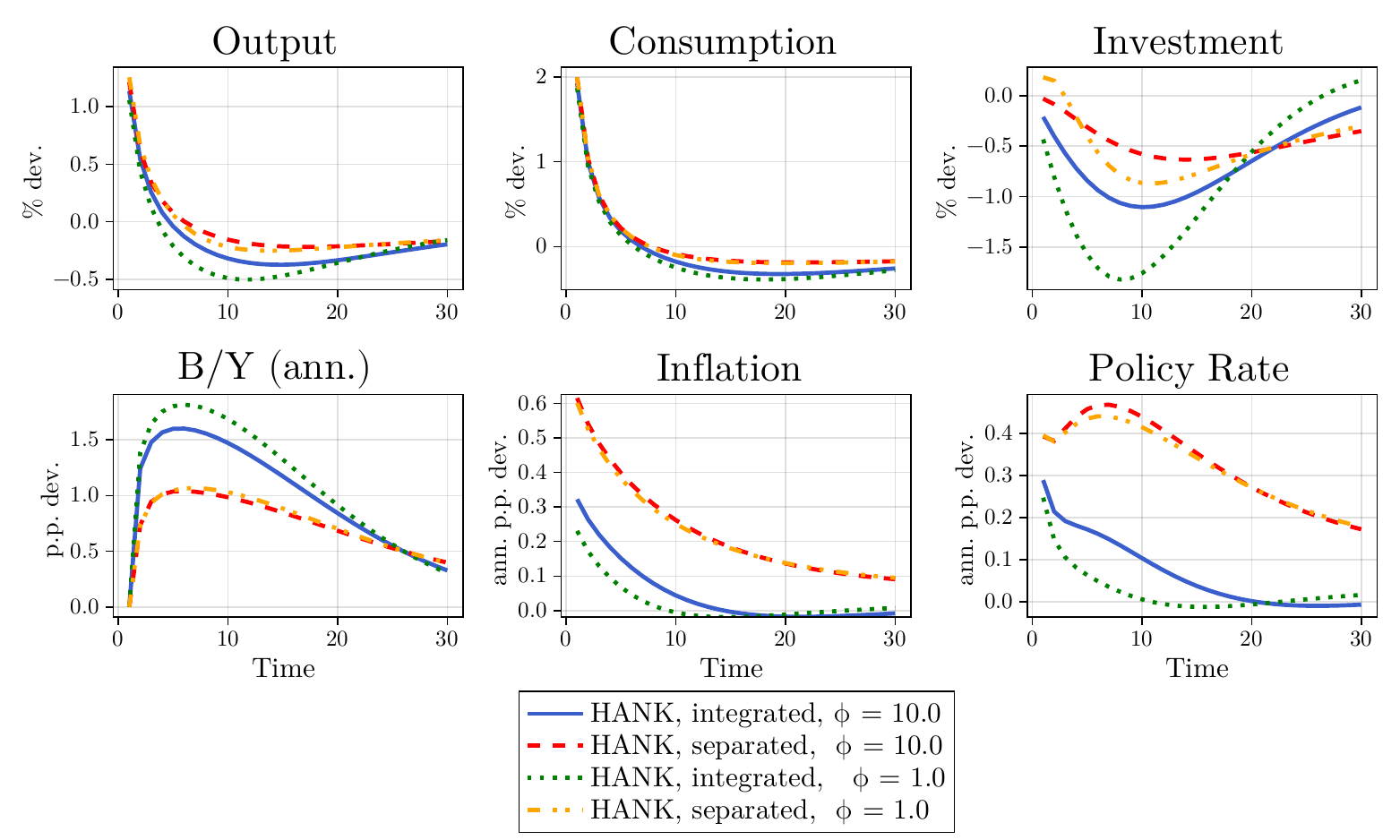}
    \fignote{ $B/Y$ represents the real market value of public debt $B^g$ over 4 times (=annualized) GDP.}
    \label{fig:T_phik}
 \end{figure}

\begin{figure}
    \centering
    \caption{IRFs to transfer shock: Varying $\delta_2/\delta_1$}
    \includegraphics[scale = 0.55]{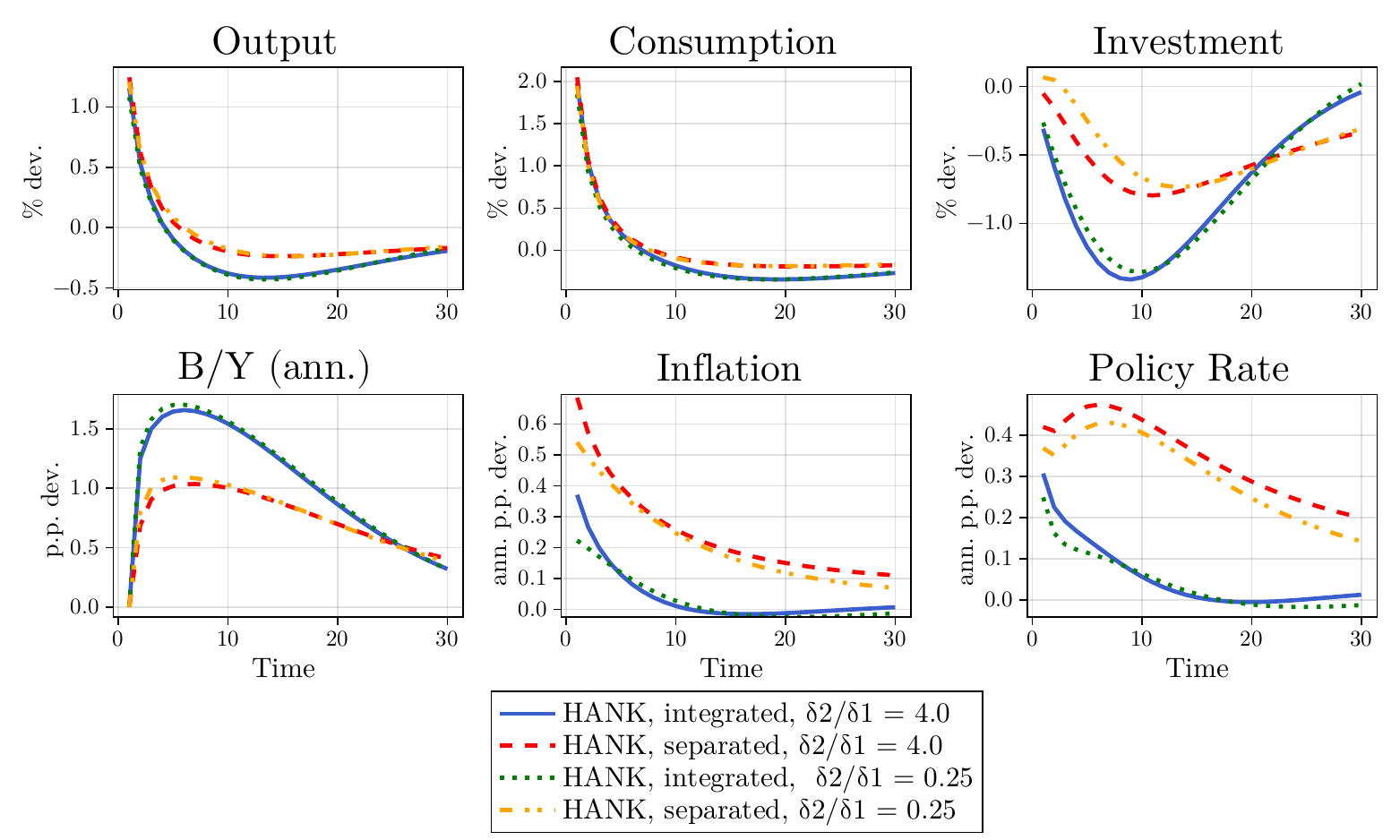}
    \fignote{ $B/Y$ represents the real market value of public debt $B^g$ over 4 times (=annualized) GDP.}
    \label{fig:T_util}
 \end{figure}

Allowing for higher or lower price stickiness by varying the NKPC slope $\kappa$ affects the absolute size of the inflation responses to the fiscal shock, but not so much the relative importance of the asset market structure. The same is the case for wage stickiness (c.f. Appendix Figures \ref{fig:T_kappay} and \ref{fig:T_kappaw}). Results are similarly robust to considering alternatives values for the investment adjustment- and utilization parameters, which affect the responses of inflation and real aggregates only moderately (Appendix Figures \ref{fig:T_phik} and \ref{fig:T_util}). 

\subsection{Sticky Expectations}

It is well known that heterogeneous agent (HA) business cycle models struggle to match empirically estimated IRFs of aggregate consumption to some business cycle shocks: In the data, such responses are often found to be very smooth, while HA models generate initial jumps due to household's high MPCs and the absence of ``habit formation'' assumptions.
However, several authors argued that HA models extended with ``sticky expectations'', in that households only infrequently update their information about macroeconomic aggregates, can resolve the tension \citep[see][]{carrollEtAl2020,auclertEtAl2020a}.\\
Here, I address the potential concern that my baseline model's ``debt inflation'' dynamics may hinge on its (perhaps less desirable) full information rational expectation assumptions and introduce sticky expectations as in \cite{auclertEtAl2020a}.
I assume that households update their information with a quarterly probability of 10\%, close to the estimation result in \cite{auclertEtAl2020a} and more sticky than in the calibration of \cite{carrollEtAl2020}. The corresponding results for the transfer shock are displayed in Appendix Figure \ref{fig:T_sticky}: The responses are quite similar to the baseline in Section \ref{sec:fisc_imps} and the key insight on the relevance of the asset market structure is essentially unaffected. 

\section{Alternative illiquid asset adjustment}\label{app:alt_illiquid}

Section \ref{subsec:heterogeneity_role} found a tension in the baseline 2-asset HANK model with separated asset markets: Generating empirically plausible MPCs requires a substantial return gap between liquid and illiquid assets, but this gap in turn implies a strong sensitivity of the natural and neutral rate to public debt supply. This appendix asks whether richer household-level frictions can break this link and reconcile micro-outcomes with the macro-sensitivity. To investigate this possibility, I extend the model in the following way: If a household $i$ gets to adjust its illiquid asset stock $k_i$, it additionally faces the cost 
 \begin{align}
  \chi(k_{it+1},k_{it})  =  \frac{\chi_1}{\chi_2} \left | \frac{k_{it+1} - k_{it}}{k_{it} + \chi_0} \right |^{\chi_2} (k_{it} + \chi_0) ~~
 \end{align}
 for doing so. The specific functional form follows \cite{auclertssj} and the costs occur \emph{in addition} to the random adjustment frictions parameterized by $\lambda$.\footnote{I note that the seminal model of \cite{kaplanEtAl2018} provides for an endogenous inaction region for illiquid asset adjustment and thus effectively also features both infrequent adjustment and a convex adjustment cost.}

 In principle, the additional parameters introduced this way could be used to directly target additional household moments such as the average MPC and the aggregate interest sensitivity with respect to public debt, or other related moments. However, it is not clear that these can be matched simultaneously with all the other moments targeted by my baseline calibration, so evaluating the overall fit would require taking a stance on which moments to add with which relative weights. 
 To avoid this, I instead fix a grid of SS return gaps and adjustment cost parameters. For each of these combinations, I match the same micro moments as for the baseline model by re-calibrating the same set of parameters, which mapped clearly to their respective targets. After that, I evaluate further micro moments and the interest rate sensitivity in the resulting steady states. This makes transparent whether the adjustment costs allow for regions in the parameter space in which all relevant moments are simultaneously well-matched, without taking a stance on specific weighting choices or optimization schemes. However, my exercise could also be seen as an extra grid search over the parameters in question.

 \subsection{Details of the analysis}\label{app:alt_illiquid_approx}

 My analysis focusses on the case with separated asset markets, as the effects of public debt supply didn't noticeably vary with household heterogeneity in the integrated baseline model. Specifically, I proceed as follows: In line with previous literature \citep[e.g.,][]{auclertssj,debortoliEtAl2025}, I fix $\chi_2 = 2$ (quadratic costs). I initially set $\chi_0 = 0.25$ following \cite{auclertssj}, but will vary this parameter in a later step.\footnote{Following \cite{kaplanEtAl2018}, the idea is that costs are approximately proportional to the fraction of illiquid assets transacted, treating $\chi_0$ as small regularity parameter that avoids infinite adjustment costs at $k_{it}=0$.}
 The main parameters for the grid evaluation will in turn be $\chi_1$ and the SS rate gap, which is implemented by adjusting $\varphi$. 

 For my analysis, I calibrate and evaluate 100 steady states with $\chi_1$ between 0 and 5 as well as SS rate differentials between 1 and 4 annual percentage points. Taking these combinations as given, I use the same parameters $(\beta, \zeta, \lambda, \bar{R})$ to target the same ``micro targets'' as in the baseline model, while all other relevant parameters are not changed. This converges in all cases except for certain combinations of very low rate gaps and high adjustment costs.\footnote{For these, there is no $\lambda \leq 1$ large enough to make households' liquidity holdings sufficiently small.}
 As these micro parameters map clearly to their respective targets, this also avoids taking a stance on relative weights. 
 For each of the resulting calibrations, one can then evaluate the additional outcomes of interest.

 While it is in principle feasible to non-linearly solve for a new SS with higher public debt for the different calibrations, doing so adds considerable computational costs on top of the calibration for a 2-HANK model with reasonably detailed grids over all dimensions of household heterogeneity. To economize this aspect, I instead approximate the interest rate sensitivity as follows: Solving the household problem for higher and lower returns on liquid assets (keeping all else constant) provides the Partial Equilibrium (PE) demand curve for public debt. 
 Inverting this relationship then traces out what liquid returns households require to hold a given amount of liquid assets. The idea is now that for moderate increases in public debt, its GE effects through capital crowd-out will typically be moderate, so this PE rate sensitivity should provide a useful approximation of the GE one under the assumption of separated asset markets (under which public debt supply has to equal private liquidity holdings).
 The benefit of this procedure is that it replaces the multidimensional non-linear solve for a new SS with computing stationary household policies and -distribution just a fixed number of times, which is much simpler.

 \begin{figure}
\centering
\caption{Evaluation of the PE approximation}\label{fig:rate_sensitivities}
    \begin{subfigure}{.5\textwidth}
    \centering
    \caption{Baseline}
    \includegraphics[width=\linewidth]{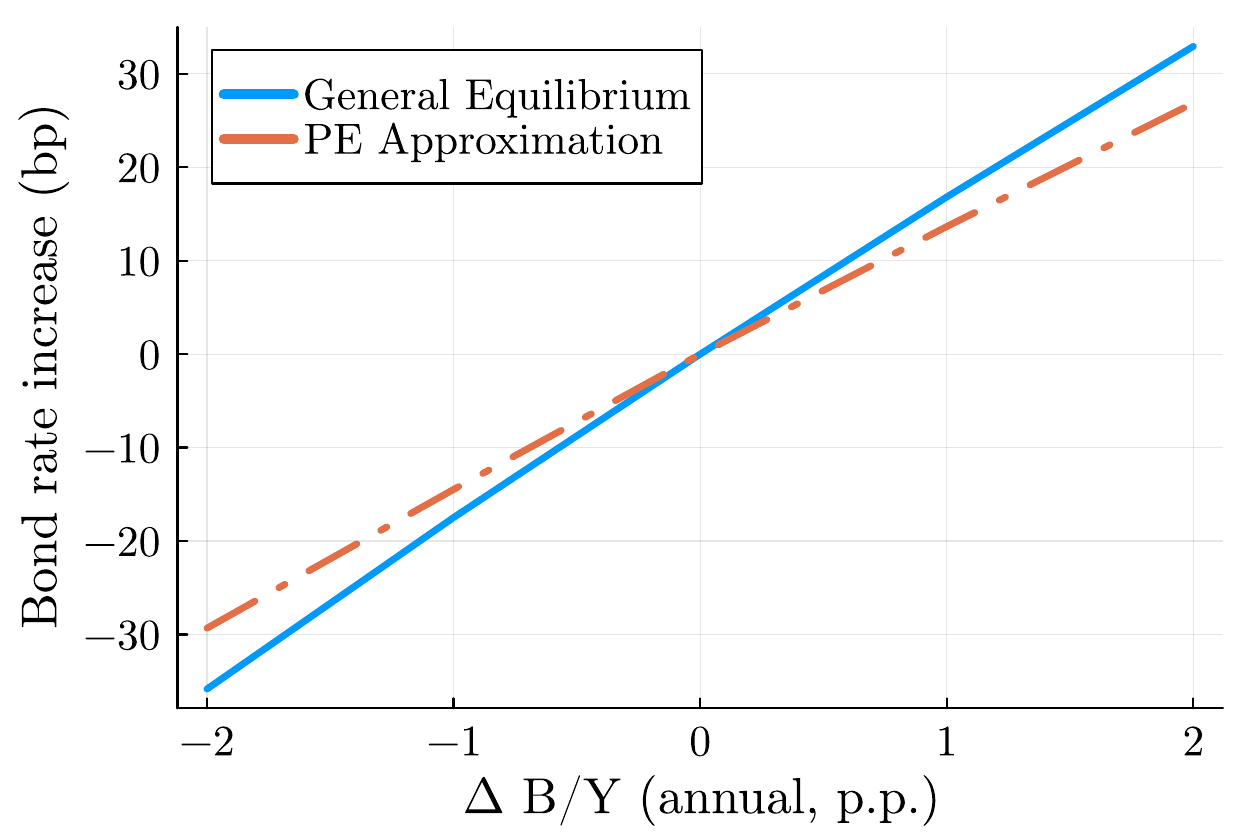}
    \label{fig:sensitivity_base}
    \end{subfigure}%
    \begin{subfigure}{.5\textwidth}
    \centering
    \caption{$\chi_1 = 5$}
    \includegraphics[width=\linewidth]{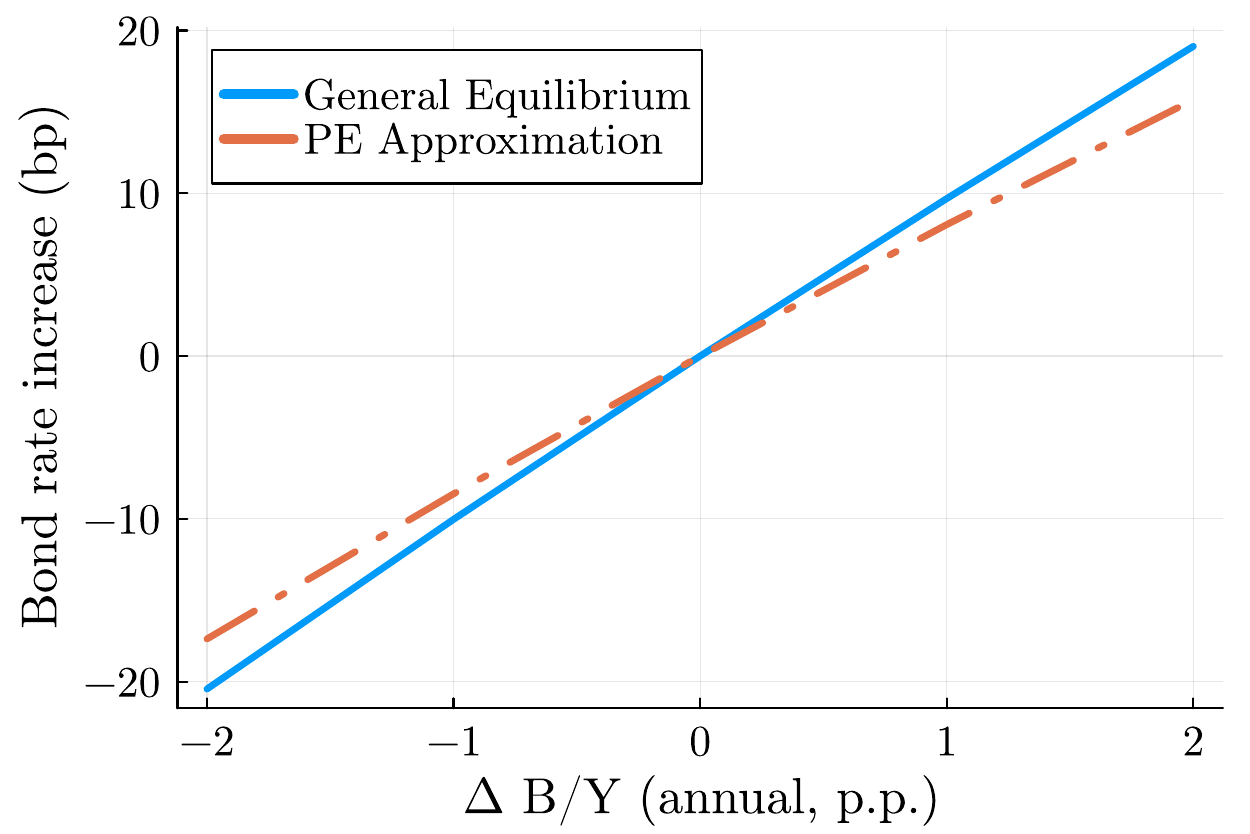}
    \label{fig:sensitivity_chi5}
    \end{subfigure}
    \fignote{The x-axis displays the long-run change in the annual debt-to-GDP ratio in percentage points (compared to the baseline calibration). The y-axis displays the effects of this change on the steady state return to public debt in basis points, both for true (blue line) GE counterfactuals and the Partial Equilibrium approximation (orange dashed line) described in Appendix \ref{app:alt_illiquid_approx}.}
\end{figure}

 The results of this procedure are illustrated in Figure \ref{fig:rate_sensitivities}, both for the baseline model and an alternative with adjustment costs $\chi_1 = 5$. The approximation seems reasonably accurate for small changes in the long-run level of public debt, but naturally becomes less accurate for larger ones. 
 Overall, the approximated bond return curve has a somewhat flatter slope than the GE one, which is economically intuitive: 
 In GE, higher public debt additionally increases the return on illiquid assets through capital crowd-out, as a lower capital stock increases its marginal return. Additionally, disposable labor income is negatively affected, as a lower capital stock decreases the marginal product of labor and the fiscal authority might have to levy higher taxes for debt service. These effects will additionally decrease households' desired liquid asset holdings in the long run, so the PE liquidity demand curve will be steeper and its inverse flatter. In turn, my procedure thus yields a conservative \emph{lower bound} to the true GE liquid rate sensitivity.

 \subsection{Results of the analysis}

\begin{figure}
    \centering
    \caption{Evaluating calibrations with adjustment costs }\label{fig:illiq_eval_1}
    \begin{subfigure}{.49\textwidth}
    \centering
    \caption{Avg. MPC}
    \includegraphics[width=\linewidth]{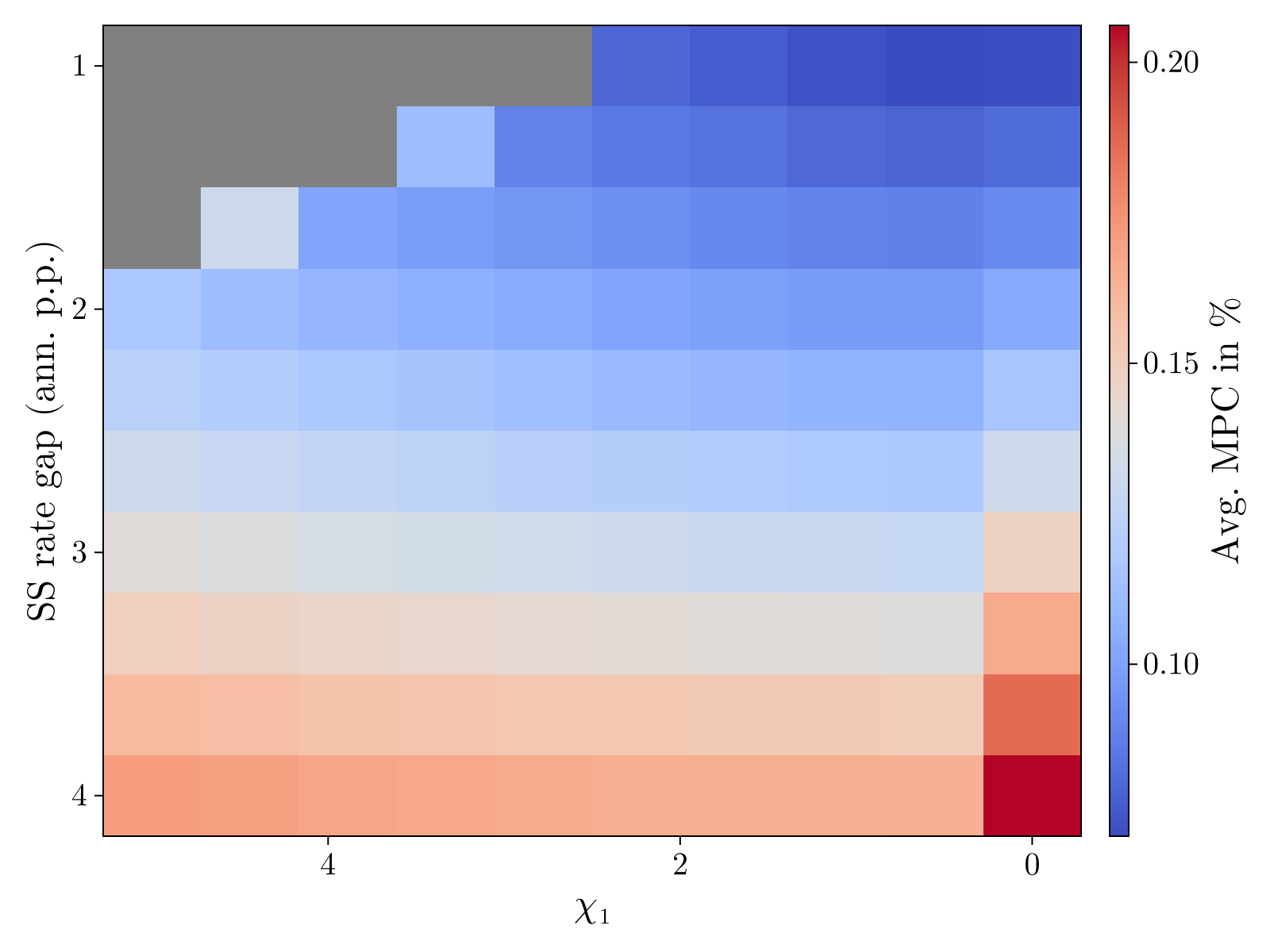}
    \label{fig:mpc_eval_1}
    \end{subfigure}%
    \begin{subfigure}{.49\textwidth}
    \centering
    \caption{Approx. rate sensitivity}
    \includegraphics[width=\linewidth]{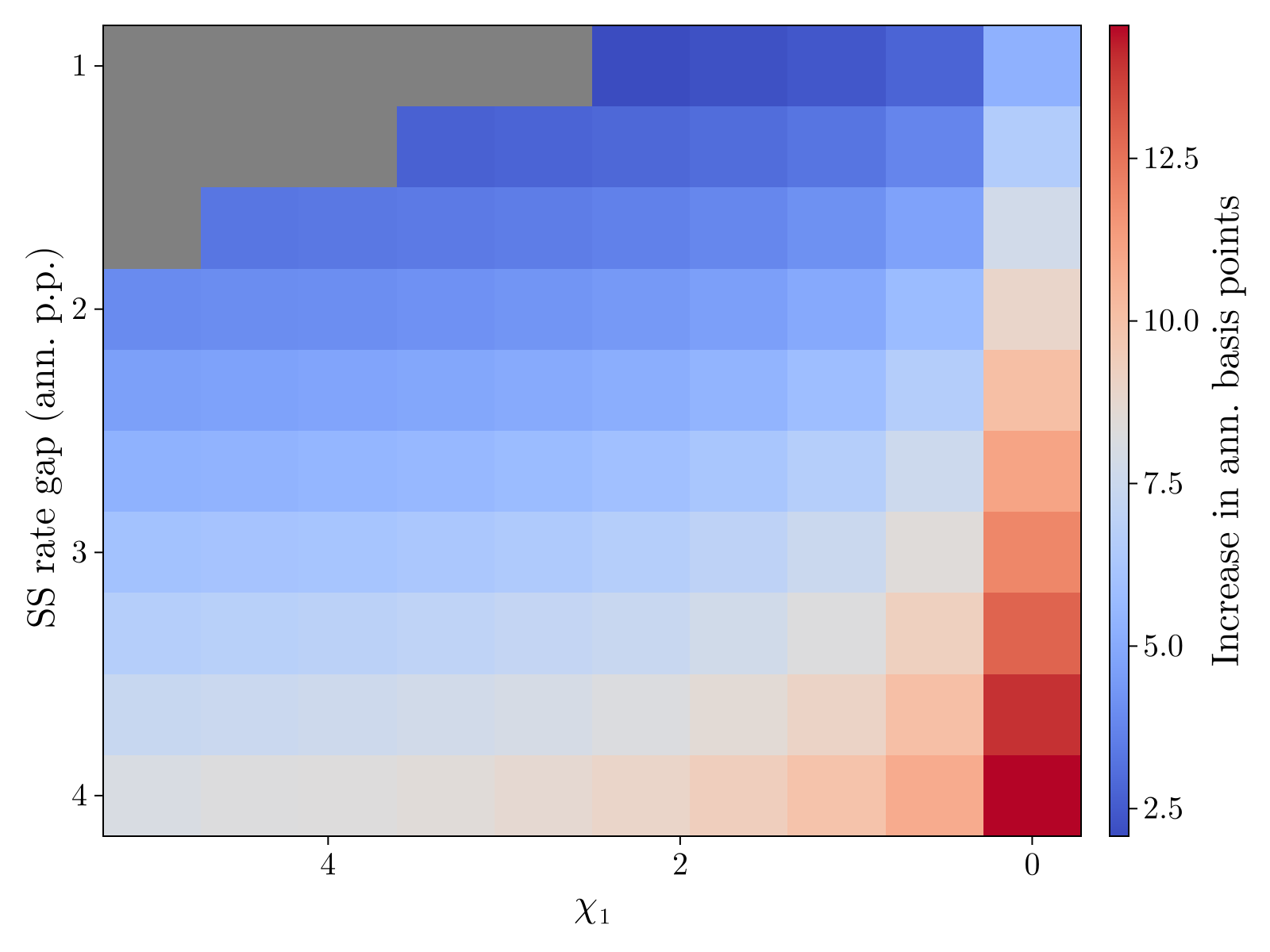}
    \label{fig:dR_eval_1}
    \end{subfigure}
     \begin{subfigure}{.49\textwidth}
    \centering
    \caption{Share of HtM households}
    \includegraphics[width=\linewidth]{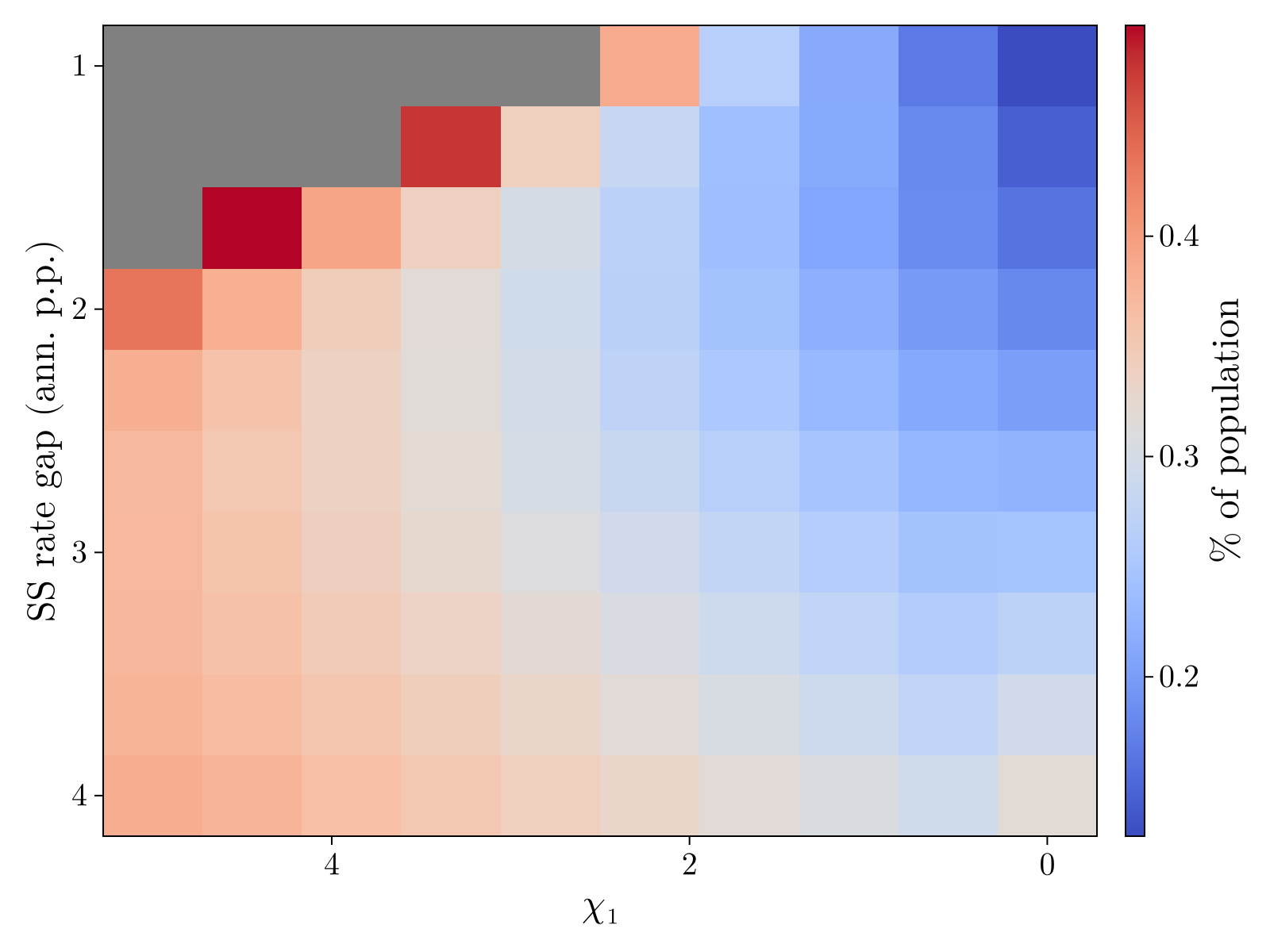}
    \label{fig:HtM_eval_1}
    \end{subfigure}
    \begin{subfigure}{.49\textwidth}
        \caption{Share of poor HtM households}
    \includegraphics[width=\linewidth]{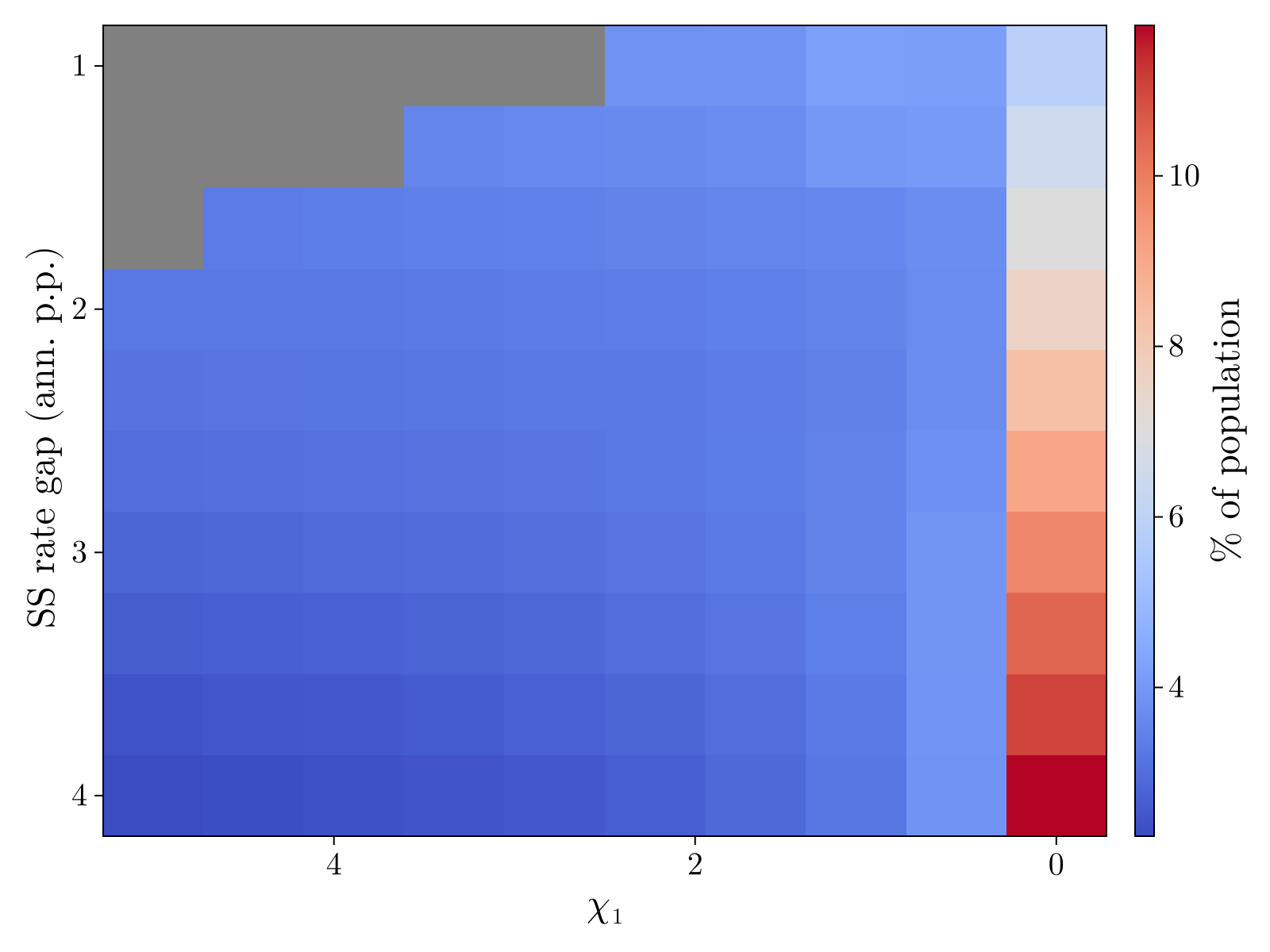}
    \label{fig:PHtM_eval_1}
    \end{subfigure}
    \fignote{Panel \ref{fig:dR_eval_1} displays the approximated effect of a 1 p.p. increase in the annual debt-to-GDP ratio on the public debt return in the long run. In the parameter region marked gray, it is not possible to match the liquid asset holding targets with $\lambda \leq 1$.}
 \end{figure}

 The results for the baseline analysis using $\chi_0 = 0.25$ are displayed in Figure \ref{fig:illiq_eval_1}.
 Overall, generating the same targets with a lower return gap and/or higher illiquid asset adjustment cost requires a higher adjustment probability $\lambda$ and lower borrowing penalty $\bar{R}$, while the differences in $\beta$ and $\zeta$ are less pronounced. Calibrations with higher return gap or higher adjustment costs require slightly higher $\beta$ values, while the calibration outcomes for $\zeta$ slightly decrease in both.\\
 Turning to the different calibrations' effect on the outcomes of interest, Panel \ref{fig:mpc_eval_1} illustrates how the average MPC in the model is shaped by the rate gap and the adjustment cost. Given that the baseline targets are matched by the other micro parameters, we see that for any given adjustment cost parameterization $\chi_1$, the average MPC clearly increases in the return gap between liquid and illiquid assets. Thus, this pattern confirms \cite{kaplanEtAl2022}'s argument of this being a key parameterization choice in 2-asset HANK models. The effects of the adjustment cost are less clear-cut: At low levels, a higher $\chi_1$ decreases the average MPC generated by the calibration, but it increases again for higher $\chi_1$.

\begin{figure}
    \centering
    \caption{Evaluation results for $\chi_0 = 2.5$}\label{fig:illiq_eval_2}
    \begin{subfigure}{.49\textwidth}
    \centering
    \caption{Avg. MPC}
    \includegraphics[width=\linewidth]{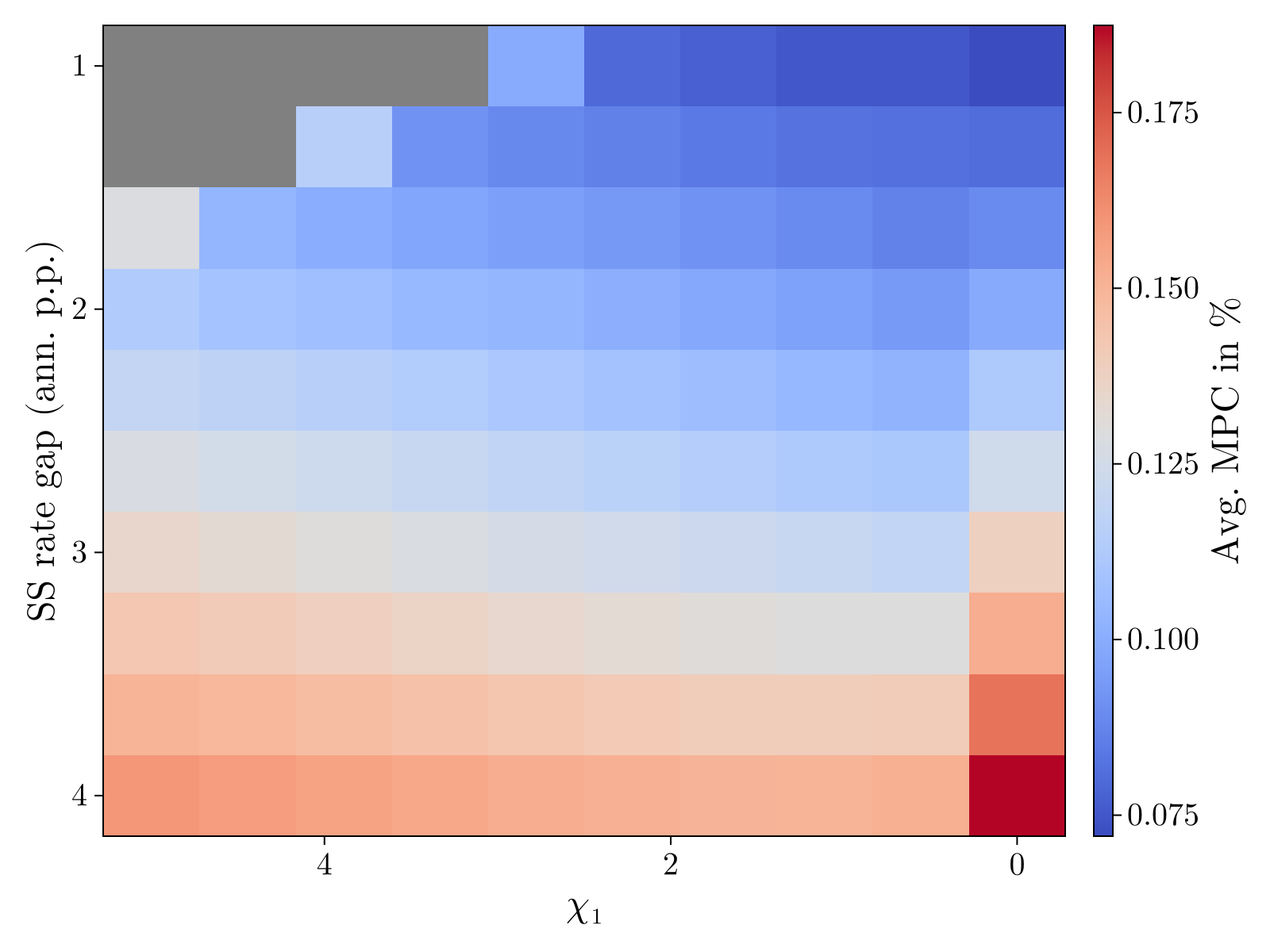}
    \label{fig:mpc_eval_2}
    \end{subfigure}%
    \begin{subfigure}{.49\textwidth}
    \centering
    \caption{Approx. rate sensitivity}
    \includegraphics[width=\linewidth]{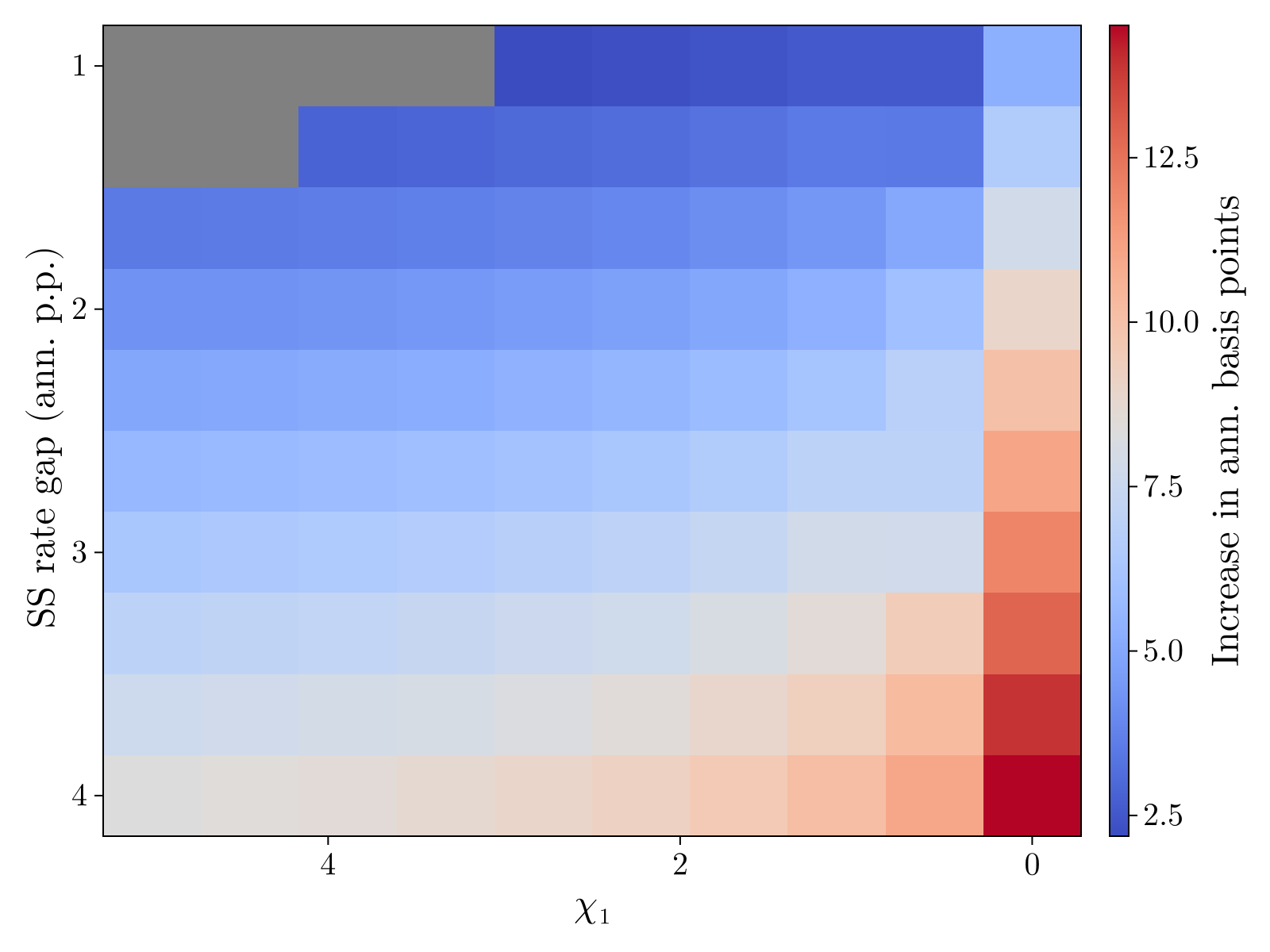}
    \label{fig:dR_eval_2}
    \end{subfigure}
     \begin{subfigure}{.49\textwidth}
    \centering
    \caption{Share of HtM households}
    \includegraphics[width=\linewidth]{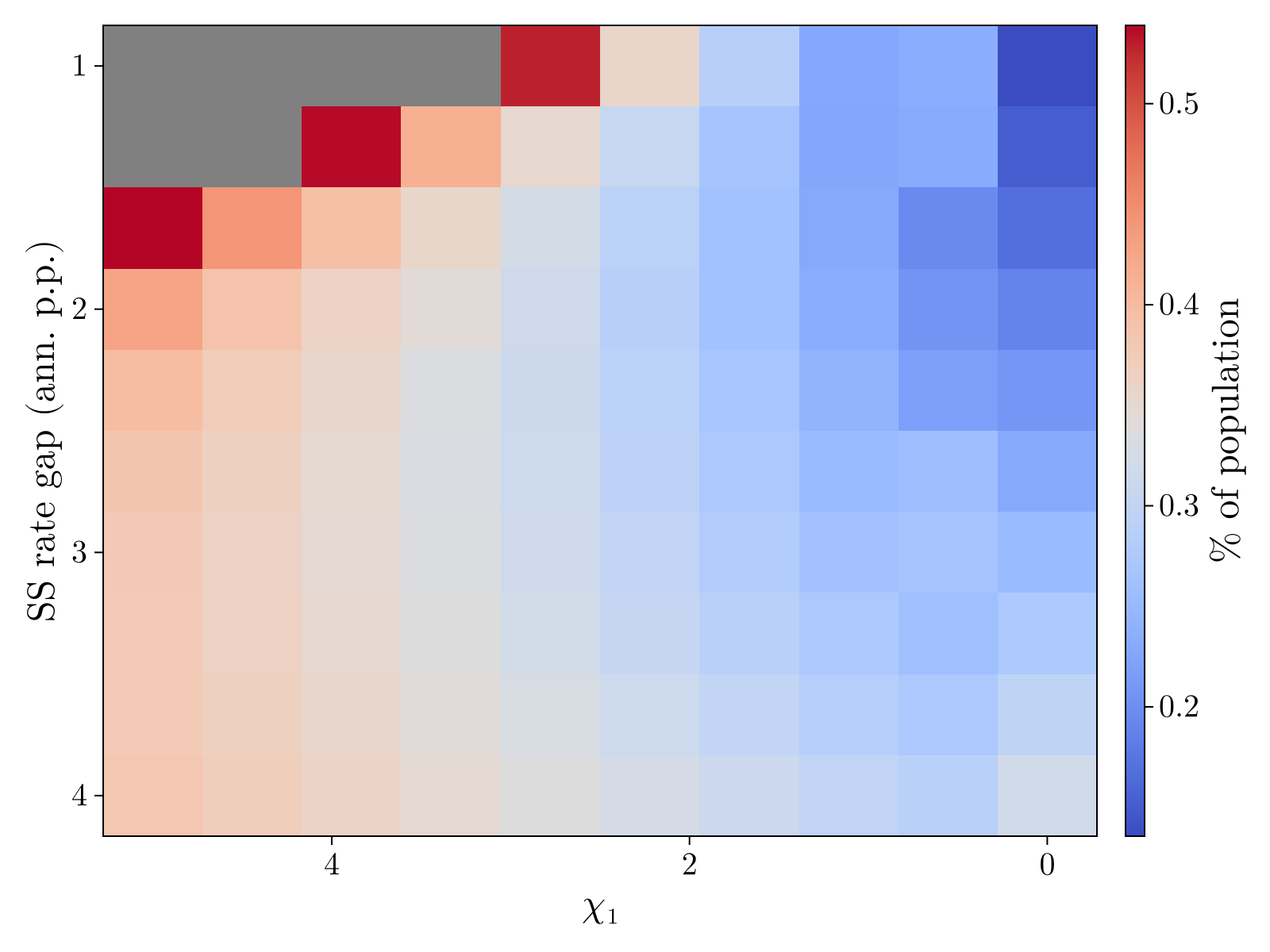}
    \label{fig:HtM_eval_2}
    \end{subfigure}
    \begin{subfigure}{.49\textwidth}
        \caption{Share of poor HtM households}
    \includegraphics[width=\linewidth]{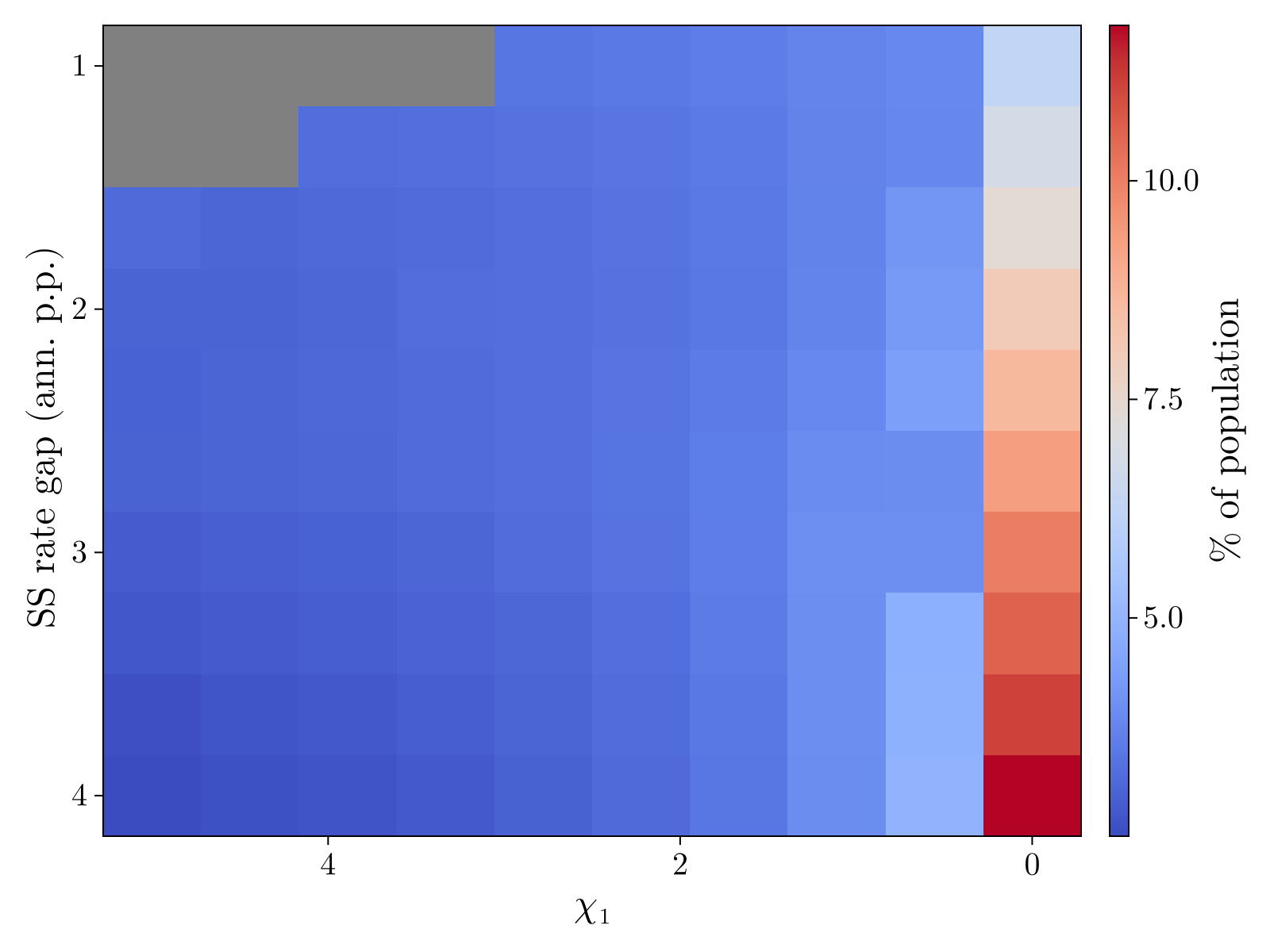}
    \label{fig:PHtM_eval_2}
    \end{subfigure}
    \fignote{Panel \ref{fig:dR_eval_2} displays the approximated effect of a 1 p.p. increase in the annual debt-to-GDP ratio on the public debt return in the long run. In the parameter region marked gray, it is not possible to match the liquid asset holding targets with $\lambda \leq 1$.}
 \end{figure}

 Regarding the liquidity effects of public debt, Panel \ref{fig:dR_eval_1} then shows clearly that the return gap and the effects of public debt supply on its return are tightly linked: For any $\chi_1$, the respective sensitivity increases in the return gap. However, for a given rate gap, the calibrations with the higher adjustment costs display a smaller effect.
 This suggests that by combining a sufficiently high $\chi_1$ with a high return gap, it should in principle be possible for the model to generate both a high average MPC and reasonable liquidity value of public debt while matching the baseline targets.\\
 However, $\chi_1$ also affects micro outcomes beyond the MPC. Given the prominence of these moments in the literature, I consider here additionally the fraction of HtM households as well as their split into ``wealthy'' and ``poor'' HtM households.\footnote{As specified in Appendix \ref{app:dist_moments}, I define ``HtM'' households by their liquid asset holdings. In line with the empirical literature, Households are classified as HtM if their liquid asset stock amount is less than 1/6th (=2 weeks) worth of the household's labor income above either 0 or the borrowing constraint. HtM households are considered ``poor'' if they don't hold illiquid assets and ``wealthy'' otherwise.}
 Considering Panel \ref{fig:PHtM_eval_1}, we see that for the relevant ``high $\chi_1$'' calibrations, there are much fewer ``poor'' HtM households than typically reported by empirical analyses such as \cite{kaplanEtAl2014} (around 10-15\% in the US). 
 \\
 Additionally, in Panel \ref{fig:HtM_eval_1}, we see that even though the overall fraction of HtM households generated by the calibrations are mostly reasonable, they are no longer monotonically related with the MPC in the presence of substantial adjustment costs. Especially for moderate return gaps, one can have a big mass of HtM households but quite moderate MPCs.  
 It thus seems that if one postulates the main friction for illiquid adjustment to be a smooth adjustment cost, the fraction of HtM households would no longer be an informative moment about the average MPC by itself.
 \\
 Overall, I conclude that although the introduction of smooth adjustment costs could in principle square high MPCs with a reasonable liquidity value of public debt while the baseline targets are matched, this comes at the cost of a poor fit with other moments emphasized by the literature and is thus no panacea.
 \\
 Finally, to what extent do these conclusions depend on the parameter $\chi_0$, which is also used by some authors to explicitly match micro-level targets? Figure \ref{fig:illiq_eval_2} repeats the exercise for the value $\chi_0 = 2.5$ instead. With that alternative parameterization, we find that the calibrations with low rate gaps generate somewhat larger masses of HtM households. In the big picture, though, not much changes otherwise. In particular, the higher $\chi_0$ value doesn't seem to do much to address the issue of the ``high MPC, low rate sensitivity''-calibrations generating unrealistically few ``poor'' HtM households.

\section{The asset market structure and short run dynamics}\label{app:irf_comp}

The aim of this Appendix is to provide insights on a) whether the long-run calibration of $\Psi$ (or other model variants) allow generating reasonable short-run variation of public debts' liquidity value, as measured by interest rates or spreads, and on b) whether standard business cycle moments can easily discriminate between the different asset market calibrations. 
\begin{figure}
    \centering
    \caption{Fit with \cite{bayerEtAl2023b} fiscal policy IRFs}\label{fig:emp_irfs}
    \includegraphics[scale = 0.6]{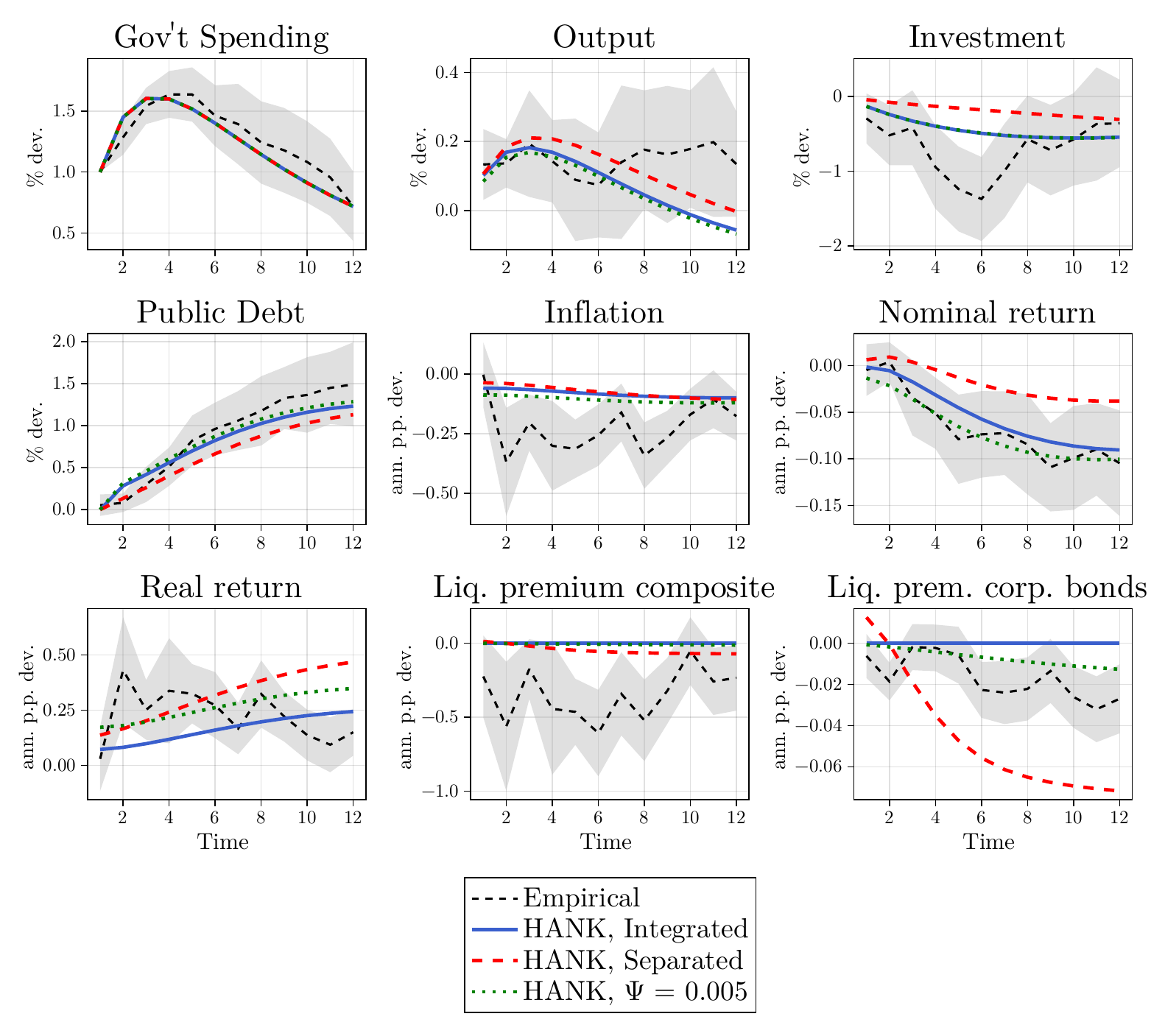}
    \fignote{The grey-shaded areas indicate 90\% confidence bounds for the empirical IRFs. ``Real return'' refers to the real return on government bonds. ``Liq. premium composite'' refers to a composite of the return premia of capital, housing, corporate bonds and equities to public debt. ``Liq. prem. corp. bonds'' refers to AAA corporate bond yields minus the government bond rate. See \cite{bayerEtAl2023b} for details on the IRF estimation and variable construction. The HANK models feature augmented policy rules as specified in Appendix \ref{app:irf_comp_adapt}.}
\end{figure}
For this purpose, I require identified variation of these outcomes as well as public debt. A recent paper that provides those is \cite{bayerEtAl2023b}, who estimate Local Projection IRFs of macro outcomes as well as various spreads to US government spending shocks.\footnote{I refer to the respective paper for details on variable construction and empirical methodology. The IRF estimates were available at \url{https://github.com/BASEforHANK/BASEtoolbox_IRFmatching.jl/blob/main/src/irf_data_0706_inclPC.csv} as of April 2026.}

\subsection{Adapting the HANK models}\label{app:irf_comp_adapt}

\begin{figure}
    \centering
    \caption{IRFs to transfer shock: Public debt reaction}
    \includegraphics[scale = 0.55]{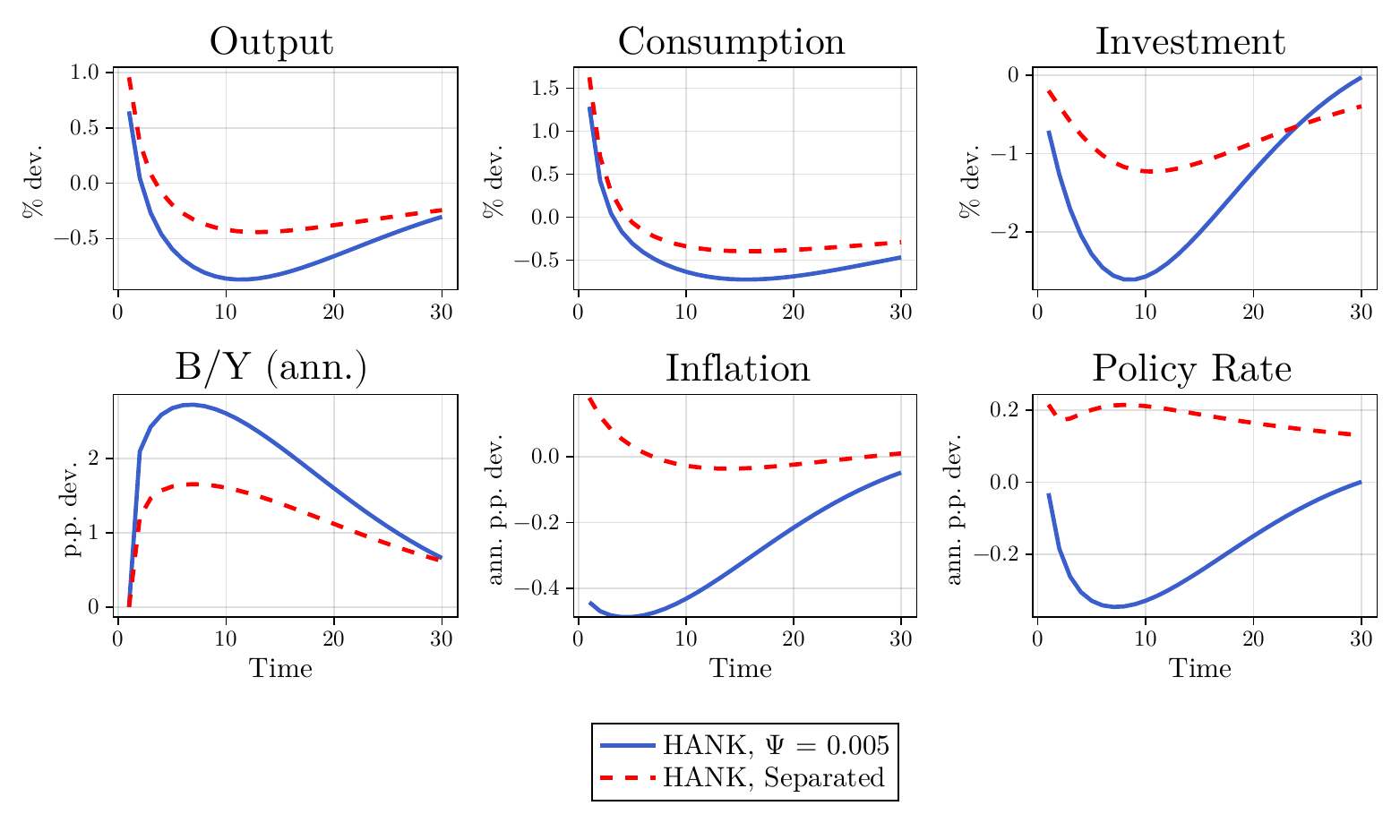}\label{fig:T_thetaB}
    \fignote{Results under monetary policy rule \eqref{Taylor_eq_app} with $\theta_B = 0.02$. $B/Y$ represents the real market value of public debt $B^g$ over 4 times (=annualized) GDP.}
 \end{figure}

To be able to meaningfully relate my HANK models to their evidence, I add two features on the policy side: Firstly, an autonomous AR(2)- law of motion of government consumption $G_t$
\begin{align*}
   \frac{G_t}{G_{SS}} = \left( \frac{G_{t-1}}{G_{SS}}\right)^{\rho_{G1}} \left( \frac{G_{t-2}}{G_{SS}}\right)^{\rho_{G2}}\epsilon_t^{G},
\end{align*}
where $\epsilon_t^{G}$ denotes the exogenous shock. As in \cite{bayerEtAl2023b}, the AR(2) model can generate the empirical hump shape for $G_t$ visible in Figure \ref{fig:emp_irfs} and I parameterize $\rho_{G1}=1.44$ and $\rho_{G2} = -0.477$ based on these authors' model estimation.\\
Additionally, the \cite{bayerEtAl2023b} estimates display the ``fiscal price puzzle'' that inflation declines in response to the fiscal shock, which typical New Keynesian models cannot generate \citep[see][]{jorgensenEtAL2022}. Several ways to address this have been employed in the literature and I again follow \cite{bayerEtAl2023b} by adding a central bank reaction directly to the level of public debt, i.e., I modify \eqref{Taylor_eq} to
\begin{align}
   R_{t+1}^R = R^R_{SS}\left(\frac{ R_{t}^R }{R^R_{SS}}\right)^{\rho^R}\left[\left(\frac{\pi_{t}}{\pi_{SS}}\right)^{\theta_\pi}\left(\frac{Y_{t}}{Y_{t-1}}\right)^{\theta_y} \left(\frac{B^g_{t}}{B^g_{SS}}\right)^{\theta_B}\right]^{1-\rho^R} \label{Taylor_eq_app}
\end{align}
where $\theta_B$ parameterizes the debt reaction. In principle, a central bank reaction to the public debt stock can be motivated by assuming the central bank wishes to offset its effect on the neutral rate (compare Footnote \ref{footnote:rate}) but \cite{bayerEtAl2023b} choose it solely based on fitting the short-run IRFs. 
Since the $\theta_B$ necessary to get a reasonable fit with the nominal rate, inflation and also public debt differs a lot depending on $\Psi$, I set different values for the different model versions to level the playing field and better isolate the role of the asset market structure.\footnote{Matching these responses seems to require a reaction to public debt that is substantially stronger than what public debt's neutral rate effects would call for, as the flipped inflation response also suggests. Note that for any given $\theta_B$, the asset market structure still strongly influences the aggregate effects of fiscal policy, as Appendix Figure \ref{fig:T_thetaB} illustrates.}
In particular, I consider $\theta_B = 0.015$ for the integrated asset market HANK, $\theta_B = 0.0315$ for the separated asset market HANK and $\theta_B = 0.02$ for the $\Psi = 0.005$ baseline. All other parameters are unchanged compared to the baseline calibration.

\subsection{Comparison with empirical IRFs}

Figure \ref{fig:emp_irfs} now compares the responses of the HANK models to a shock $\epsilon_t^G$ to their empirical counterparts. Either model version seems to be able to generate a reasonable fit for aggregate outcomes such as output, the stock of public debt, the nominal rate and in fact also the real return to liquid assets following the shock. 
It arguably makes sense that the short-run responses would be driven a lot by channels present in either model version, such as the aggregate wealth effect of the shock to $G$, households' consumption behavior as well as strong central bank reactions to $B$. We see that the HANK model with separated asset market undershoots the investment response and eventually overshoots the return IRF, consistent with too little crowding out of capital. However, at least the former could be mitigated in IRF matching exercises by different investment adjustment costs. This suggests that it may be tricky to pin down $\Psi$ based on short-term macro IRFs alone. And since investment dynamics being driven by adjustment costs or capital crowd-out has different policy implications, it motivates the long-run rate sensitivity as additional target.

\begin{figure}
     \centering
     \caption{IRF fit: Comparison with $\theta_B = 0.0$}\label{fig:emp_irfs_adv}
     \includegraphics[scale = 0.6]{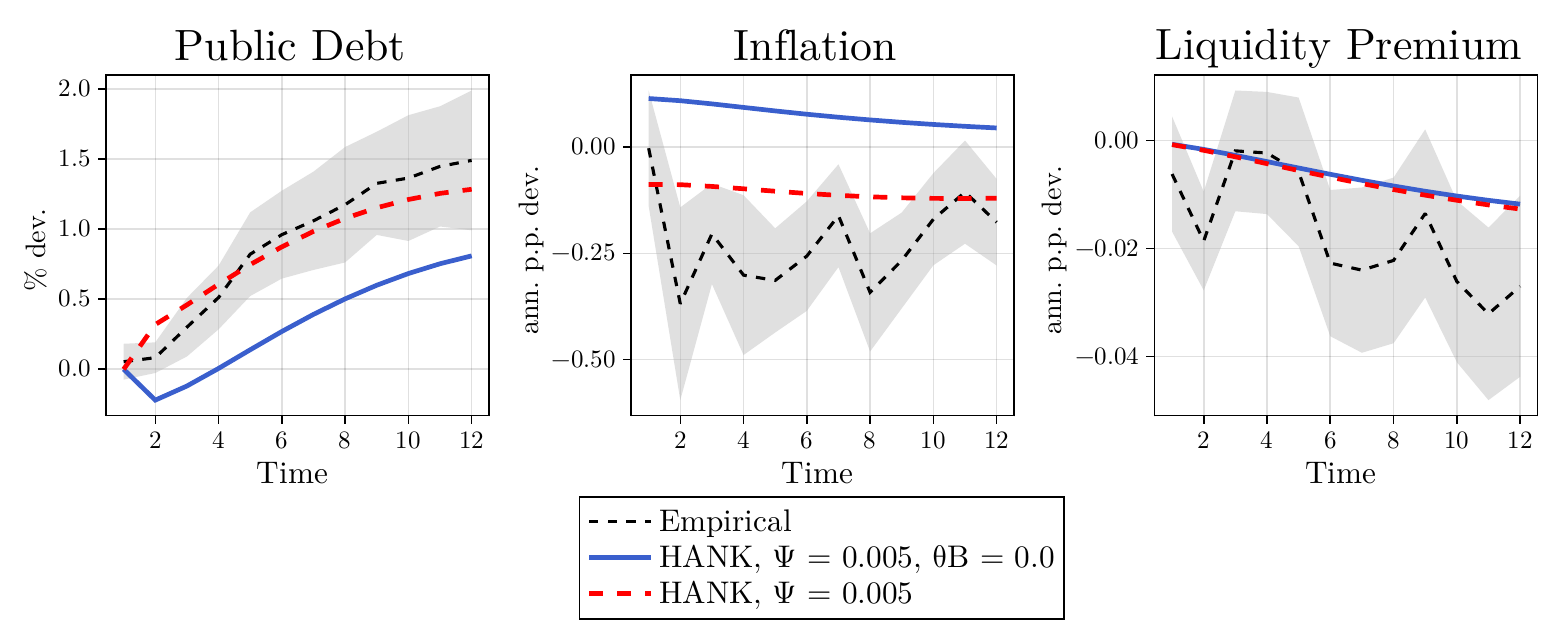}
     \fignote{The grey-shaded areas indicate 90\% confidence bounds for the empirical IRFs. The empirical ``Liquidity Premium'' refers to AAA corporate bond yields minus the government bond rate. See \cite{bayerEtAl2023b} for details on the IRF estimation and variable construction.}
 \end{figure}

The question remains whether the calibrated model with $\Psi=0.005$ (or the other model versions) generate reasonable short-run variation in public debts' liquidity value. In the bottom row of Figure \ref{fig:emp_irfs}, we see that a reasonable fit with real liquid returns can be achieved under very different asset market structures, consistent with the interest rates being driven mostly by other channels immediately after the shock. Hence, I turn to different types of spreads. 
First, I compare the IRFs of the model-generated liquidity premium (the spread between capital and government bond returns) with the one for \cite{bayerEtAl2023b}'s composite measure that considers the returns to capital, housing and equity. Clearly, none of the HANK variants can match the substantial magnitude of the empirical IRFs. 
The composite spread is likely driven by channels beyond the scope of linearized HANK models, such as time-varying risk premia.
\\
A more appropriate counterpart for the HANK models' liquidity channel is arguably the spread between AAA-rated corporate bonds and public debt. While such bonds may not seem like an obvious counterpart to ``capital'' in the 2-asset HANK model, this spread is a standard measure of public debt's liquidity value in the finance literature \citep[c.f.][]{krishnamurthyEtAl2012}.  As the asset market calibration aims to discipline precisely public debt's liquidity value, it seems a more relevant measure than the composite spread likely driven by economic forces absent in typical HANK models. We see that the calibrated model with $\Psi=0.005$ does well in matching also the short-run dynamics of this spread, without this being specifically targeted.
In contrast, the HANK with segmented asset markets overshoots this measure of the liquidity premium substantially, while it is constant by construction in the setup with integrated asset markets.\\
It is important to note that the good fit with public debt's liquidity value is unrelated to the unconventional $\theta_B$ parameter. Figure \ref{fig:emp_irfs_adv} briefly contrasts responses with and without this parameter. With $\theta_B = 0$, the fit with inflation (and other macro outcomes) deteriorates, but the one for this liquidity premium hardly changes --- it is mostly driven by the persistent increase in public debt, which reaches similar levels in the medium run regardless of $\theta_B$. 

\section{Post-2020 US experiment}\label{app:experiment}

As described previously in Appendix \ref{app:two_asset_hank}, the HANK model contains 5 shocks, two supply shocks (investment technology, ``cost push''/markup), a demand shock (discount factor) and two policy shocks (monetary policy, transfers). 
Under the assumption that the economy was in SS in 2019, these exogenous disturbances are used to replicate the subsequent evolution of aggregate output, investment, inflation, the central bank's policy rate as well as government transfer spending.
For the real variables subject to trend growth, this refers to the relative deviation from their pre-pandemic trends instead of levels.\footnote{Strictly speaking, I cannot match transfer spending's relative deviation from trend in my model as transfer payments are zero in its stationary equilibrium. Instead, I match the deviation of transfers relative to trend output.} 
For further information on the construction of the targeted variables, please refer to Appendix \ref{app:data}.

The choice of the 5 business cycle shocks mentioned above is motivated as follows: The discount factor shock is supposed to induce pandemic-related consumption restraints as in \cite{bardoczyEtAl2024}, which requires additional investment shocks to not give rise to counterfactually higher investment.
The monetary policy- and transfer shocks are needed to replicate the time-paths of the policy rate and transfer payments while the ``cost-push'' shock introduces another source of supply side inflation.
I intentionally do not model other Covid-related spending programs such as support for corporations as it is less clear to whom these should be assigned in my framework. Given that this will result in my model generating less public debt than in the data, I view this as a conservative choice.

Since the number of shocks equals the number of target variables, the \cite{mckayEtAl2021} filtering method does not require me to take a stance on the variance of the business cycle shocks. The method can account for an ELB and is described in more detail in Appendix \ref{app:numerics}.\\
However, assuming that all shocks follow $AR(1)$-processes, I still need to make assumptions on their persistence in order to compute the necessary IRFs. My calibration choice is presented in Table \ref{tab:rhos}: The values for the supply shocks ($\mu, Z^I$) and the discount factor disturbance ($A$) are set to salient values in the ballpark of \cite{bayerEtAlaer}'s estimates.
Since many of the big transfer expansions during the pandemic period were designed to be short-lived, I further assume the autoregressive parameter of the transfer shock to be a low $0.5$. As the rate smoothing term of the monetary policy rule \eqref{Taylor_eq} already provides for a persistent impact of the rate shock $\epsilon^R$, the latter does not depend on its previous value.
\begin{table}
    \centering
    \begin{tabular}{c | c c c c c }
        \hline \hline
      & $Z^I$ & $\mu$ &  $\epsilon^R$ & $A$ &  $T$ \\ \hline 
      AR(1) persistence & 0.75 & 0.9 & 0.0 & 0.9 & 0.5 \\ \hline \hline
    \end{tabular}
    \caption{Assumed persistence of model shocks} \label{tab:rhos}
\end{table}
To ensure that the short-lived transfer shocks result in a long-lived debt response, I  furthermore reduce fiscal responsiveness to $(\rho_\tau, \psi_B) = (0.94, 0.5)$. These values were selected so that public debt does not noticeably decline before 2026 in the filtered baseline scenario. 
\\
Naturally, the resulting analyses have caveats. Firstly, my model doesn't feature any Covid-related features such as lock-downs but rather assigns the observed aggregate dynamics during 2020-2021 to standard business cycle shocks. While in line with other DSGE model-based investigations of the post-pandemic inflation \citep[e.g.][]{galiardoneEtAl2023,bianchiEtAl2023}, the model does thus only provide a simplistic account of various pandemic-specific phenomena. This includes the nature of transfer payments made by the government, which are assumed to consist solely of uniform lump-sum payments for the purpose of this exercise: Since transfers specifically aimed at poor agents with high MPCs tend to have larger effects in HANK models, this simplification can again be seen as a conservative choice.
Secondly, the fact that the analysis is based on a linearized model means that we will miss out on non-linearities that may be relevant for the large shocks occurring during the period under consideration. Again, my analysis shares this reservation with numerous other studies of the US post-2020 period (including the two cited previously).
Finally, all the results obviously depend on the assumption of underlying policy rules: Under different ones, e.g., a partly active fiscal policy regime as in \cite{bianchiEtAl2023} instead of the active Taylor rule, the same aggregate dynamics might be assigned to different shocks.

\subsection{Post-2020 aggregate dynamics}

 \begin{figure}
    \centering
    \caption{Aggregate dynamics using filtered shocks}
    \includegraphics[scale = 0.55]{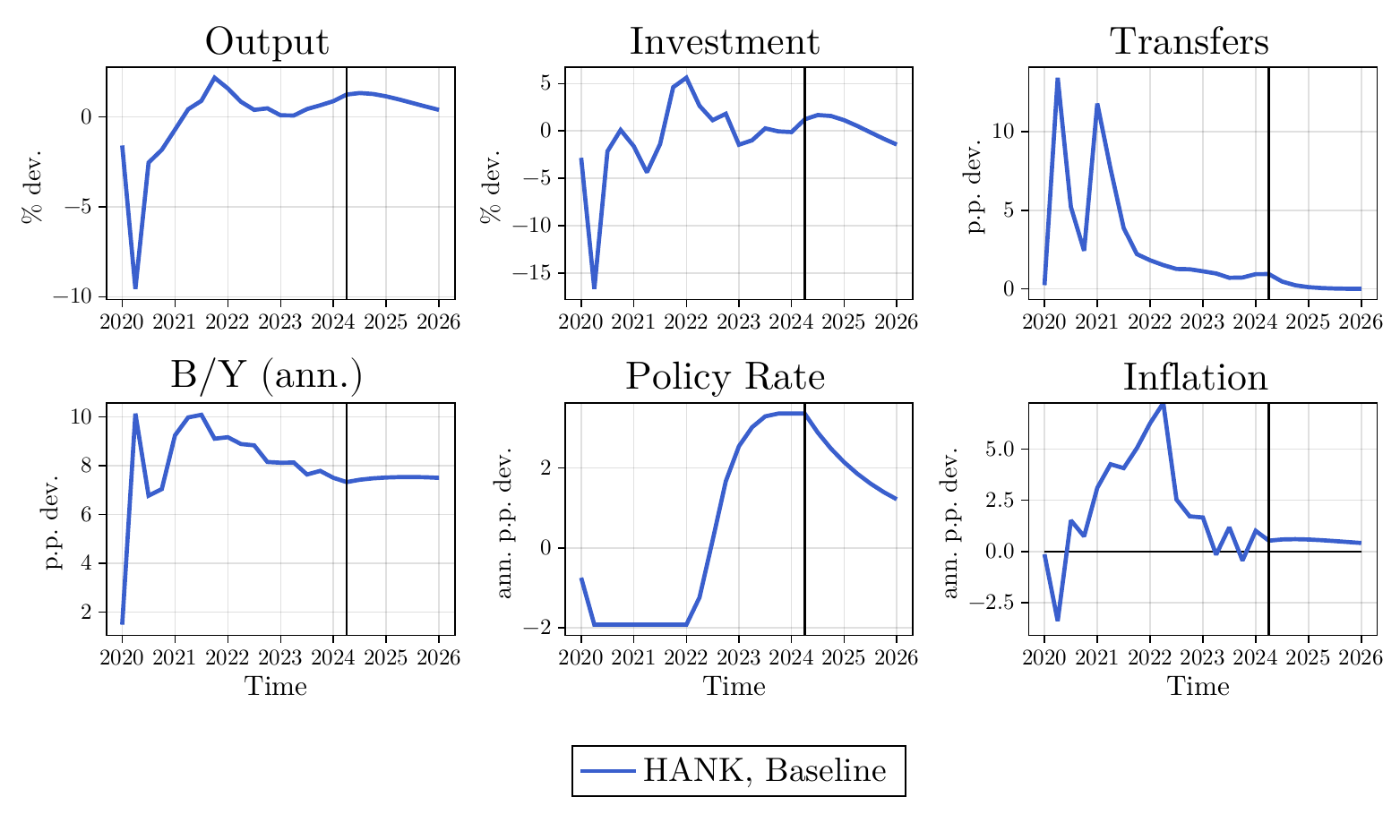}
    \fignote{$B/Y$ represents the market value of public debt $B^g$ over annualized GDP.}
    \label{fig:postcov_base}
 \end{figure}

 Using the resulting set of aggregate shocks obtained as described, one can simulate the model from 2020:Q1 forward: 
 The dynamics of several key macroeconomics aggregates are displayed in Figure \ref{fig:postcov_base}. Note that by construction, the time paths of all displayed variables except the public debt variable ($B/Y$) equal their counterparts in the data until my sample ends in 2024:Q2 (indicated by the black vertical line). Beyond that point, the model is simulated forward without any additional shocks hitting, implying the targeted variables deviate from their empirical counterparts.

 In the beginning of 2020, we see real variables such as output and investment deteriorate sharply, accompanied by declining inflation and the central bank's policy rate hitting its ELB. Through the lens of the model, this is due to a combination of households' Covid-related consumption restraints (a discount factor shock) and an unfavorable investment technology disturbance. 
 At the same time, public debt relative to GDP jumps up, both due to the decline in the denominator and a big shock to transfers, the targeted time path of which is displayed in the top-right panel. Of course, as I also illustrate in Appendix \ref{app:non_tar}, the model increase in debt is smaller than its relevant counterpart in the data, given that I targeted only transfer spending and not other fiscal variables.

 Afterwards, the economy recovered quite quickly: In the HANK model, this is partly due to the initial shocks easing quicker than expected, i.e., expansionary discount factor- and investment technology innovations, and also aided by accommodative monetary policy as well as another spike of transfers at the beginning of 2021 (the American Rescue Plan Act).
 Eventually, inflation started to rise precipitously amid output and consumption above their pre-covid trends, inducing the Federal Reserve (Fed) to start raising its nominal rate in the beginning of 2022.
 Price pressures eased quickly initially, but inflation remains elevated above target and according to the model's out-of-sample prediction, it would be for quite some time into the future, as it subsequently turned out to be.
 At the same time, the model's value of public debt relative to GDP has not substantially declined since its peak at the beginning of the pandemic.\footnote{Under the fiscal rule in place, public debt would be expected to start declining in late 2026. For a graph with a longer simulation horizon, see Appendix Figure \ref{fig:postcov_long}.} 
 
 Are the implied model dynamics plausible considering non-targeted model moments and subsequent (non-matched) macroeconomic dynamics for the US? A respective analysis is provided in the subsequent Appendix \ref{app:non_tar}. To briefly summarize its results, the model has difficulties relating to some labor market moments during the initial Covid-pandemic and -recovery phase. This is arguably not surprising, as its simple labor-market setup is ill-suited to capture the many specifics of the Covid unemployment surge. Nevertheless, the fact that it matches later developments in aggregate labor compensation suggests that it captures important economic drivers of the later inflationary period. Additionally, the model's unconditional forecasts of the time paths of inflation and interest rates beyond 2024:Q2 seem roughly in line with expectations at the time.

\subsection{Comparison for untargeted variables}\label{app:non_tar}

    In this Appendix, I gauge how well the model relates to some non-targeted moments to be able to judge what aspects of the US economy it does or does not capture well for the post-2020 exercise. Again, the construction of the additional data used here is specified in Appendix \ref{app:data}.
    \begin{figure}
    \centering
    \caption{Non-targeted variables: Model vs. Data}
    \includegraphics[scale = 0.575]{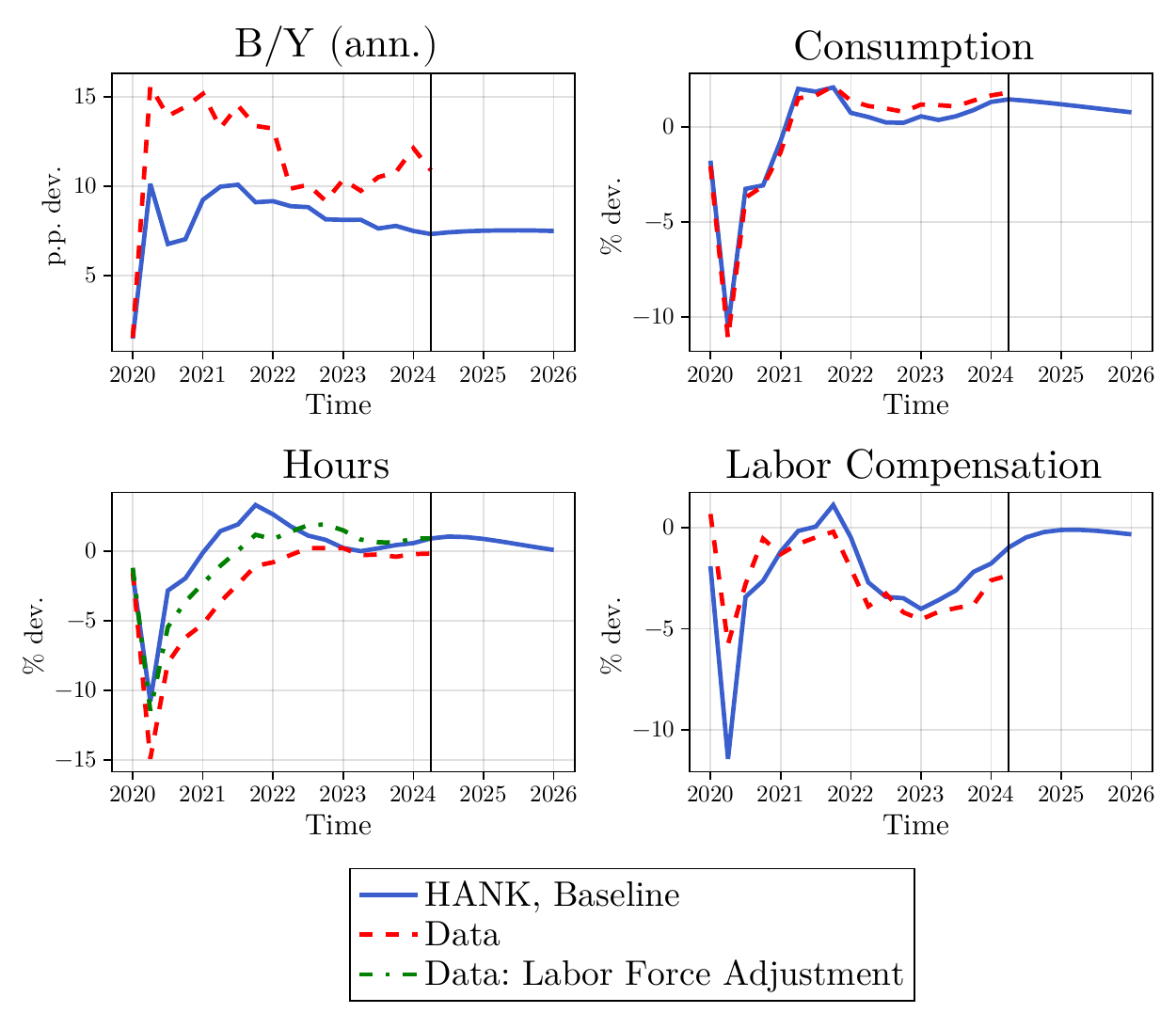}
    \fignote{Note: $B/Y$ represents the market value of public debt $B^g$ over annualized GDP. The construction of the data series is specified in Appendix \ref{app:data}.}
    \label{fig:nontar}
    \end{figure}
    The top-left panel of Figure \ref{fig:nontar} compares the relative value of public debt in the model with an approximation of the market value of domestically held US federal debt in the data. As anticipated, the model generates a smaller expansion, implying that if anything, my exercise may have under-estimate the amount of ``debt inflation'' for a given calibration.

    While aggregate consumption was not directly targeted in the construction of the shocks, the resulting fits the data very well: This is not surprising as in the model (and reality), the by far most important components of GDP are private investment and consumption. Having targeted both output and investment, a good fit for consumption is to be expected.\footnote{The fact that the match is not perfect is due to the variables being de-trended separately and the dynamics of government consumption not being targeted.}

    A relevant and more interesting set of non-targeted model variables relates to labor supply.
    Here, the success of my model partly depends on which data moment one would consider the most relevant real-world counterpart for the simple set-up in my HANK model: In the center panel of Figure \ref{fig:nontar}, we observe that if one follows the common convention of using Hours Worked divided by the aggregate population (red-dashed line), my model substantially over-estimates the labor supply recovery after the pandemic. In contrast, if adjusting by the size of the civilian labor force (green dot-dashed line), it does a better job. 

    Given that the conventional DSGE labor market set-up in my model is arguably too simplistic to capture details such as time-varying participation and composition-effects important during the pandemic recovery, a perhaps more reasonable demand is that it matches well the overall amount of labor compensation paid ($h_t H_t$ in the model), which directly matters for the model households' aggregate consumption- and savings decisions.
    While my model overstates the initial drop at the beginning of the pandemic, it matches its subsequent dynamics well, inspiring confidence that the HANK framework captures the relevant economic forces at work at least after 2020.\footnote{The smaller initial drop in labor compensation despite the strong fall in hours at the beginning of the Covid-pandemic is due to composition effects, with much more low-income than high-income workers being laid off in early 2020.}

    Finally, one may be wondering whether the model's forecast of elevated inflation and a decreasing federal funds rate is at odds with the developments after 2024:Q2. On this issue, recall that the forward simulation does not account for subsequent shocks hitting the economy and any further disturbances could of course explain the difference. 
    Indeed, from the perspective of 2024:Q2, inflation expectations in my model do not seem unreasonably out of touch with various measures of inflation expectations at the time: 
    According to the Cleveland Fed's model of inflation expectations, 1- and 2-year expected inflation was approx. 2.7\% and 2.6\% in June 2024, respectively. The May 2024 Survey of Professional Forecasters suggests an expected inflation of $3.1\%$ for 2024 and 2.5\%  for the period 2024 - 2028, while FOMC member's inflation expectations ranged between $2.5 - 3.0\%$ for 2024 and $2.2 - 2.5\%$ for 2025 at the time \citep[c.f.][]{soep_june2024}. For comparison, my model's predicted inflation rate is approx. 2.7\% for 2024 and $2.6\%$ for 2025.
    There is less publicly available data for expectations about the federal funds rate. The model-implied values, approx. $3.2\%$ for the end of 2025 and $2.9\%$ for the end of 2026, fall within the ranges of the June 2024 FOMC Summary of Economic Projections (SOEP), although only at the lower end for 2025 \citep[again, see][]{soep_june2024}.

\section{Data Construction}\label{app:data}

\subsection{Data for model exercise}

This Appendix describes the construction of the data used for the model exercises in Section \ref{sec:filtering}. The analysis uses the following aggregate variables, the data for which were again obtained from the Federal Reserve Economic Data (FRED) platform:
\begin{itemize}
    \item \underline{\textbf{Nominal Rate}}: The variable corresponds to the federal funds rate (FRED series FEDFUNDS).
    \item  \underline{\textbf{Inflation}}: (Gross) Inflation corresponds to the growth of the GDP deflator (GDPDEF) compared to the previous quarter.
    \item \underline{\textbf{Output}}: (Real) Output corresponds to the sum of the following variables divided by the GDP deflator and current population (CNP16OV):
    \begin{itemize}
        \item Personal Consumption Expenditures: Non-durable Goods (PCND)
        \item Personal Consumption Expenditures: Durable Goods (PCDG)
        \item Personal Consumption Expenditures: Services (PCESV)
        \item Gross Private Domestic Investment (GPDI)
        \item Government Consumption Expenditure and Gross Investment (GCE)
    \end{itemize}
    \item \underline{\textbf{Investment}}: Gross Private Domestic Investment (GPDI) divided by the GDP deflator and current population (CNP16OV).
    \item \underline{\textbf{Transfers}}: (Real) Transfer payments consist of the sum of the following variables divided by the GDP deflator and current population (CNP16OV).
    \begin{itemize}
        \item Federal government current transfer payments: Government social benefits to persons (B087RC1Q027SBEA)
        \item Federal government current transfer payments: Grants-in-aid to state and local governments (FGSL)
    \end{itemize}
    The definition of this variable follows \cite{bianchiEtAl2023}.
    \item \underline{\textbf{Hours worked:}} Nonfarm Business Sector Hours Worked for All Workers (HOANBS) divided by either the level of the civilian population (CNP16OV) or the civilian labor force (CLF16OV).
    \item \underline{\textbf{Labor Compensation:}} Compensation of Employees (W209RC1) divided by the GDP deflator (GDPDEF) and current population (CNP16OV).
\end{itemize}
The pre-covid trends for Output, Government Consumption, Investment and Transfers are taken to be linear time trends estimated for the respective variables over the period 2014Q1 to 2019Q4.

Finally, for the comparison between model-implied and actual public debt, I use an approximation of the market value of treasury debt held by the domestic public: To the best of my knowledge, there is no publicly available breakdown of the market value of US treasury debt into domestic and foreign holdings throughout the entire period 2020Q1-2024Q2. 
Instead, I calculate a ``foreign share'' $s^f$ as Federal Debt Held by Foreign and International Investors (FDHBFIN) over Federal Debt held by the Public (FYGFDPUN) and then take the market value of domestically-held public debt to be $(1-s^f)$ times the Market Value of Marketable Treasury Debt (MVMTD027MNFRBDAL) minus Federal government checkable deposits and currency as reported in Federal Reserve's Financial Accounts of the United States (FL313020005.Q): The latter would reduce the governments net liquid asset supply from the perspective of the model.

Implicitly, the approximation being correct requires the treasury debt portfolios held by domestic and foreign agents to not differ systematically in terms of maturity structure etc.
The US \cite{treasury2024} reports foreign holdings to have weighted average maturity of 6.3 years, a bit but not overwhelmingly higher than the overall average.

\section{Additional Tables}\label{app:tables}

This Appendix contains the auxiliary Table \ref{tab:alt_calib} referred to in the main text. 

\begin{table}
    \centering
    \begin{tabular}{ c |  c c c c  }
        \hline \hline
     Rate gap  & $\beta$ & $\zeta$ & $\lambda$ & $\bar{R}$  \\
    \hline
    3.75\% (Baseline model)  &0.9836 & 0.0005 & 0.0404 & 0.0356  \\
     3.06\%   & 0.9837 & $0.0005$ &0.0511 & 0.0326 \\
     2.04\%  & 0.9838 & 0.0005 & 0.0708 & 0.02604\\
     1.03\%  & 0.9839 & 0.0004 & 0.1062 & 0.0171  \\
     \hline \hline
    \end{tabular}
    \caption{Alternative calibrations considered in Section \ref{subsec:heterogeneity_role}}\label{tab:alt_calib}
\end{table}

\section{Additional Figures}\label{app:figures}

This Appendix contains auxiliary Figures referred to in the main text or previous appendices.

\begin{figure}
    \centering
    \caption{Model IRFs to fiscal shock}
    \includegraphics[scale = 0.55]{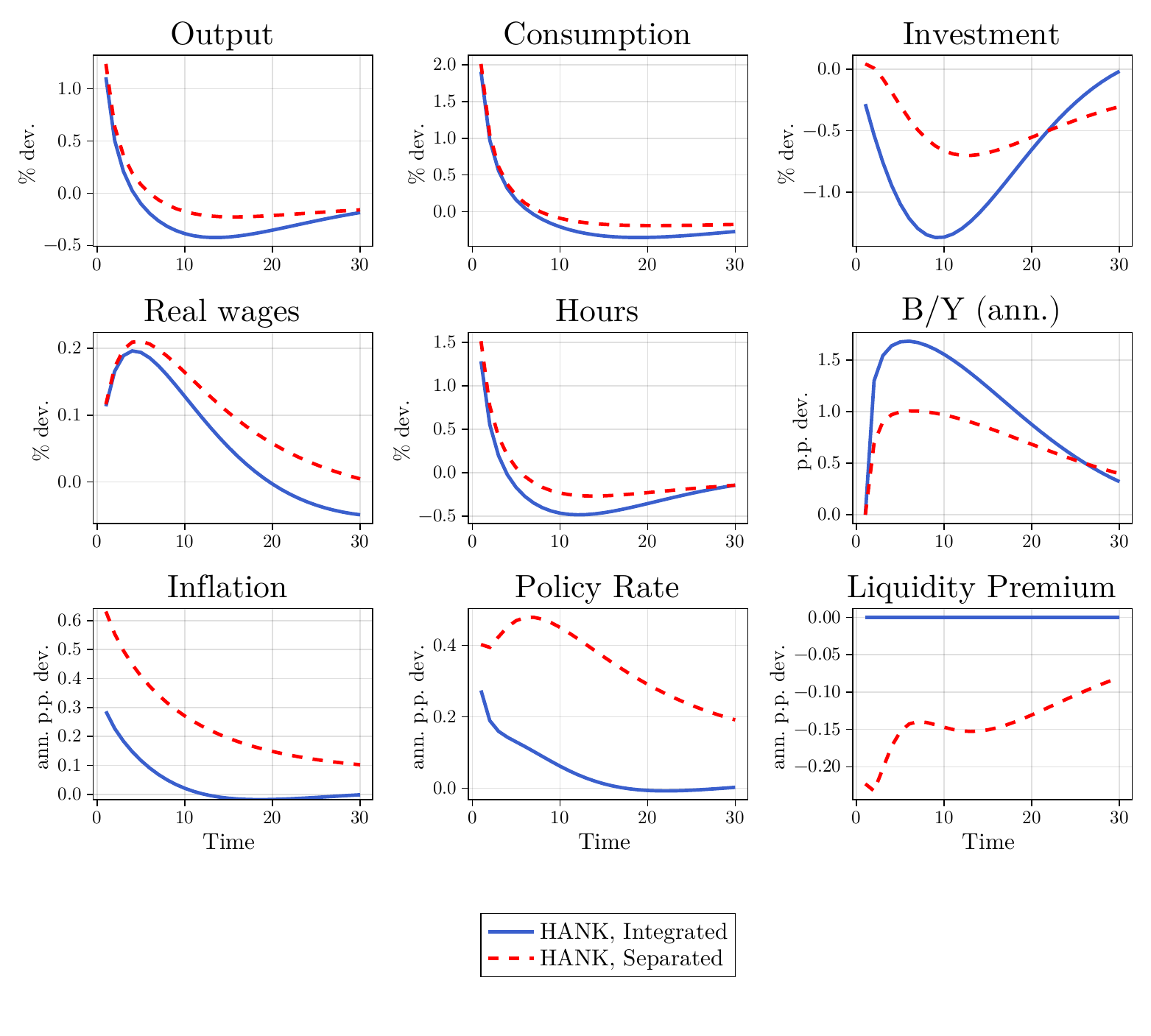}
    \fignote{$B/Y$ represents the real market value of public debt $B^g$ over 4 times (=annualized) GDP. The liquidity premium is defined as $\mathbb{E}_t \left(\frac{q_{t+1}+r_{t+1}^k}{q_{t}}-\frac{R^r_t}{\pi_{t+1}}\right)$}
    \label{fig:T_shock9}
 \end{figure}

 \begin{figure}
    \centering
    \caption{Model IRFs to Transfer shocks: $G$ adjusts}
    \includegraphics[scale = 0.6]{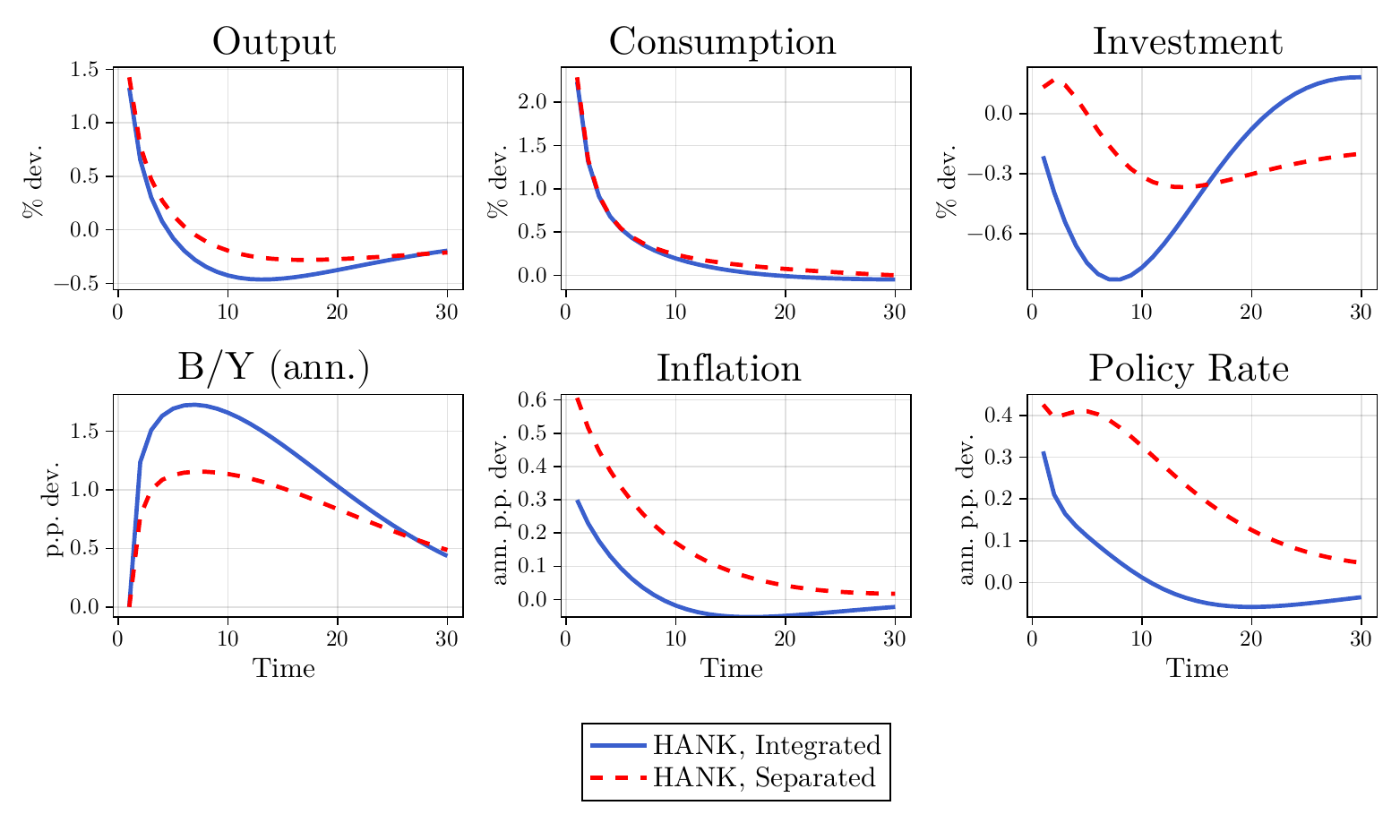}
    \fignote{$B/Y$ represents the real market value of public debt $B^g$ over annualized GDP. Figures display relative (in \%) or percentage point (p.p.) deviations from Steady State. Instead of adjusting taxes, the government is assumed to consolidate its finance by following the rule $\log G_t =  \rho_G\log G_{t-1} + (1-\rho_G)(\log G_{SS} - 0.85*(\log B^g_t- \log B_{SS}))$ with $\rho_G = 0.9$.
    }
    \label{fig:G_adj}
 \end{figure}


  \begin{figure}
    \centering
    \caption{IRFs to transfer shock: Sticky Expectations}
    \includegraphics[scale = 0.55]{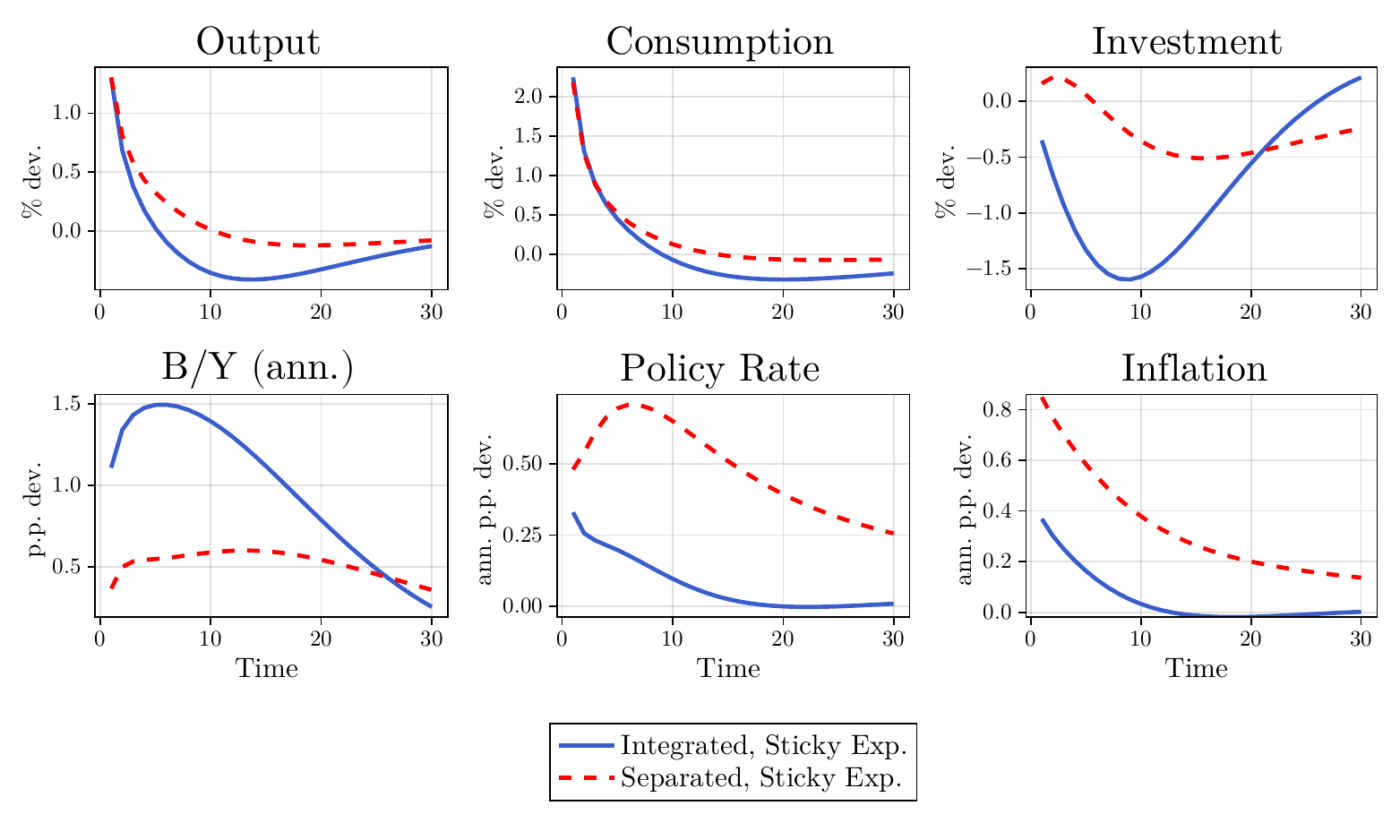}
    \fignote{ $B/Y$ represents the real market value of public debt $B^g$ over 4 times (=annualized) GDP.}
    \label{fig:T_sticky}
 \end{figure}

\begin{figure}
    \centering
    \caption{Post-2020 exercise: Aggregate dynamics using filtered shocks}
    \includegraphics[scale = 0.55]{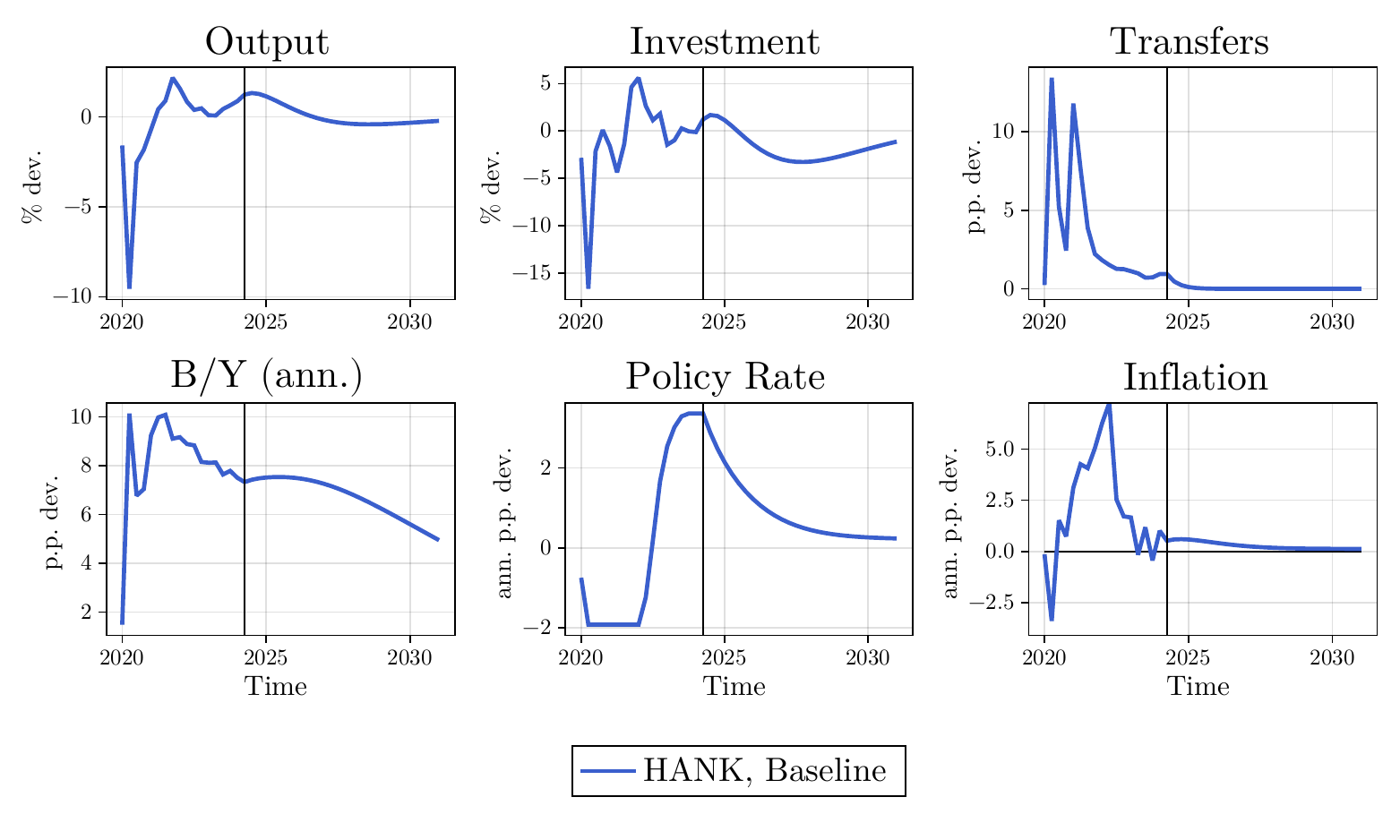}
    \fignote{$B/Y$ represents the market value of public debt $B^g$ over annualized GDP. From 2024:Q2 forward, the economy is no longer hit by shocks but just simulated forward.}
    \label{fig:postcov_long}
 \end{figure}

\end{document}